\documentclass[acmlarge]{acmart}
\AtBeginDocument{%
  \providecommand\BibTeX{{%
    \normalfont B\kern-0.5em{\scshape i\kern-0.25em b}\kern-0.8em\TeX}}}

\setcopyright{acmcopyright}
\copyrightyear{2023}
\acmYear{2023}
\acmDOI{XXXXXXX.XXXXXXX}

\acmJournal{POMACS}
\acmVolume{42}
\acmNumber{1}
\acmArticle{1}
\acmMonth{12}

\usepackage{multirow}
\usepackage{adjustbox}
\usepackage{float}

\begin{document}

\title{Exploring the Effects of User-Agent and User-Designer Similarity in Virtual Human Design to Promote Mental Health Intentions for College Students}

\author{Pedro Guillermo Feijóo-García}
\email{pfeijoogarcia@gatech.edu}
\orcid{0000-0002-3024-1257}
\affiliation{%
  \institution{Georgia Institute of Technology}
  \city{Atlanta}
  \state{Georgia}
  \country{USA}
}

\author{Chase Wrenn}
\email{chasewrenn@ufl.edu}
\orcid{0000-0002-5684-1349}
\affiliation{%
  \institution{University of Florida}
  \city{Gainesville}
  \state{Florida}
  \country{USA}
}

\author{Alexandre Gomes de Siqueira}
\email{agomesdesiqueira@ufl.edu}
\orcid{0000-0002-9213-9602}
\affiliation{%
  \institution{University of Florida}
  \city{Gainesville}
  \state{Florida}
  \country{USA}
}

\author{Rashi Ghosh}
\email{rashighosh@ufl.edu}
\orcid{0000-0002-7297-9911}
\affiliation{%
  \institution{University of Florida}
  \city{Gainesville}
  \state{Florida}
  \country{USA}
}

\author{Jacob Stuart}
\email{jpstuar@emory.edu}
\orcid{0000-0003-2103-5782}
\affiliation{%
  \institution{Emory University}
  \city{Gainesville}
  \state{Florida}
  \country{USA}
}

\author{Heng Yao}
\email{hengyao1993@ufl.edu}
\orcid{0000-0002-7830-4114}
\affiliation{%
  \institution{University of Florida}
  \city{Gainesville}
  \state{Florida}
  \country{USA}
}

\author{Benjamin Lok}
\email{lok@cise.ufl.edu}
\orcid{0000-0002-1190-3729}
\affiliation{%
  \institution{University of Florida}
  \city{Gainesville}
  \state{Florida}
  \country{USA}
}

\renewcommand{\shortauthors}{Feijóo-García, et al.}

\begin{abstract}
Virtual humans (i.e., embodied conversational agents) have the potential to support college students' mental health, particularly in Science, Technology, Engineering, and Mathematics (STEM) fields where students are at a heightened risk of mental disorders such as anxiety and depression. A comprehensive understanding of students, considering their cultural characteristics, experiences, and expectations, is crucial for creating timely and effective virtual human interventions. To this end, we conducted a user study with 481 computer science students from a major university in North America, exploring how they co-designed virtual humans to support mental health conversations for students \textit{similar to them}. Our findings suggest that computer science students who engage in co-design processes of virtual humans tend to create agents that closely resemble them demographically--agent-designer demographic similarity. Key factors influencing virtual human design included age, gender, ethnicity, and the matching between appearance and voice. We also observed that the demographic characteristics of virtual human designers, especially ethnicity and gender, tend to be associated with those of the virtual humans they designed. Finally, we provide insights concerning the impact of user-designer demographic similarity in virtual humans' effectiveness in promoting mental health conversations when designers' characteristics are shared explicitly or implicitly. Understanding how virtual humans' characteristics serve users' experiences in mental wellness conversations and the similarity-attraction effects between agents, users, and designers may help tailor virtual humans' design to enhance their acceptance and increase their counseling effectiveness.
\end{abstract}

\ccsdesc[500]{Computing methodologies~Intelligent agents}

\keywords{embodied conversational agents, virtual humans, mental health, similarity-attraction, participatory design}

\maketitle

\section{Introduction} \label{introduction}
This paper discusses the role of the similarity-attraction effect \cite{bernier2010} when users act as co-designers of virtual humans (i.e., embodied conversational agents), particularly its effects on their perceptions and intentions in sensitive areas like mental health counseling. We also delve into how demographic similarities between users, virtual humans, and designers influence computer science students' willingness to engage in mental health conversations. Virtual humans are computer-generated characters that resemble humans in appearance, behavior, and ability to communicate using natural language \cite{guadagno2007virtual}. They have been successfully employed in various scenarios, such as training medical students \cite{31,43} as well as providing mental health support \cite{30,43,33}. In scenarios of mental health support, the lack of rapport between counselors and patients can result in premature patient withdrawal \cite{hatcher2005, zlotnick1998}. Designing virtual humans for mental health support can be challenging, as these scenarios also require the development of rapport between users and virtual humans \cite{blow2008, wintersteen2005}. Therefore, understanding the users' identities, needs, and expectations when designing virtual humans to assist in mental health contexts is critical \cite{pedro, iva2022, TAP} and can help in increasing rapport among users and agents \cite{lucasDisclosure}.

Undergraduate college students have been found to struggle with mental health problems such as anxiety \cite{brazil1, brazil2}, depression \cite{brazil2}, as well as the impostor phenomenon \cite{porter}: i.e., highly-accomplished individuals perceiving themselves as frauds \cite{clance1978imposter}. Similarly, research has reported on graduate students experiencing up to six times higher rates of mental health problems compared to individuals in their same age range \cite{graduateHealth}. Within college populations, students who major in Science, Technology, Engineering, and Mathematics (STEM) are at higher risk for mental health problems \cite{brazil1}. For instance, undergraduate engineering students experience higher rates of anxiety and depression than general undergraduate students \cite{brazil3}, and among all STEM majors, \textbf{computer science  students} have been found to experience the highest rates of anxiety and depression \cite{brazil3}, with female computer science students being particularly more at risk \cite{porter, brazil1}.

Prior research in social psychology has examined the similarity-attraction effect \cite{bernier2010} in counselor-patient matches in mental health contexts, focusing on personality \cite{67, 70,71}, attitude \cite{68,69}, gender \cite{72, 74}, race \cite{73}, and ethnicity \cite{73,74}. Studies have also explored this effect in virtual human design, considering aspects like gender \cite{72, 74, TAP}, ethnicity \cite{86, 89, TAP}, and language \cite{pedro, iva2022, TAP}. For example, accent similarity influences computer science students' perceptions of virtual humans' expertise \cite{pedro} and conversational abilities \cite{iva2022}. However, research on virtual human design related to agent-user similarity is limited. For instance, as of 2018, only four papers from the Association of Computing Machinery (ACM) International Conference on Intelligent Virtual Agents (IVA) addressed agents' identities concerning cultural aspects \cite{36}.

Literature on the designer as a product cue has reported its advantages in various markets and products: Customers' perceptions of products can be shaped by the information they have about their creative actors and the design processes behind them \cite{DaSilva, Fuchs}. As a result, having the designer as a product cue could improve customers' perceptions of a product's overall quality \cite{idemen}. Nonetheless, research on virtual human design has not yet explored the designer's role as a product cue and the extent to which virtual humans' characteristics reflect the creative actors behind their designs--i.e., the designer's legacy. Also, although the designer as a product cue has been used in recent years as an extrinsic cue to promote products in various markets (e.g., "designed by") \cite{idemen}, it remains unclear how extrinsic this cue may be in the context of virtual human design and to what degree end users expect virtual humans to resemble their designers. Therefore, this paper reports on the influence of the designer's legacy (i.e., designer-agent demographic similarity) on virtual humans created to encourage computer science students' intentions toward mental health conversations. Our work explored how computer science students perceived designers' legacies (i.e., designer-agent demographic similarity) on virtual humans, when asked to describe the designers behind the virtual humans they interact with. We aimed to determine how designers' characteristics were perceived and manifested through the virtual human as an interactive interface, and to understand the similarity-attraction effect \cite{bernier2010} by examining whether students' motivation and interest in a virtual human, when students acted as co-designers, increased when the agent's designer was demographically more similar to the end user student.

Our research builds upon the research conducted by Feijóo-García et al. \cite{TAP}, who explored how computer science students co-designed virtual humans to support mental wellness conversations for target users \textit{similar to them}. Co-design has proven effective in enhancing user adoption \cite{good, muller, participatoryDesignLiteratureReview}, improving user input \cite{wobbrock, participatoryDesignLiteratureReview}, and empowering user communities by giving them more control over solutions that impact their lives \cite{good, vasalou, simonsen}. Feijóo-García et al. \cite{TAP} report on students leaning toward user-agent (or agent-designer in this case) demographic similarity in regard to gender, age, ethnicity, and their places of origin when they participate in virtual human co-design. Moreover, the authors provided insights on how the interplay between the virtual human's appearance and voice impacted computer science students' intentions toward gratitude journaling \cite{TAP}. 

The findings by Feijóo-García et al. \cite{TAP} contributed to the understanding of the similarity-attraction effect in co-design scenarios of virtual humans, especially in regard to user-agent demographic similarity. However, their work had limitations concerning the size of their sample population (their first user study had a total of 73 computer science students acting as co-designers of virtual agents) and reported missing opportunities concerning the exploration of user rationale in the selection of virtual humans' appearances and voices, as well as how both characteristics interplay concerned designers' demographic similarity expectations \cite{TAP}. Therefore, continuing their efforts, our user study went beyond and far extended their research with the aim of better understanding how computer science students co-designed virtual humans to support. Our research had a larger and more representative sample population (481 computer science students) and extended the protocol used by Feijóo-García et al. \cite{TAP} to include participants' rationale behind the selection of the agents' demographic characteristics, as well as their appearances and voices, to analyze the interplay between appearance, voice, and agents' desired and perceived demographics.

The research we report in this paper involved a two-part user study that extended the work by Feijóo-García et al. \cite{TAP}. The first part of this user study investigated how virtual humans designed by computer science students were influenced by demographic similarity expectations--students willing their agents to be like them, reporting insights on the interplay among virtual humans' characteristics, appearances, and voices, when co-designed by computer science students to support someone similar to them in their mental health. The second part of this user study explored whether end users' attraction to virtual humans increased with a designer more demographically similar to them when designers' characteristics were implicitly or explicitly used as product cues. Additionally, the second part of our user study explored how end users perceived designers' demographic characteristics through their agents, reporting insights on end users' expectations of agent-designer demographic similarity. 

Our research aimed to respond to the following research questions:

\begin{itemize}
    \item\textbf{RQ-``Similarity Elucidation"}: \textit{How do similarity expectations influence the design decisions made by computer science  students when co-designing virtual humans for mental health support?}
    \item\textbf{RQ-``Designer Cues"}: \textit{Which characteristics of a virtual human's designer can be identified by computer science  students during their interactions with the agent?} 
    \item\textbf{RQ-``Designer's Legacy Effectiveness"}: \textit{How effective are virtual humans in promoting computer science   students’ intentions toward mental wellness conversations when co-designed by computer science students with similar demographic characteristics to their end users?} 
\end{itemize}

\section{Related Literature}

\subsection{The Designer as a Product Cue}
Prior research in marketing has reported the impacts of having the designer as a product cue. That is, to allow end consumers to know who is behind the products they use \cite{idemen}. Studies have reported on how customers' perceptions of products are shaped by the information they have about their creative actors and the design processes behind them \cite{DaSilva, Fuchs}. Fuchs et al. \cite{Fuchs} conducted a set of studies that compared handmade and machine-made products, evaluating how the stated mode of production impacted products' attractiveness. The authors found that consumers indicated stronger purchase intentions and paid more for handmade products when buying gifts for their loved ones, suggesting that customers had a special appreciation for the human factor in production. Similarly, Da Silva et al. \cite{DaSilva} conducted a set of studies to evaluate how the knowledge of designers' intentions (or rationale) impacted users' appreciation of products. The authors argued that a product could present designers' intentions through the product itself or via marketing and advertisement. Their findings suggest customers' appreciation of a product can increase when they know the designers' intentions. However, customers' appreciation may decrease when the revealed intentions mismatch customers' initial perceptions of the product. Designers must identify and decide what intentions should be shared and how to the end users.

Having the designer as a product cue can increase how customers perceive the overall quality of a product \cite{idemen}. Idemen et al. \cite{idemen} conducted a study to identify customers' insights concerning the use of the designer as a product cue. Their work was interview-based and used a discovery-oriented grounded theory approach. The authors' found a designer cue can affect how customers perceive a product's quality, as information about the designer and the design process can reveal the effort and intellectual background behind a product. The authors present that designer cues can also signal product quality and support customers' reflective level of the experiences they expect with the products. Likewise, their findings suggest designer cues can signal a product's social value and status and expose the design process and involvement behind a product.

Research has found customers care more about the creative actors behind products in fashion \cite{fashion}, music \cite{music}, cinema \cite{cinema}, art \cite{art1, art2}, and wine-making industries \cite{wine}. Wu et al. \cite{wu}  found customers evaluate products more positively when they know about the process and effort invested in their designs \cite{wu}. The authors also found that customers appreciated highly aesthetic products more, as these products indicated more effort in their design--these findings are similar to those by Idemen et al. \cite{idemen}. Nevertheless, research on virtual human design has yet to explore the role of the designer as a product cue, as literature about this topic is still missing. Also, research is needed to understand how virtual humans' characteristics reflect the creative actors behind their designs--i.e., \textbf{the designer's legacy}, and the impact of the designer's legacy on users who interact with virtual humans in sensitive contexts such as mental health.

\subsection {User Participation in Virtual Human Design}
Several studies on virtual human design have adopted diverse methods in shaping the conversational design of agents. In the context of virtual human patients,  approaches have gone from using persona-based design \cite{whittaker2021} to characterizing agents' dialogues based on their narrative, the problems they aim to address \cite{bearman2001}, as well as promoting agents' credibility by using corpus-based design for question-answering scenarios \cite{halan2014}.

Previous research has also actively involved users in the design of virtual humans. Halan et al. \cite{halan2015} investigated how racial similarity impacted the design of virtual human patients in the scenario of promoting empathetic abilities among medical trainees. They had participants design and interview the virtual human patients. Remarkably, they found a notable difference in empathy ratings between groups with user-agent racial concordance and discordance when the virtual patient was Afro/Black American and the user Caucacian, suggesting that the effect of creating a virtual human patient could be more pronounced when there is user-agent racial discordance. Similarly, Ghosh et al. \cite{rashi} explored how gender is modeled and designed in virtual humans with computer science students engaging in their co-design process. Their findings suggest that gender cues in virtual human faces, such as age, hair, makeup, and facial features, are often stereotypically applied for binary genders (male and female) and inconsistently used for non-binary virtual humans. This inconsistency suggests a lack of clear guidelines or norms for representing non-binary individuals in virtual human design, highlighting the need for more inclusive and varied representations of gender in virtual environments \cite{rashi}.

A recent study by Feijóo-García et al. \cite{TAP} had computer science  students co-design virtual humans to assist students \textit{similar to them} in their mental well-being. Their research found that computer science students leaned toward creating agents closely resembling their own demographic characteristics, with Black, Indigenous, and People of Color (BIPOC) participants preferring user-agent ethnic similarity and female participants preferring user-agent gender similarity. Age was also found critical, as it was found that participants negatively perceived a virtual human whose age exceeded their expected maximum age. Additionally, Feijóo-García et al. reported on how the interplay between a virtual agent's appearance and voice could impact the agent's effectiveness in promoting computer science students' intentions toward gratitude journaling \cite{TAP}. These findings relate to those by Yin et al. \cite{bickmoreLinguistics}, who assessed the interplay between virtual human appearance ("Latin American"--dark-skinned or "Anglo-American"--light-skinned)   and language (English or Spanish) and its effectiveness on user persuasion. As they report, "Many subjects mentioned that the agent's language strongly influenced their perception of the agent's cultural background" \cite{bickmoreLinguistics}.

Our user study builds upon the foundational work of Feijóo-García et al. \cite{TAP}, delving into the dynamics of user-agent similarity in the design of virtual humans, particularly in scenarios where users partake in the co-creation process--agent-designer demographic similarity in this case. By investigating how computer science students co-design virtual agents that reflect aspects of their own demographics - such as gender, ethnicity, and age - our research contributes to the existing body of knowledge on the application of virtual humans in mental health support scenarios, particularly in regard college students needs. Our research not only echoes the findings of Feijóo-García et al. \cite{TAP} about the tendency of designers to create agents resembling themselves but also amplifies the understanding of how these virtual entities can be optimally designed to foster mental wellness conversations among college student populations.

\subsection{The Similarity-Attraction Effect on Virtual Human Design}
Previous research on virtual human design has used different angles to understand the impact characteristics such as gender, culture, and language have on users' interactions with computer systems such as virtual humans.

\subsubsection{Gender similarity} \label{HCI Gender}
Research in human-computer interaction reveals that gender significantly shapes user preferences towards interactive systems like robots and virtual humans. Studies, such as those by Kulms et al. \cite{19}, and Stal et al. \cite{RaganGender}, often show that female virtual agents are viewed more positively than male agents. However, the picture is more complex, as research by Guadagno et al. \cite{90}, and Kim and Baylor \cite{91} indicates a general preference for virtual humans of the same gender as the user, with perceptions varying based on both the user's and the agent's gender. Despite some positive views of female agents, Baylor et al. \cite{86} observed that they are often perceived as less knowledgeable than male agents. Regarding gender recognition, Nag and Yalçın \cite{40} found a tendency to categorize androgynous virtual agents as male.

Expanding on these findings, the idea of user-agent similarity-attraction based on gender is clearly visible in this field. Lee et al. \cite{84} reported a preference among users for systems that match their own gender, though male voices were generally favored, reflecting societal stereotypes. This intersection of culture and gender is further highlighted in Koda and Takeda's \cite{88} study, where both gender and cultural gaze behaviors were found to influence Japanese participants' views on virtual agents. Finally, Baylor et al. \cite{89} noted in an educational context that male students preferred male pedagogical agents, emphasizing gender's significant influence in virtual learning environments.

\subsubsection{Race and ethnicity similarity} \label{HCI Race}
Research has extensively explored the roles of race, culture, and ethnicity in human-computer interaction, particularly in the design of virtual agents and interfaces \cite{18, 85, 86, 17, 89, 61, 20, 38, 88}. Koda and Takeda \cite{88} found that Japanese participants' impressions of virtual agents varied based on cultural familiarity with the agents' gaze behaviors. Lucas et al. \cite{38} reported that cultural background (Japanese or American) influenced rapport-building with agents, particularly in responses to ice-breakers and dialogue errors. Wang et al. \cite{61}, applying Hofstede's six-dimension model \cite{80}, observed cultural differences in reactions to agents' feedback, with American users showing more resistance to negative feedback.

User studies on race and ethnicity in educational contexts \cite{86, 89} revealed that the agents' ethnicity influenced participants' learning outcomes. For instance, Baylor and Yanghee \cite{86} found better learning transfer with Black (i.e., dark-skinned) virtual humans compared to White (i.e., light-skinned or Caucasian) ones. Iacobelli and Cassell \cite{17} observed that ethnicity recognition in virtual agents could occur through language use, as African American children identified and preferred agents speaking African American Vernacular English. Pratt et al. \cite{85} and Baylor et al. \cite{89} also found that students preferred agents of their own ethnicity, indicating a similarity-attraction effect.

However, despite these insights, only a few studies have directly addressed the similarity-attraction effect and its impact on user interaction with culturally relevant virtual humans \cite{17,85,89}. Further research is necessary to fully comprehend this aspect of human-computer interaction, particularly for virtual human design.

\subsubsection{Language similarity} \label{HCI Language}
Language is critical in human-computer interaction (HCI), influencing user preferences and perceptions in various domains. In social robotics, language plays a crucial role in creating culturally relevant systems \cite{11, 2, 9}, with humor modeling \cite{2} enhancing users' positive perceptions. Cultural adaptation, based on models like Hofstede's \cite{80}, is also significant in human-robot interaction (HRI) \cite{11}. Additionally, studies show users prefer robots with similar personalities to their own \cite{9}.

For speech assistants, research indicates a preference for human voices over synthetic ones \cite{14}, with multiple factors like gender, age, accent, and vocal characteristics influencing user interaction \cite{13}. Technical limitations still exist in adequately addressing varied speech patterns \cite{15}. Parviainen and Søndergaard \cite{3} examined how whispering affects voice assistant experiences, highlighting the importance of voice tone and proxemics. Braun et al. \cite{6} found personalized voice assistants enhance trust, particularly when matching the user's personality.

In virtual human research, Iacobelli and Cassell \cite{17} used language to model non-visual ethnicity, demonstrating that African American participants could identify agent ethnicity based on language use, with a preference for agents that matched their own ethnic background. Studies on spoken interfaces like Dahlbäck et al. \cite{76} reveal a similarity-attraction effect, with users preferring accents similar to their own.

Despite these findings, there's limited research on the impact of language and its properties, such as accent, on virtual human design, particularly concerning the similarity-attraction effect. Our work extends the work by Feijóo-García et al. \cite{TAP} and aims to explore language's role in user engagement with virtual humans, particularly in sensitive contexts like mental health. This paper reports on how users as co-designers of virtual humans influence perceptions and intentions toward coping techniques such as gratitude journaling \cite{gratitude}.

\subsection{Counselor-Patient Matching in Mental Health Scenarios}
The principle of similarity-attraction states that people prefer to interact with others they perceive as similar to themselves \cite{bernier2010}. Previous research in social psychology has reported on the effects of similarity-attraction concerning mental health setups among patients and therapists, considering multiple factors such as 1) personality \cite{67, 70,71}, 2) attitude \cite{68,69}, 3) gender \cite{72, 74}, 3) race \cite{73}, and 4) ethnicity \cite{73,74}.

\subsubsection{Counselor-patient ethnic matching} \label{ethnic matching}
Cabral and Smith \cite{73} conducted a meta-analysis of over 80 references about patients' preferences towards counselors regarding their gender and ethnicity. They reported that the enhancement (i.e., increasing the number of common characteristics) of counselor-patient similarity had no impact on mental health treatment. Treatment outcomes were not found to be different when patients experienced a process with a therapist who shared their same race or ethnicity. However, the authors argued that it could be due to biases according to interpersonal-similarity beliefs: an assumption of more remarkable similarity than the actual one. Their analysis suggests that patients preferred having a therapist with the same race and ethnicity: racial and ethnic matching could initiate counseling. The authors found that once patients had initiated the therapeutic treatment, race, and ethnicity became secondary and did not generally impact how much benefit they received. However, the authors also found interesting insights concerning African Americans. Different from other ethnic and racial groups, African American patients strongly preferred African American counselors.

\subsubsection{Counselor-patient gender matching} \label{gender matching}
Concerning gender and ethnicity with Asian participants, Zane and Ku \cite{74} studied whether ethnic and gender similarity impacted how patients self-disclosed (about feelings, habits, and sexual issues) with counselors. Their population consisted of 110 Asian American students from a North-American college. The authors did subgroup analysis on different Asian ethnic subgroups: characterized as Chinese Americans (45\%), Japanese Americans (18\%), Korean Americans (26\%), and Vietnamese Americans (11\%). The authors' methodology asked participants to listen to an audio recording of a counselor, which could display the characteristics of an Asian male or female or a White (i.e., Caucasian or of European origin) male or female. The study reports a similar position to Cabral and Smith \cite{73}: Ethnic matching did not affect patients' self-disclosure. However, gender matching led patients to self-disclose more.

Gender was also explored by Bathi \cite{72}, who followed a naturalistic (or in-situ) methodology and hypothesized gender to be meaningless towards therapeutic alliance once treatment was initiated. Their study assessed 92 adults diagnosed with a mental disorder. Participants were primarily White Americans (93.5\%) and females (76.1\%). Participants were assigned to counselors that matched or did not their gender: 49 pairs matched, and 43 did not. The study \cite{72} was conducted for 15 weeks, and data was collected every three to five sessions. Gender similarity was found to impact therapeutic alliance between patients and counselors positively. Moreover, the authors found that female patients matched with female counselors reported higher therapeutic alliance ratings than male counselor-patient pairs.

\subsubsection{Counselor-patient attitude matching} \label{attitude matching}
Social psychologists have also explored the similarity-attraction effect concerning attitude and interpersonal agreement \cite{68,69,71}. Clore and Baldridge \cite{68} reported results from one study designed to evaluate interpersonal attraction. Their study assessed 84 participants from a North-American college who were asked to respond to a 56-item attitude survey that included 4-point scales that measured participants' interest per item. Then, the researchers randomly distributed participants' anonymously answered surveys, asking them to judge their peers' responses based on criteria such as intelligence, knowledge, morality, and how likely they would like to pair with their matching peers. The authors found that participants could be attracted to a stranger when they happened to agree more \cite{68}.

Similarly, Condon and Crano \cite{69} evaluated how attitude-attraction between individuals varied on how they assessed others. They recruited 260 first-year undergraduate students from a North American college. They asked them to answer a questionnaire that collected their ratings on religion, political party, premarital sex, etc. Then, the authors randomly assigned participants' anonymously responded surveys among them, asking participants to judge their peers' interests. As Clore and Baldrige \cite{68} report, Condon and Crano \cite{69} found Agreement to affect how individuals felt attracted to each other ---these findings can contribute to the design of virtual humans that provide feedback to users \cite{1, 5}. Fang and Kenrick \cite{71} evaluated similarity-attraction and dissimilarity-repulsion concerning group membership---e.g., belonging to the same political party. Two of their studies considered attraction and repulsion on 1) political affiliation (Democrat or Republican) and 2) sexual orientation (Heterosexual or Homosexual). The authors found that participants experienced a higher attraction towards groups they shared similarities with. Additionally, they found that groups that participants knew were different fostered higher participants' repulsion.

\subsubsection{Counselor-patient personality matching} \label{personality matching}
Personality has also been considered when evaluating the similarity-attraction effect in counselor-patient setups \cite{67, 70}. Byrne et al. \cite{67} assessed 450 participants---male and female undergraduates from a North American college, asking them to respond to the Repression-Sensitization (R-S) Scale \cite{52} to measure their personality. As done in studies for attitude-similarity \cite{68,69}, Byrne et al. \cite{67} asked participants to respond to a survey that collected their interests on specific topics. Participants then had to judge their peers' responses, agreeing or disagreeing with them. Their results suggested that personality correlated with attitude, as participants were more attracted to individuals with similar personality attitudes. Similarly, Sing \cite{70} evaluated the role of affective states concerning similarity-attraction, assessing 143 male and 127 female undergraduate first-year students from a North American college. Their personality was measured with the scale proposed by Byrne \cite{52}, and participants also had to judge their peers based on the limited information provided by an already filled survey \cite{70}. The authors' findings also suggest that personality-similarity increases the attraction between individuals, as also attitude-similarity does \cite{70}.

\section{Study Design, Part 1: Co-Design of Virtual Humans for Mental Health Support} \label{W3:part1}

The user study took place in the Spring of 2023. It was conducted online and asynchronously, and consisted of two distinct parts. Both parts were separated by a period of over four days, allowing sufficient time to create tailored virtual humans for the second part of the study. The user study was approved by the Institutional Review Board (IRB) prior to its execution (Protocol \#17817).  

\begin{figure}[!]
 \centering 
 \includegraphics[scale=0.40]{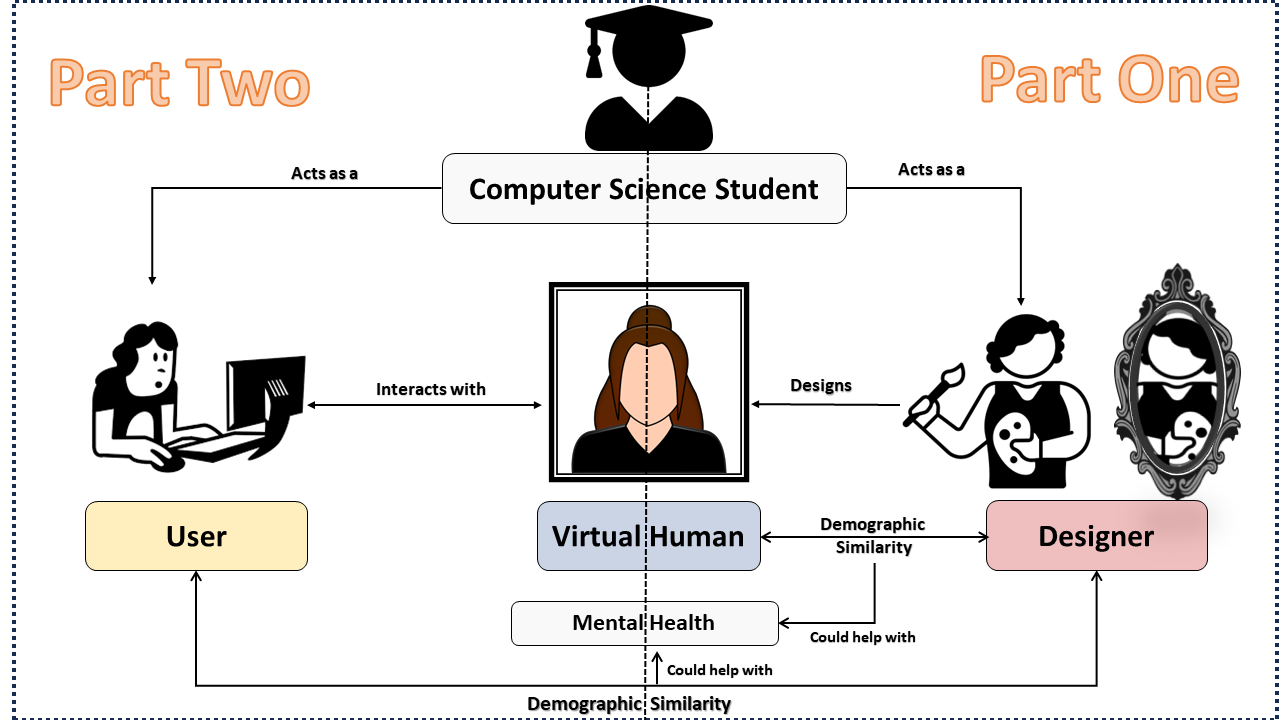}
 \caption{\centering Two-Part User Study Overview: User-Designer-Agent Demographic Similarity in Mental Health}
 \label{Work3_Part1_Similarity}
\end{figure}

The first part of this user study helped explore designer-agent demographic similarity when computer science students engaged in virtual human co-design intending to design virtual humans to assist toward mental health scenarios (Figure \ref{Work3_Part1_Similarity}). This first part addressed the following research question:

\textbf{RQ-``Similarity Elucidation"}: \textit{How do similarity expectations influence the design decisions made by computer science students when co-designing virtual humans for mental health support?}

\subsection{Participants} \label{W3:Part 1 Participants}
The first part of this user study had a sample population that consisted of adult graduate and undergraduate computer science students from the University of Florida. A total of 481 adult (18 years of age or older) participants were recruited regardless of gender, age, language, ethnicity, or place of origin. Participants were enrolled in the Spring of 2023 and came from a diverse range of courses: two graduate-level Human-Centered Computing courses, as well as six undergraduate computer science courses, including two freshman, one sophomore, and three senior computer science courses.

\begin{table}[!]\centering
\caption{\centering\label{W3:participantsDemographics}Ethnic Groups per Gender}
\begin{tabular}{|llccc|}
\hline
\multicolumn{5}{|c|}{\textit{\textbf{Ethnic Groups per Gender}}} \\ \hline
\multicolumn{1}{|l|}{\multirow{2}{*}{\textit{Ethnic Group}}} & \multicolumn{1}{l|}{\multirow{2}{*}{\textit{Total}}} & \multicolumn{3}{c|}{\textit{Gender}} \\ \cline{3-5} 
\multicolumn{1}{|l|}{} & \multicolumn{1}{l|}{} & \multicolumn{1}{c|}{\textit{Female}} & \multicolumn{1}{c|}{\textit{Male}} & \textit{Other} \\ \hline
\multicolumn{1}{|l|}{\textit{Asian}} & \multicolumn{1}{c|}{\textit{n=217}} & \multicolumn{1}{c|}{\textit{n=94}} & \multicolumn{1}{c|}{\textit{n=118}} & \textit{n=5} \\ \hline
\multicolumn{1}{|l|}{\textit{Afro/Black American}} & \multicolumn{1}{c|}{\textit{n=28}} & \multicolumn{1}{c|}{\textit{n=11}} & \multicolumn{1}{c|}{\textit{n=15}} & \textit{n=2} \\ \hline
\multicolumn{1}{|l|}{\textit{Latin American}} & \multicolumn{1}{c|}{\textit{n=100}} & \multicolumn{1}{c|}{\textit{n=38}} & \multicolumn{1}{c|}{\textit{n=57}} & \textit{n=5} \\ \hline
\multicolumn{1}{|l|}{\textit{Middle-Eastern/North-African}} & \multicolumn{1}{l|}{\textit{n=14}} & \multicolumn{1}{l|}{\textit{n=4}} & \multicolumn{1}{l|}{\textit{n=9}} & \multicolumn{1}{l|}{\textit{n=1}} \\ \hline
\multicolumn{1}{|l|}{\textit{Native American/Alaska Native}} & \multicolumn{1}{l|}{\textit{n=2}} & \multicolumn{1}{l|}{\textit{n=1}} & \multicolumn{1}{l|}{\textit{n=1}} & \multicolumn{1}{l|}{\textit{n=0}} \\ \hline
\multicolumn{1}{|l|}{\textit{Pacific Islander}} & \multicolumn{1}{l|}{\textit{n=8}} & \multicolumn{1}{l|}{\textit{n=3}} & \multicolumn{1}{l|}{\textit{n=4}} & \multicolumn{1}{l|}{\textit{n=1}} \\ \hline
\multicolumn{1}{|l|}{\textit{White (non-Latin American)}} & \multicolumn{1}{c|}{\textit{n=202}} & \multicolumn{1}{c|}{\textit{n=63}} & \multicolumn{1}{c|}{\textit{n=126}} & \textit{n=13} \\ \hline
\end{tabular}
\end{table}

Participants’ reported ages were: 18-20 years of age (n=339), 21-25 years of age (n=119), 26-30 years of age (n=8), and over 30 years of age (n=13). Two participants (n=2) did not disclose their age. Participants also reported the number of languages they were familiar with--monolingual (n=197) or multilingual (n=284), and their place of origin--country and city. Participants reported being from 34 different countries: from the United States (n=354) or other places (n=127)--see Appendix \ref{study1:additional}, Table \ref{W3:Places}. Participants also reported being domestic (n=402) or international (n=79) students.

Participants were also asked to report their gender and ethnicity. The gender distribution among participants included female (n=182), male (n=281), and other (n=18)—e.g., gender-queer, non-binary, transgender. Table \ref{W3:participantsDemographics} presents the participants' ethnic groups categorized by their gender. Each row should be considered independently, as there was a group of multi-ethnic participants (n=76) who selected more than one ethnic group.

\begin{figure}[!]
 \centering 
 \includegraphics[scale=0.50]{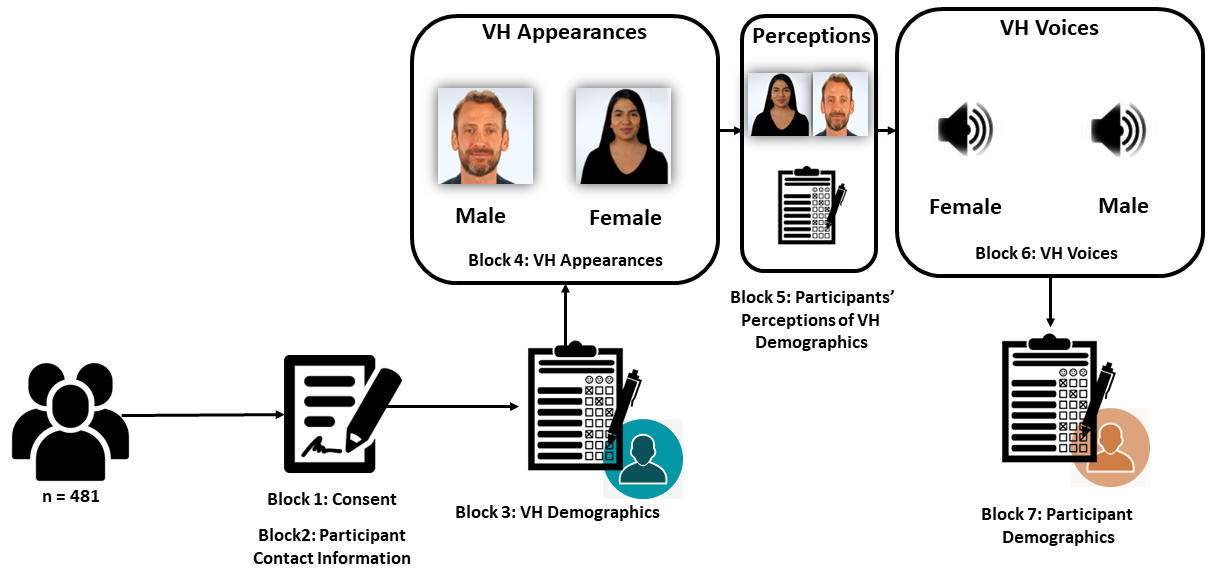}
 \caption{\centering Part 1: Procedure}
 \label{fig:W3Part1}
\end{figure}

\subsection{Procedure} \label{W3 Part 1:procedure}
The first part of the user study was conducted online and asynchronously, and took up to 30 minutes for participants to complete it. Participants were asked to respond to an online questionnaire, in which they were tasked with designing a mental health virtual human to assist someone \textit{"similar to them"}. Figure \ref{fig:W3Part1} illustrates the procedure of the first part of this user study. Questions were organized into seven blocks:

\textbf{Block 1-\textit{Informed Consent}} provided an informed consent document with details of the study and two multiple-choice questions. Participants were asked to read it and respond 1) if they agreed to participate and 2) if they were 18 years of age or older. Participants were informed and agreed to share institutional emails, instructors' names, and courses' names.

\textbf{Block 2-\textit{Participant's Information}} included three questions that asked participants to input their institutional email, institutional emails, instructors' names, and courses' names.

\textbf{Block 3-\textit{Virtual Human's Demographics}} asked participants to describe the virtual human they wanted to design. The block included four questions concerning the virtual human's demographic characteristics: age (open-ended numeric input), ethnicity (multiple-choice close-ended question based on \cite{pedro}), place of origin (open-ended input as country-city), and gender (close-ended multiple-choice question based on \cite{gender}).

\textbf{Block 4-\textit{Virtual Human's Appearance and Voice}} focused on the appearance of the virtual human, asking participants to respond to six questions to select a male and a female appearance. The catalog of virtual humans featured 12 male and 15 female appearances from agents available in \textit{Synthesia} \cite{synthesia} that could be used with the Synthesia API \cite{synthesiaAPI}. The first three questions focused on selecting a male appearance. The first question asked participants to select the top three male appearances they considered would encourage students similar to them to engage in mental wellness conversations. The second question asked participants to select the voice considered most suited from the previous three selected voices for the top male they decided before. Finally, the last question asked them to explain their choice. The last three questions of this block followed the same format to select a female appearance.

\textbf{Block 5-\textit{Participant's Perceptions of the Virtual Humans' Demographics}} presented four close-ended questions. Participants were first asked to indicate the perceived age (open-ended numeric input) and ethnicity (multiple-choice close-ended question based on \cite{pedro}) for the top male appearance they selected. Next, they did so for the top female appearance they chose. 

\textbf{Block 6-\textit{Virtual Human's Voice}} centered on the virtual human's voice and presented six questions for participants to select a male and a female voice. The catalog of voices featured 12 male and 12 female voices available in \textit{Synthesia} \cite{synthesia} that could be used with the Synthesia API \cite{synthesiaAPI}--see Table \ref{W3:maleVoicesTable} for male voices, and Table \ref{W3:femaleVoicesTable}- for female voices. All voices featured a five-second utterance (to limit mental load) and were labeled with their nationality (e.g., Canada). The first three questions focused on selecting a female voice. The first question asked participants to select the top four female voices they considered would encourage students similar to them to engage in mental wellness conversations. The second question asked them to select the top choice from the previous four selected options. Finally, the last question asked them to explain their top choice. The last three questions of this block followed the same format to select a male voice.

\textbf{Block 7-\textit{Participants' Demographics}} focused on participants' demographics, presenting six questions. First, it asked about participants' age (open-ended numeric input), ethnicity (multiple-choice close-ended question based on \cite{pedro}), and gender (close-ended multiple-choice question based on \cite{gender}). It followed by asking about participants' international student status (USA domestic or international) and their place of origin (open-ended input as country-city). Finally, it asked one open-ended question to have participants list all languages used to communicate.

\subsection{Data analysis} \label{W3 Part 1:analysis}

Data from the first part of this user study was analyzed with a focus on agent-designer demographic similarity, comparing the characteristics of the virtual humans selected by the participants to the participants' own demographic characteristics reported after participating in the first part of the user study. The analysis considered gender, ethnicity, age (with a similarity window of \textpm{5} years \cite{TAP}), and place of origin. We used the Chi-Squared Test categorical data to evaluate the statistical association between 1) participants' self-reported genders and the gender preferences for the virtual agents and 2) participants' self-reported places of origin and the places of origin preferred for the virtual agents concerning their countries. We used the Chi-Squared Test on these categorical responses as they corresponded to questions in which participants could only select one answer \cite{chisquare}.

The Pearson correlation coefficient was used to analyze user-agent age similarity, assessing the relationship between participants' self-reported ages and the preferred ages of their virtual agents. We used this statistical method on these responses as they were numerical and corresponded to questions in which participants could only select one answer \cite{pearson}.

We also analyzed the relationship between participants' self-reported ethnic groups and those ethnicities selected for their virtual agents by calculating the Jaccard similarity coefficient to assess the similarity among response groups. We decided to use this method due to the multi-label format of participants' responses, as participants could select more than one ethnic group for themselves, as well as multiple ethnic groups for their virtual agents \cite{jaccard}.

Moreover, the study examined how participants selected virtual humans' appearances and voices when prompted to choose options that would encourage "someone similar to them" to engage in mental wellness conversations. Top preferences on appearances and voices were analyzed using descriptive statistics, cross-referenced with participants' demographics. We used the Pearson correlation coefficient to compare participants' self-reported ages with the perceived ages of the agents' appearances they selected. Additionally, we calculated the Jaccard similarity coefficient to evaluate the relationship between participants' self-reported ethnic groups and the ethnic groups they perceived the agents' appearances they selected. Likewise, we used the Chi-Squared Test to evaluate the relationship between participants' places of origin and their voices' selection \cite{chisquare}. We centered our analysis on participants who indicated being originally from India (n=57) or the United States (n=354), as the catalog of female and male voices explicitly featured American and Indian-accented voices (see Appendix \ref{study1:additional}: Tables \ref{W3:maleVoicesTable} and \ref{W3:femaleVoicesTable}).

Finally, multi-label thematic analysis was employed to collect and examine participants' insights from two open-ended questions regarding their design rationale for selecting appearances and voices. Cohen's Kappa \cite{cohen1, cohen2} and the Jaccard similarity coefficient were used to assess the inter-rater reliability of the themes that came from the qualitative analysis \cite{jaccard}.

\subsection{Findings: Agent-designer matching of demographic characteristics}\label{W3:agentDesignerDemographicMatching}

Of the four characteristics matched between participants and virtual humans they co-designed, age was the only demographic characteristic for which agent-participant (i.e., agent-designer) matching was not the preferred choice among most participants when designing a virtual human to provide mental health assistance to someone similar to them. Three participants (n=3) did not report a counselor's age. Among the remaining participants who self-reported their age (n=476), 66.8\% of participants preferred an older virtual human (more than five years older), 0.4\% preferred a younger virtual human (less than five years younger), and 32.8\% designed a virtual human as old as they reported to be (having a window of \textpm{5} years of age for similarity): virtual humans' average age was of 28.6 years of age, reporting a maximum of 60 years of age. The Pearson correlation coefficient revealed a significant association between participants' self-reported ages and the virtual agents'  (\(r=0.257, n=476, p<0.01\)), suggesting a positive relationship between them. This correlation suggests that as the participant's age increases, the agent's age also tends to increase, which supports the existence of user-agent age similarity when participants design agents to support individuals similar to them in mental well-being.

For the other three demographic characteristics, most participants (n=481) made design decisions that matched their demographics with those of the agents they designed. Ethnic groups were alike among participants and their virtual agents, with a Jaccard index of 0.62 suggesting that more than 60\% of the ethnic categories were shared between them \cite{jaccard}. These observations highlight potential cultural influences on participants' preferences based on expectations of ethnic similarity. Likewise, the Chi-Squared test revealed a significant association between participants' self-reported genders and virtual agents' genders (\(\chi^{2}=293.4, df=4, p<0.01\)), as well as participants' self-reported countries of origin and virtual agents' countries of origin (\(\chi^{2}=5839.7, df=1054, p<0.01\)).

From our sample population (n=481), 61.1\% of participants decided on ethnic matching and 76.9\% on matching their country of origin: 24.1\% indicated the same city of origin. The sample population had male (n=281), female (n=182), and participants from other gender groups (n=18)--e.g., non-binary, transgender, etc. From the general population, 53.0\% of them preferred a female agent, 37.2\% a male one, and 9.8\% did so for an agent with another gender identity. See Appendix \ref{study1:additional}: Table \ref{W3:ethnicitiesAndGenders} shows preferences on gender concerning participants' ethnic groups.

Looking at gender similarity, 72.8\% of the general population decided on gender matching. Female participants (n=182) highly preferred a female agent: 90.1\% of the female population. Similarly, 83.3\% of participants who identified with other gender identities (n=18--e.g., non-binary, gender-queer, transgender) preferred an agent who gender-matched them. Male participants (n=281) also leaned toward a gender-matching agent, although not as pronounced as participants from other gender groups (61.9\% of males). Therefore, gender similarity is critical in the design of virtual humans to support non-male minorities in mental health.

These findings add to those from Feijóo-García et al., \cite{TAP}, indicating that when co-designing virtual humans to provide mental health assistance to someone similar to themselves, participants generally choose demographic characteristics for the virtual humans that match their own.

\subsection{Findings: Participants' rationale on virtual human appearance selection} \label{W3:AppearancesRationale}

Participants (n=481) were asked to provide information about their design decisions concerning virtual human appearance selections. Table \ref{table:appearanceCategories} displays the ten themes identified through the thematic analysis conducted on participants' criteria.

\begin{table}[!]\centering
\caption{\centering\label{table:appearanceCategories}Themes for Appearance Selection}
\begin{tabular}{|lll|}
\hline
\multicolumn{3}{|c|}{\textit{\textbf{Themes for Appearance Selection}}} \\ \hline
\multicolumn{1}{|c|}{\textit{\textbf{ID}}} & \multicolumn{1}{c|}{\textit{\textbf{Theme}}} & \multicolumn{1}{c|}{\textit{\textbf{Description}}} \\ \hline
\multicolumn{1}{|l|}{\textit{A1}} & \multicolumn{1}{l|}{\textit{Age}} & \textit{The participant considered age as a selection criterion.} \\ \hline
\multicolumn{1}{|l|}{\textit{A2}} & \multicolumn{1}{l|}{\textit{Approachability}} & \textit{\begin{tabular}[c]{@{}l@{}}The participant expressed their choice was due to \\ the agent's friendly, welcoming, and/or non-judgmental attitude.\end{tabular}} \\ \hline
\multicolumn{1}{|l|}{\textit{A3}} & \multicolumn{1}{l|}{\textit{Attire}} & \textit{The participant referred to the attire of the agent.} \\ \hline
\multicolumn{1}{|l|}{\textit{A4}} & \multicolumn{1}{l|}{\textit{Ethnicity}} & \textit{The participant considered ethnicity or race for their selection.} \\ \hline
\multicolumn{1}{|l|}{\textit{A5}} & \multicolumn{1}{l|}{\textit{Expertise}} & \textit{\begin{tabular}[c]{@{}l@{}}The participant referred to the perceived knowledge, \\ experience, or skill of the agent.\end{tabular}} \\ \hline
\multicolumn{1}{|l|}{\textit{A6}} & \multicolumn{1}{l|}{\textit{Gender}} & \textit{The participant considered gender as a selection criterion.} \\ \hline
\multicolumn{1}{|l|}{\textit{A7}} & \multicolumn{1}{l|}{\textit{Non-verbal Language}} & \textit{\begin{tabular}[c]{@{}l@{}}The participant referred to the facial expression \\ and/or body language of   the agent.\end{tabular}} \\ \hline
\multicolumn{1}{|l|}{\textit{A8}} & \multicolumn{1}{l|}{\textit{Random}} & \textit{\begin{tabular}[c]{@{}l@{}}The participant did not elaborate on their \\ rationale or provided a nonsensical response.\end{tabular}} \\ \hline
\multicolumn{1}{|l|}{\textit{A9}} & \multicolumn{1}{l|}{\textit{Self-similarity}} & \textit{\begin{tabular}[c]{@{}l@{}}The participant expressed their choice was \\ based on finding the agent similar to themselves.\end{tabular}} \\ \hline
\multicolumn{1}{|l|}{\textit{A10}} & \multicolumn{1}{l|}{\textit{Similarity to Trusted Agent}} & \textit{\begin{tabular}[c]{@{}l@{}}The participant's choice was based on a former \\ relationship and trust with a previous \\ familiar actor (e.g., a counselor, a family member).\end{tabular}} \\ \hline
\end{tabular}
\end{table}

Multiple labels were assigned to participants' responses based on the themes presented in Table \ref{table:appearanceCategories}. Inter-rater reliability was assessed separately for male and female appearance selections. To achieve this, two independent researchers coded 20\% of the responses \cite{cohen1}. Following this, Cohen's kappa \cite{cohen1} (for each theme) and Jaccard indexes \cite{jaccard} were calculated for both appearance selections: male and female--see Table \ref{appearance:Cohen} in Appendix \ref{study1:additional} for inter-rater reliability scores for male and female appearance selections. Inter-rater reliability was high for each theme, with Cohen's kappa scores exceeding 0.80 \cite{cohen2}. Additionally, multi-label inter-rater reliability was also high among raters, with Jaccard indexes surpassing 0.70 in both cases \cite{jaccard}.

\subsection{Findings: Preferences on encouraging virtual human male appearances} \label{W3:maleAppearances}
\begin{figure}[!]
 \centering 
 \includegraphics[scale=0.50]{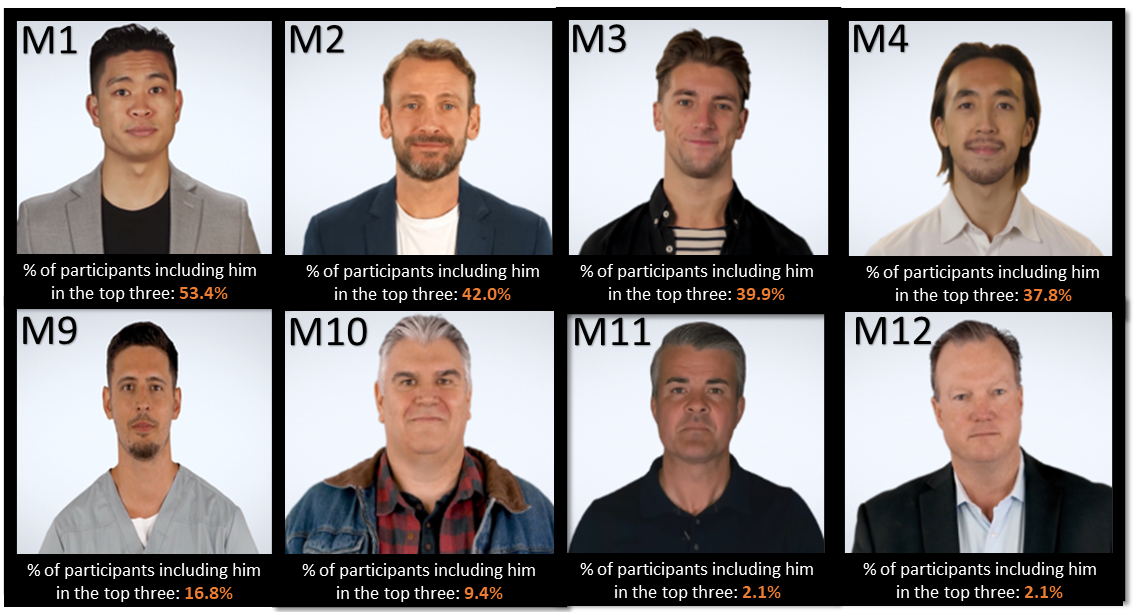}
 \caption{\centering Most and Least Selected Male Appearances}
 \label{fig:W3maleChoices}
\end{figure}

Participants (n=481) were asked to choose up to three male appearances based on their potential to encourage someone to engage in mental wellness conversations. Subsequently, they selected their top male appearance and answered two questions regarding their perceptions of the agent's age and ethnicity. Figure \ref{fig:W3maleChoices} displays the four male appearances most frequently included in participants' top three selections, as well as the four least selected male appearances. Figure \ref{fig:W3topmales} presents the appearances that participants most often chose as their top choice. The four appearances from Figure \ref{fig:W3maleChoices} were also the most preferred top choices, as shown in Figure \ref{fig:W3topmales}.

\begin{table}[H]\centering
\caption{\centering\label{table:maleRaces}Participants' Perceptions of Ethnicity and Age for Their Top Male Appearances}
\begin{tabular}{lllllll}
 &  &  &  &  &  &  \\ \cline{1-5}
\multicolumn{1}{|c|}{\textit{}} & \multicolumn{2}{c|}{\textit{\textbf{Perceived Age}}} & \multicolumn{2}{c|}{\textit{\textbf{Perceived Ethnicity}}} &  &  \\ \cline{1-5}
\multicolumn{1}{|c|}{\textit{\textbf{Agent}}} & \multicolumn{1}{c|}{\textit{\textbf{\begin{tabular}[c]{@{}c@{}}Average\\ {[}years of age{]}\end{tabular}}}} & \multicolumn{1}{c|}{\textit{\textbf{\begin{tabular}[c]{@{}c@{}}Standard Deviation\\ {[}years of age{]}\end{tabular}}}} & \multicolumn{1}{c|}{\textit{\textbf{Group}}} & \multicolumn{1}{c|}{\textit{\textbf{\begin{tabular}[c]{@{}c@{}}\% of responses\\ out of n\end{tabular}}}} &  &  \\ \cline{1-5}
\multicolumn{1}{|l|}{\textit{M1 (n=115)}} & \multicolumn{1}{c|}{\textit{30.3}} & \multicolumn{1}{c|}{\textit{3.6}} & \multicolumn{1}{l|}{\textit{Asian}} & \multicolumn{1}{c|}{\textit{96.5\%}} &  &  \\ \cline{1-5}
\multicolumn{1}{|l|}{\textit{M2 (n=70)}} & \multicolumn{1}{c|}{\textit{38.2}} & \multicolumn{1}{c|}{\textit{5.2}} & \multicolumn{1}{l|}{\textit{White (non-Latin American)}} & \multicolumn{1}{c|}{\textit{94.3\%}} &  &  \\ \cline{1-5}
\multicolumn{1}{|l|}{\textit{M3 (n=70)}} & \multicolumn{1}{c|}{\textit{25.5}} & \multicolumn{1}{c|}{\textit{3.0}} & \multicolumn{1}{l|}{\textit{White (non-Latin American)}} & \multicolumn{1}{c|}{\textit{92.9\%}} &  &  \\ \cline{1-5}
\multicolumn{1}{|l|}{\textit{M4 (n=67)}} & \multicolumn{1}{c|}{\textit{31.7}} & \multicolumn{1}{c|}{\textit{3.9}} & \multicolumn{1}{l|}{\textit{Asian}} & \multicolumn{1}{c|}{\textit{98.5\%}} &  &  \\ \cline{1-5}
\end{tabular}
\end{table}

\subsubsection{Perceptions of age and ethnicity of virtual human male appearances} \label{perceptionsMaleRace}
Using the Pearson correlation coefficient, we found that participants' self-reported ages (n=479, as two participants did not report their age) significantly correlated with the ages of the male agents they selected as their top-one choice (\(r=0.116, n=479, p<0.05\)). This positive correlation suggests the presence of user-agent age similarity when participants selected the appearances of their male virtual agents.

\begin{figure}[!]
 \centering 
 \includegraphics[scale=0.50]{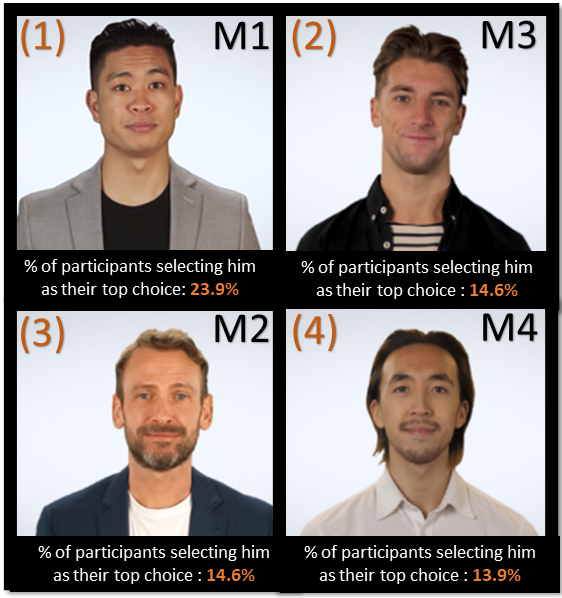}
 \caption{\centering Top Preferences of Male Appearances}
 \label{fig:W3topmales}
\end{figure}

Participants' perceptions of the four top male appearances' ages and ethnic groups are shown in Table \ref{table:maleRaces}. As observed in Table \ref{table:maleRaces}, the top agents were between 25 and 40 years of age, with the top two (M1 and M3) close to the general average age of 28.6 years previously reported (Section \ref{W3:agentDesignerDemographicMatching}). Concerning the ethnicity of the male appearances, two of the top four male appearances (Figure \ref{fig:W3topmales}) were perceived as White (non-Latin American)--M2 and M3, while the other two (M1 and M4) were considered Asian. 

Ethnic groups were moderately alike among participants and their male virtual agents, with a Jaccard index of 0.44 suggesting that more than 40\% of the ethnic categories were shared between them \cite{jaccard}. Upon examining the ethnic groups more closely, it was observed that participants' preferences leaned towards ethnic similarity. For example, among the Afro/Black American population (n=28), 75.0\% of them chose a top appearance they perceived as non-White, with 67.9\% selecting a male appearance perceived as  Afro/Black American. Similarly, 53.5\% of the White participants (n=202) chose an appearance they perceived as White (non-Latin American), 55.3\% of the Asian participants (n=217) favored an Asian-looking male appearance, and 62.5\% of the Pacific Islander participants (n=8) selected an ethnically matching appearance. Additionally--although not to be generalized due to the small sample population, both Native American/Alaska Native participants (n=2) chose a male appearance they perceived as matching their ethnicity. 

These observations suggest potential cultural influences on participants' preferences based on expectations of ethnic similarity. However, it's important to note that the Jaccard index concerning male appearances' selection is 20\% below the one reported in Section \ref{W3:agentDesignerDemographicMatching}. We consider this may be due to a limited catalog of male appearances. We encourage further research to explore options that provide an open canvas to help elucidate and observe participants' ethnic expectations when designing virtual humans to support individuals similar to them.

Of the seven ethnic groups considered, two of them did not tend toward ethnic similarity concerning male appearances. Among Latin American participants (n=100), only 33.0\% of them selected a male appearance they perceived as Latin American. Likewise, only 21.4\% of Middle Eastern/North African participants (n=14) leaned toward ethnically similar male appearances. As 93.0\% of the Latin American and 92.3\% of the Middle Eastern/North African population were domestic, further research is needed to understand if this phenomenon is due to limitations on the catalog of appearances or familiarity aspects regarding the under-representation of these ethnic groups in counseling scenarios in the United States--counselors in the United States are predominantly White (non-Latin American) \cite{demographics}.

\subsubsection{Participants' rationale on male appearance selection}  \label{W3:AppearancesMaleRationale}
Participants' preferences on male appearances were highly led by three of the ten themes used to label data (Section \ref{W3:AppearancesRationale}): \textit{Approachability}, \textit{Age}, and \textit{Self-similarity}. 
 \textit{Approachability} (see Table \ref{table:appearanceCategories} for detailed descriptions) was the most prevalent theme among ethnic groups, with participants indicating to perceive a welcoming or friendly look from the male appearance they selected: e.g., \textit{``The counselor looked friendly which will help me to share things without hesitation"} [MA028], \textit{``The only explanation I can give is that the person I selected that the most empathetic and caring look to them"} [MA208]. See Appendix \ref{study1:additional}, Table \ref{table:appearanceThemesPerEthnicity} for details on the thematic analysis per ethnic group for male appearances' selection.

Next to \textit{Approachability}, \textit{Age} prevailed in all ethnic groups but the Native American/Alaska Native one: it could have been due to the small number of participants from that group (n=2). Also, \textit{Self-similarity} was a top three theme among our Asian (n=217), Afro/Black American (n=28), Latin American (n=100), and Pacific Islander (n=8) participants. It is important to note that, out of the n=262 responses labeled under \textit{Age}, n=70 also referred to age-matching between the virtual human and the user, comprising 58.8\% of responses labeled with \textit{Self-similarity}: e.g., ``\textit{Looks closer to my age; can relate to people that are my age. Attaching a specific archetype to a virtual counselor is difficult}" [MA125], ``\textit{I feel like that virtual counselor is more close to my age which would help them relate to what I'm going through}" [MA280].

These findings suggest that the agent's age plays a crucial role in designing virtual humans tailored to assist computer science students with their mental health needs. Moreover, these findings indicate that user-agent similarity is an important factor for encouraging computer science  students to engage in mental health discussions with male virtual humans.

\subsection{Findings: Preferences on encouraging virtual human female appearances} \label{W3:femaleAppearances}
\begin{figure}[!]
 \centering 
 \includegraphics[scale=0.50]{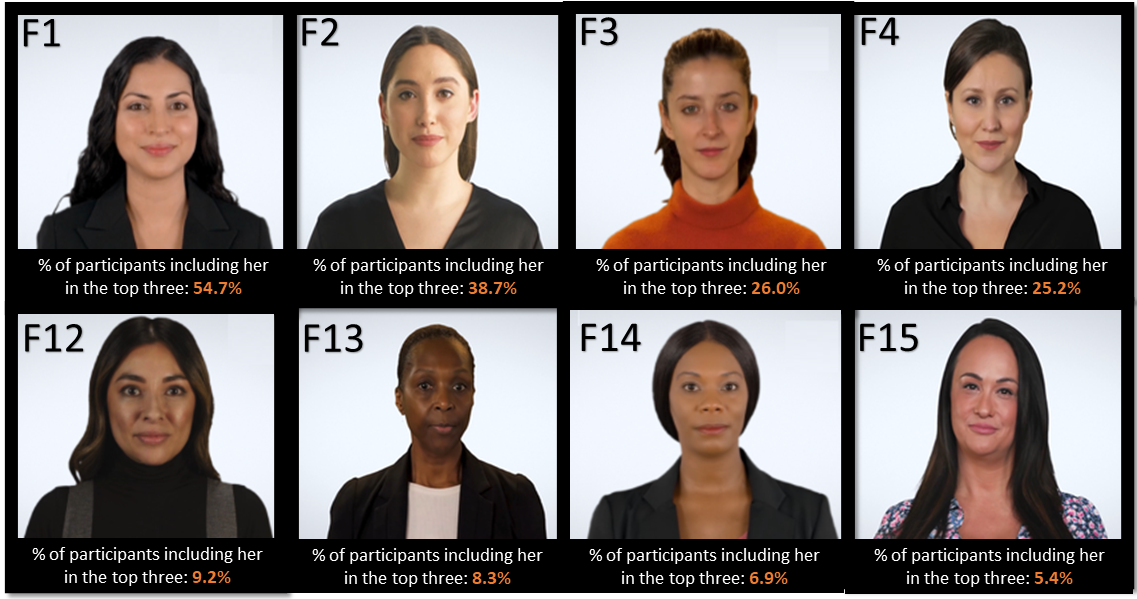}
 \caption{\centering Most and Least Selected Female Male Appearances}
 \label{fig:W3femaleChoices}
\end{figure}

Participants (n=481) were asked to choose up to three female appearances based on how they could encourage someone similar to them to engage in mental wellness conversations. Subsequently, they selected their top female appearance and answered two questions regarding their perceptions of the agent's age and ethnicity. Figure \ref{fig:W3femaleChoices} displays the four female appearances most frequently included in participants' top three selections, as well as the four least selected female appearances. Figure \ref{fig:W3topfemales} presents the appearances that participants most often chose as their top choice. Three of the four appearances from Figure \ref{fig:W3femaleChoices} were also mostly preferred as top one choices, as shown in Figure \ref{fig:W3topfemales}. Figure \ref{fig:W3topfemales} features a new female appearance (F7). She was the seventh most selected in participants' top three selections of female appearances, with 21.4\% of participants selecting her.

\subsubsection{Perceptions of age and ethnicity of virtual human female appearances} \label{perceptionsFemaleRace}
Using the Pearson correlation coefficient, we found that participants' self-reported ages (n=479, as two participants did not report their age) significantly correlated with the ages of the female agents they selected as their top-one choice (\(r=0.187, n=479, p<0.01\)). This positive correlation suggests the presence of user-agent age similarity when participants selected the appearances of their female virtual agents.

Participants' perceptions of the four top female appearances' ages and ethnic groups are shown in Table \ref{table:femaleRaces}. As observed in Table \ref{table:femaleRaces}, the top agents were between 26 and 35 years of age, with the top three (F1, F2, and F3) close to the general average age of 28.6 years previously reported (Section \ref{W3:agentDesignerDemographicMatching}). 

Ethnic groups were moderately alike among participants and their female virtual agents, with a Jaccard index of 0.33 suggesting that more than 33\% of the ethnic categories were shared between them \cite{jaccard}. This index was lower than the index calculated for male appearance selection (0.44).

\begin{figure}[!]
 \centering 
 \includegraphics[scale=0.50]{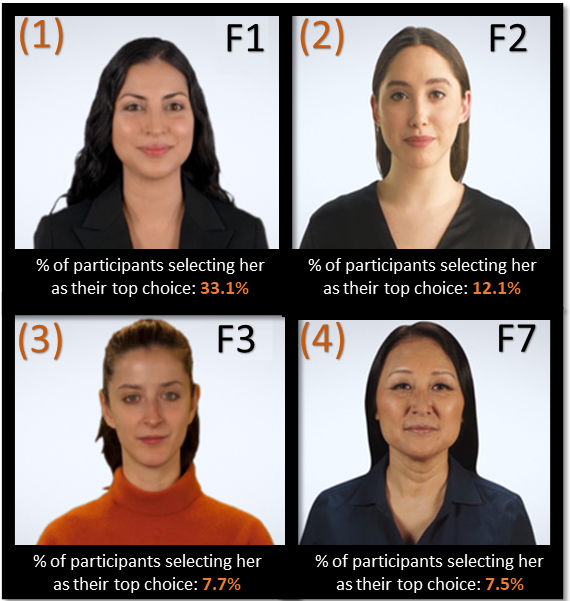}
 \caption{\centering Top Preferences of Female Appearances}
 \label{fig:W3topfemales}
\end{figure}

Concerning the ethnicity of the female appearances, two of the top four female appearances (Figure \ref{fig:W3topfemales}) were perceived as White (non-Latin American)--F2 and F3, while F1 was perceived as Latin American and F7 as Asian. Analyzing participants' preferences based on their ethnic groups, it was observed that ethnic matching was not a tendency as it was for male appearances. Although 82.1\% of the Afro/Black American population (n=28) selected a non-White female appearance, only a 32.1\% of them preferred an ethnic-matching female appearance. Similarly, 42.6\% of White (non-Latin American) participants (n=202), 14.3\% of Middle Eastern/North African participants (n=14), and 12.5\% of Pacific Islander participants (n=8) selected a female appearance that matched them ethnically. 

Comparing these findings to those from Section \ref{perceptionsMaleRace}, we note that the gender of the virtual human affected how much participants desired the agent to match them ethnically. This could be due to participants' general preference toward female agents (Section \ref{W3:agentDesignerDemographicMatching}). These findings reveal an interesting interplay between ethnicity and gender that should be addressed with further research.

\begin{table}[!]\centering
\caption{\centering\label{table:femaleRaces}Participants' Perceptions of Ethnicity and Age for Their Top Female Appearances}
\begin{tabular}{lllllll}
 &  &  &  &  &  &  \\ \cline{1-5}
\multicolumn{1}{|c|}{\textit{}} & \multicolumn{2}{c|}{\textit{\textbf{Perceived Age}}} & \multicolumn{2}{c|}{\textit{\textbf{Perceived Ethnicity}}} &  &  \\ \cline{1-5}
\multicolumn{1}{|c|}{\textit{\textbf{Agent}}} & \multicolumn{1}{c|}{\textit{\textbf{\begin{tabular}[c]{@{}c@{}}Average\\ {[}years of age{]}\end{tabular}}}} & \multicolumn{1}{c|}{\textit{\textbf{\begin{tabular}[c]{@{}c@{}}Standard Deviation\\ {[}years of age{]}\end{tabular}}}} & \multicolumn{1}{c|}{\textit{\textbf{Group}}} & \multicolumn{1}{c|}{\textit{\textbf{\begin{tabular}[c]{@{}c@{}}\% of responses\\ out of n\end{tabular}}}} &  &  \\ \cline{1-5}
\multicolumn{1}{|l|}{\textit{F1 (n=159)}} & \multicolumn{1}{c|}{\textit{27.2}} & \multicolumn{1}{c|}{\textit{3.3}} & \multicolumn{1}{l|}{\textit{Latin American}} & \multicolumn{1}{c|}{\textit{93.1\%}} &  &  \\ \cline{1-5}
\multicolumn{1}{|l|}{\textit{F2 (n=70)}} & \multicolumn{1}{c|}{\textit{27.1}} & \multicolumn{1}{c|}{\textit{3.0}} & \multicolumn{1}{l|}{\textit{White}} & \multicolumn{1}{c|}{\textit{82.8\%}} &  &  \\ \cline{1-5}
\multicolumn{1}{|l|}{\textit{F3 (n=70)}} & \multicolumn{1}{c|}{\textit{26.7}} & \multicolumn{1}{c|}{\textit{2.8}} & \multicolumn{1}{l|}{\textit{White}} & \multicolumn{1}{c|}{\textit{91.9\%}} &  &  \\ \cline{1-5}
\multicolumn{1}{|l|}{\textit{F7 (n=67)}} & \multicolumn{1}{c|}{\textit{34.2}} & \multicolumn{1}{c|}{\textit{9.4}} & \multicolumn{1}{l|}{\textit{Asian}} & \multicolumn{1}{c|}{\textit{86.1\%}} &  &  \\ \cline{1-5}
\end{tabular}
\end{table}

Among the other ethnic groups, Latin American participants (n=100) were those who mostly leaned toward ethnic similarity with their selected female appearances (77.0\%), followed by Asian participants (n=217): 55.3\% of them selected an Asian-looking female agent. Also, both Native American/Alaska Native participants (n=2) preferred a female appearance that matched them ethnically--although further research with a broader sample population is needed concerning this ethnic group. These findings suggest that the interplay of gender and ethnicity may affect user-agent expectations concerning ethnicity. Further research is needed to understand this phenomenon fully.

\subsubsection{Participants' rationale on female appearance selection}  \label{W3:femaleAppearancesRationale}
As it was for participants' preferences on male appearances (Table \ref{table:appearanceThemesPerEthnicity}), the \textit{Approachability} theme prevailed as the top criterion on participants' selections of female appearances. Table \ref{table:femaleAppearanceThemesPerEthnicity} shows \textit{Approachability} (see Table \ref{table:appearanceCategories} for detailed descriptions) being the most prevalent theme among ethnic groups, with participants indicating to perceive a welcoming or friendly look from the female appearance they selected: e.g., \textit{``The selected virtual counselor appears to be more approachable as she looks like someone who may be able to empathize with younger age groups and has an overall welcoming look to herself"} [FA038], \textit{``This counselor also appears to be laid-back, wanting to listen to my problems and help me improve"} [FA476].

Different from the rationale on the selection of male appearances (see Appendix \ref{study1:additional}--Table \ref{table:appearanceThemesPerEthnicity}), \textit{Self-similarity} did not showcase in the top three themes (see Appendix \ref{study1:additional}--Table \ref{table:femaleAppearanceThemesPerEthnicity}). Only the Latin American population (n=100) had it in third place, with 25.0\% of them referring to it in their responses. Nevertheless, age still was the second most relevant theme among ethnic groups, with a total of 254 responses (52.8\% of participants) including it as relevant: e.g., \textit{``I look for age in a counselor, someone who is more experienced, so that is why I chose that counselor"} [FA106], \textit{``she looked old enough to have been through some stuff so she could talk me through something but also not too old and had an inviting look"} [FA459].

Out of the responses referring to age (n=254), 21.7\% of them (55 responses) referred to \textit{Self-similarity}: this, compared to the 58.8\% reported for male appearance selection. While not as prevalent as with male selections, some participants expressed a preference for engaging with a female agent whose age was closer to their own: e.g.,\textit{ ``They look young which mean they would understand the things I might have to say"} [FA224], \textit{``She looks youngest among the counselors. So, it is easy to approach and share our problems with the people who are in the same age group"} [FA048].

These findings suggest that age is an essential factor when designing virtual humans to assist computer science students in mental health conversations, regardless of the preferred gender or ethnicity of the agent. However, participants' preferences on gender do seem to affect user-agent ethnic similarity, as \textit{Self-similarity} was less prevalent among participants' responses concerning female appearance selections. As reported earlier (Section \ref{W3:agentDesignerDemographicMatching}, and Table \ref{table:femaleRaces}), female virtual humans were generally preferred among ethnic groups. The selection of male virtual humans could have triggered participants to try to resemble agents ethnically; however, it was not the case for female agents. Further research is needed to better understand the interplay between gender and ethnicity concerning computer science students' gender preferences. These findings showcase an exciting phenomenon that may help virtual human designers better tailor agents in sensitive contexts such as mental health.

\subsection{Findings: Participants' rationale on virtual human voice selection} \label{W3:VoicesRationale}
Participants (n=481) were asked to provide information about their design decisions concerning virtual human voice selections. Table \ref{table:voiceCategories} displays the seven themes identified through the thematic analysis conducted on participants' criteria.

Participants' responses were assigned multiple labels based on the themes presented in Table \ref{table:voiceCategories}. Inter-rater reliability was assessed separately for male and female voice selections. To achieve this, two independent researchers coded 20\% of the responses \cite{cohen1}. Following this, Cohen's kappa \cite{cohen1} (for each theme) and Jaccard indexes \cite{jaccard} were calculated for both voice selections: male and female. Inter-rater reliability was high for each theme (see Appendix \ref{study1:additional}--Table \ref{voice:Cohen}), with Cohen's kappa scores exceeding 0.80 \cite{cohen2}. Additionally, multi-label inter-rater reliability was also high among raters, with Jaccard indexes surpassing 0.70 in both cases \cite{jaccard}.

\begin{table}[H]
\centering
\caption{\centering\label{table:voiceCategories}Themes for Voice Selection}
\begin{tabular}{|l|l|l|}
\hline
\multicolumn{1}{|c|}{\textit{\textbf{ID}}} & \multicolumn{1}{c|}{\textit{\textbf{Theme}}} & \multicolumn{1}{c|}{\textit{\textbf{Description}}} \\ \hline
\textit{V1} & \textit{Clarity} & \textit{\begin{tabular}[c]{@{}l@{}}The participant referred \\ to the clarity of the voice.\end{tabular}} \\ \hline
\textit{V2} & \textit{Comfort} & \textit{\begin{tabular}[c]{@{}l@{}}The participant indicated finding \\ the voice soothing, friendly, \\ and/or comforting.\end{tabular}} \\ \hline
\textit{V3} & \textit{Ethnicity} & \textit{\begin{tabular}[c]{@{}l@{}}The participant considered the ethnicity, \\ race, or perceived origin of the voice \\ for their selection.\end{tabular}} \\ \hline
\textit{V4} & \textit{Familiarity} & \textit{\begin{tabular}[c]{@{}l@{}}The participant expressed finding the \\ voice familiar to voices they were \\ used to listen.\end{tabular}} \\ \hline
\textit{V5} & \textit{Matching} & \textit{\begin{tabular}[c]{@{}l@{}}The participant made their selection \\ because the voice fit or matched the \\ appearance they previously selected \\ for the agent.\end{tabular}} \\ \hline
\textit{V6} & \textit{Random} & \textit{\begin{tabular}[c]{@{}l@{}}The participant did not elaborate on \\ their rationale or provided a \\ nonsensical response.\end{tabular}} \\ \hline
\textit{V7} & \textit{Self-similarity} & \textit{\begin{tabular}[c]{@{}l@{}}The participant expressed finding the \\ voice similar to theirs.\end{tabular}} \\ \hline
\end{tabular}
\end{table}

\subsection{Findings: Preferences on encouraging virtual human male voices} \label{W3:maleVoices}

Among participants' selections (n=481) on their top four male voices (see Appendix \ref{study1:additional}--Table \ref{W3:maleVoicesTable}), the two most selected voices were North-American: 62.0\% of participants selected Canada (CA\_M), and 55.9\% did so for United States (US\_M). These two voices are closely followed by the United Kingdom (GB\_M--55.1\%) and Australia (AU\_M--49.9\%) voices. North-American voices were still in the first two places concerning participants' top one male voice choices: 23.3\% of participants preferred the male voice from the United States (US\_M) while 21.2\% did so for the Canadian voices (CA\_M). However, instead of the British voice (GB\_M--9.1\%), the Philippine (PH\_M--14.6\%) voice was third among the most preferred male voices. Finally, the Australian voice (AU\_M--12.1\%) was the fourth most preferred male voice.

Similar to the findings reported by Feijóo-García et al. \cite{TAP}, there were voice preferences among Indian participants (n=57) based on user-agent matching regarding the place of origin. There was a significant positive correlation between Indian participants' place of origin and the selection of the female Indian voice (\(\chi^{2}=79.6, df=1, p<0.01\)): 63.2\% of Indian participants (n=57) selected the male Indian voice (IN\_M) in their top four most encouraging voices. Participants from the United States (n=354) were also observed to include the male American voice (US\_M) in their four most encouraging female voices (50.3\% of responses). Similarly, we found a significant correlation between American participants' place of origin and their selection of the male American voice (\(\chi^{2}=11.5, df=1, p<0.01\)). However, when asked to select the most encouraging voice among each participant's top four male voices, the male Indian voice (IN\_M) was not favored among Indian participants (n=57): Only 5.3\% of them preferred it as their top choice. It was the same for American participants (n=354), with only 21.5\% of them selecting the male American voice (US\_M) as the most preferred voice among the top four male voices each participant selected. 

Among the themes described in Table \ref{table:voiceCategories}, participants (n=481) generally indicated \textit{Comfort} (50.9\% of responses), \textit{Matching} (45.9\% of responses), and \textit{Ethnicity} (25.6\% of responses) as their top three criteria to select male voices: e.g., \textit{``Out of the voices I selected the Canada voice best fits my counselor. It sounds the least choppy out of the selection and is a bit deep as I imagine the vc having just by looking at the image. It also sounds confident and soothing which translates to non intimidating which I feel that other students would agree that would be an attraction point for a vc"} [MV009], `\textit{`I chose the voice because I think it suits the picture better and I can clearly understand what they are trying to say"} [MV420].

Regardless of the catalog of male voices (see Appendix \ref{study1:additional}--Table \ref{W3:maleVoicesTable}), there were no accent similarity expectations among international participants (n=79), as 53.2\% of them preferred a North-American voice: American (US\_M--31.6\%) or Canadian (CA\_M--21.5\%), and their third most preferred voice was the British one (GB\_M--17.7\%). Considering the themes identified in Table \ref{table:voiceCategories}, international participants (n=79) made their choices due to agent-voice \textit{Matching} (44.3\% of responses), and on voice \textit{Clarity} (32.9\% of responses) and \textit{Comfort} (49.4\%). Although 20.3\% of the international population (n=79) referred to the voice's \textit{Ethnicity}, \textit{Self-similarity} was not a prevalent criterion among them--8.9\% of them referred to it.

\textit{Self-similarity} neither was a prevalent criterion among the domestic population (n=402): Only 7.2\% of them referred to it. Although \textit{Comfort} (51.2\% of responses) and \textit{Matching} (46.3\% of responses) were still two of the themes mostly considered, \textit{Clarity} (22.6\% of responses) did not make it to the top three of themes. Instead, \textit{Ethnicity} appeared to be more relevant (26.6\% of responses) as participants referred to the ethnicity, race, or perceived place of origin of the voice to be matched with their preferred male appearance: e.g., \textit{``It matches his appearance and fits his nationality in the most natural way"} [MV068].

These findings add to those reported by Feijóo-García et al., \cite{TAP}, as it was found how male appearances affected the selection of male voices in virtual human design. As seen in Table \ref{W3:maleVoicesTable}, participants' choices varied when asked to match the male appearance they previously preferred. For instance, the 55.1\% of preference toward the British voice (GB\_M) became a 9.1\% when participants selected the most preferred male voice. In general, appearance-voice matching was an essential criterion when selecting the male voice, as voice mismatches could jeopardize a student's experience with a male virtual human \cite{TAP}.

\subsection{Findings: Preferences on encouraging virtual human female voices} \label{W3:femaleVoices}

Among participants' selections (n=481) on their top four female voices (see Appendix \ref{study1:additional}--Table \ref{W3:femaleVoicesTable}), the two most selected female voices were again North American: 60.5\% of participants selected the American voice (US\_F), and 52.2\% did so for the Canadian one (CA\_F). In contrast to the male voice selections (Section \ref{W3:maleVoices}), the Irish voice (IE\_F) emerged as the third most popular choice (49.9\% of participants), followed by the British voice (GB\_F) with 41.4\% of participants favoring it. Concerning the top female voice selections, only the American (27.2\% of participants) and Irish (18.7\% of participants) made it to the first three places. Both were followed by the Philippine voice (PH\_F--15.6\% of participants). Differently from male voice selections, neither British (GB\_F) nor Canadian (CA\_F) voices were preferred for female agents.

Similar to observations on participants' choices of male voices (Section \ref{W3:maleVoices}), Indian participants (n=57) leaned toward female voices that matched their place of origin (i.e., India) when selecting the four most encouraging female voices. There was a significant positive correlation between Indian participants' place of origin and the selection of the female Indian voice (\(\chi^{2}=94.5, df=1, p<0.01\)): 80.7\% of Indians (n=57) selected the female Indian voice (IN\_F) in their top four most encouraging voices. Participants from the United States (n=354) were also observed to include the female American voice (US\_F) in their four most encouraging female voices (56.2\% of responses). However, contrary to what we observed among Indian participants, there was not a significant correlation between American participants' place of origin and their selection of the female American voice (\(\chi^{2}=3.22, df=1, p=0.07\)). On a different direction, when asked to select the most encouraging voice among the top four female voices each participant selected, Indian participants (n=57) ended up selecting voices different from the female Indian voice (IN\_F): Only 29.8\% of them preferred it. It was the same for American participants (n=354), with only 24.9\% of them selecting the female American voice (US\_F) as the most preferred voice among the top four female voices each participant selected. 

Among the themes described in Table \ref{table:voiceCategories}, participants (n=481) generally indicated \textit{Comfort} (45.9\% of responses), \textit{Matching} (40.1\% of responses), and \textit{Ethnicity} (30.8\% of responses) as their top three criteria to select female voices. As observed among international participants (n=79) on their preferences of male voices (Section \ref{W3:maleVoices}), \textit{Self-similarity} was not a prevalent criterion for their selection of female voices: only 6.3\% of responses included it. On the contrary, \textit{Comfort} (48.1\% of responses) and \textit{Matching} (43.0\% of responses) were the top reasons behind participants selections: e.g., ``\textit{I believe this voice has a calming effect and it fits accordingly to the persona of the counselor i chose. Anything else will probably make me want to go to a different counselor}" [FV026], \textit{``it looks perfect to her and also it will helps in motivation"} [FV046].

Accent similarity expectations were not observed among international participants (n=79) when selecting their top-most encouraging female voices. Among participants from this group, 36.7\% preferred the American female voice (US\_F), while 20.3\% favored the Irish (IE\_F) voice. The Indian (IN\_F) voice ranked as the third most popular choice, with 17.7\% of international participants preferring it. This could be attributed to the presence of Indian participants (n=52) in this group, where 26.9\% ranked first the Indian female voice (IN\_F) and 36.7\% preferred the American voice (US\_F). Although \textit{Ethnicity} made it in the top three main themes (21.5\% of responses) for female voice selections among international participants (n=79), \textit{Self-Similarity} was their least relevant theme: 6.3\% referred to it. Hence, user-agent voice similarity was not observed and did not emerge as a general trend among international students (n=79) when selecting female voices. 

\textit{Self-similarity} was also not a prevalent criterion among domestic participants (n=402) in the selection of female voices: only 9.0\% of them referred to it. Instead, \textit{Comfort} (45.5\% of responses) and \textit{Matching} (39.6\% of responses) were the two themes mostly considered, followed by \textit{Ethnicity} (32.6\% of responses). These themes were predominant also in their selection of male voices (Section \ref{W3:maleVoices}): e.g., \textit{``It sounded genuine and friendly and went well with the image selected"} [FV164], \textit{``she looks like American, so that I thought that voice suits her and also the grace and elegance of the voice suits her"} [FV045]. 

These findings contribute those presented by Feijóo-García et al. \cite{TAP}, concerning the interplay between female appearances and the selection of female voices in virtual human design. As presented in Table \ref{W3:femaleVoicesTable}, participants' choices diverged when asked to match their previously preferred female appearance with a female voice. This is observed in the decision-making of participants from India (n=57) and the United States (n=354). Although selecting their regional voices in their top four most encouraging female voices (expressing user-agent similarity expectations), they ultimately chose the female voice that best matched their preferred female appearance--no user-agent similarity was observed.

\section{Study Design, Part 2: User-Designer and User-Agent Similarity in the Promotion of Mental Health Intentions} \label{W3:part2}

The second part of this user study followed a structure that addressed the following research questions:

\begin{itemize}
	\item\textbf{RQ-``Designer Cues"}: \textit{Which characteristics of a virtual human's designer can be identified by computer science students during their interactions with the agent?} 
     \item\textbf{RQ-``Designer's Legacy Effectiveness"}: \textit{How effective are virtual humans in promoting computer science students’ intentions toward mental wellness conversations when co-designed by computer science students with similar demographic characteristics to their end users?} 
\end{itemize}

 This part explored user-designer demographic similarity in a scenario where users interacted with the agents they designed. This was done to increase user-designer demographic similarity as much as possible. At least four days were given between the first and second parts of the user study to prevent participant biases concerning their own designs. Also, the agent-designer similarity was assessed as participants were asked to guess the demographic characteristics of the designer behind their agent (see Figure \ref{Work3_Part1_Similarity} for reference).

\subsection{Participants} \label{W3:Part 2 Participants}
The second part of this user study reached out to the same adult population of graduate and undergraduate computer science  students from the University of Florida (n=481). However, only 240 participants completed the second part of this user study. Participants were enrolled in the Spring of 2023. They came from a diverse range of courses: two graduate-level Human-Centered Computing courses, as well as six undergraduate courses, including two freshman, one sophomore, and three senior CS courses.

Participants' reported ages were: 18-20 years of age (n=161), 21-25 years of age (n=69), 26-30 years of age (n=5), and over 30 years of age (n=5). Participants also reported the number of languages they were familiar with--monolingual (n=101) or multilingual (n=139), and their place of origin--country and city. Participants reported being from the United States (n=164) or other places (n=76): India (n=36), China (n=5), Venezuela (n=5), and from other countries (n=16). Participants indicated being domestic (n=196) or international (n=44) students.

\begin{table}[!]\centering
\caption{\centering \label{W3:part2_GenderedDemographics} Ethnic Groups per Gender}
\begin{tabular}{|lcccc|}
\hline
\multicolumn{5}{|c|}{\textit{\textbf{Ethnic Groups per Gender}}} \\ \hline
\multicolumn{1}{|c|}{\multirow{2}{*}{\textit{\textbf{Ethnic Group}}}} & \multicolumn{1}{c|}{\multirow{2}{*}{\textit{\textbf{Total}}}} & \multicolumn{3}{c|}{\textit{\textbf{Gender}}} \\ \cline{3-5} 
\multicolumn{1}{|c|}{} & \multicolumn{1}{c|}{} & \multicolumn{1}{c|}{\textit{\textbf{Female}}} & \multicolumn{1}{c|}{\textit{\textbf{Male}}} & \textit{\textbf{Other}} \\ \hline
\multicolumn{1}{|l|}{\textit{Asian}} & \multicolumn{1}{c|}{\textit{n=107}} & \multicolumn{1}{c|}{\textit{n=44}} & \multicolumn{1}{c|}{\textit{n=61}} & \textit{n=2} \\ \hline
\multicolumn{1}{|l|}{\textit{Afro/Black American}} & \multicolumn{1}{c|}{\textit{n=11}} & \multicolumn{1}{c|}{\textit{n=4}} & \multicolumn{1}{c|}{\textit{n=6}} & \textit{n=1} \\ \hline
\multicolumn{1}{|l|}{\textit{Latin American}} & \multicolumn{1}{c|}{\textit{n=55}} & \multicolumn{1}{c|}{\textit{n=17}} & \multicolumn{1}{c|}{\textit{n=36}} & \textit{n=2} \\ \hline
\multicolumn{1}{|l|}{\textit{Middle Eastern/North-African}} & \multicolumn{1}{c|}{\textit{n=7}} & \multicolumn{1}{c|}{\textit{n=2}} & \multicolumn{1}{c|}{\textit{n=4}} & \textit{n=1} \\ \hline
\multicolumn{1}{|l|}{\textit{Native American/Alaska Native}} & \multicolumn{1}{c|}{\textit{n=1}} & \multicolumn{1}{c|}{\textit{n=0}} & \multicolumn{1}{c|}{\textit{n=0}} & \textit{n=1} \\ \hline
\multicolumn{1}{|l|}{\textit{Pacific Islander}} & \multicolumn{1}{c|}{\textit{n=1}} & \multicolumn{1}{c|}{\textit{n=0}} & \multicolumn{1}{c|}{\textit{n=0}} & \textit{n=1} \\ \hline
\multicolumn{1}{|l|}{\textit{White (non-Latin American)}} & \multicolumn{1}{c|}{\textit{n=90}} & \multicolumn{1}{c|}{\textit{n=32}} & \multicolumn{1}{c|}{\textit{n=52}} & \textit{n=6} \\ \hline
\end{tabular}
\end{table}

Participants were also asked to report their gender and ethnicity for the second part of this user study. The gender distribution among participants included female (n=88), male (n=144), and other (n=8)—e.g., gender-queer, non-binary, transgender. Table \ref{W3:part2_GenderedDemographics} presents the participants' ethnic groups categorized by their gender. Each row should be considered independently, as there was a group of multi-ethnic participants (n=66) who selected more than one ethnic group.

\begin{figure}[!]
 \centering 
 \includegraphics[scale=0.50]{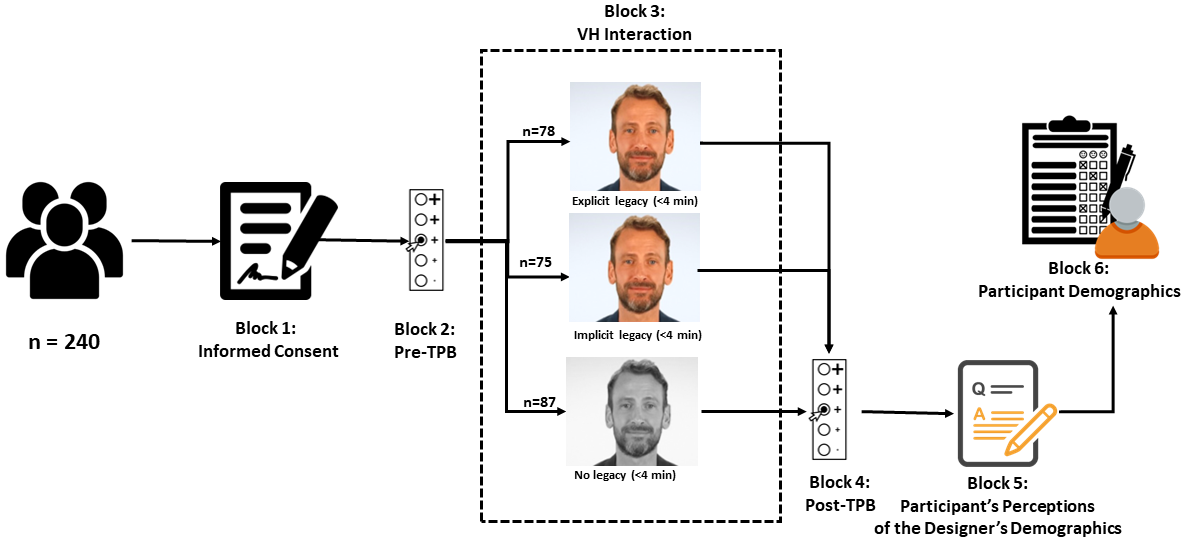}
 \caption{\centering Part 2: Procedure}
 \label{fig:W3Part2}
\end{figure}

\subsection{Procedure} \label{W3 Part 2:procedure}
The second part of this user study was conducted online, asynchronously, and was between-subjects. Participants did not take more than 30 minutes to complete it. The second part happened at least four days after participants completed the first part of the user study (Section \ref{W3:part1}).

\subsubsection{Experimental groups} \label{W3 P2: Experimental Groups}
First, participants were randomly assigned to one of three groups, each introducing one  virtual human designed to talk about gratitude journaling \cite{gratitude}. Gratitude journaling has been a mental health topic addressed in several studies concerning virtual humans design for mental well-being support \cite{TAP, pedro, mohan}.  For the following two groups, each participant interacted with the preferred of the two virtual humans (male or female) they designed in the first part (Section \ref{W3:part1}). This was done to maximize the similarity between the participant and the designer behind the virtual human--they were the designers. The two intervention groups were as follows:

\begin{itemize}
\item \textbf{Intervention group - ``Explicit legacy:"} The participant interacted with a virtual human designed by a computer science student whose demographic characteristics were similar to those of the participant. The virtual human explicitly described the demographic characteristics of the designer who created it: gender, age, place of origin, and ethnicity.

\item \textbf{Intervention group - ``Implicit legacy:"} The participant interacted with a virtual human designed by a fellow computer science student whose demographic characteristics were similar to those of the participant. The virtual human did not describe the designer who created it.

\end{itemize}

The third group was used as a control group, featuring a female virtual human whose characteristics were determined based on findings from work by Feijóo-García et al. \cite{TAP}, and participants' general preferences during the co-design process in part one (Section \ref{W3:agentDesignerDemographicMatching}). 

\begin{itemize}
\item \textbf{Control Group -``No legacy:"} The participant interacted with a generic virtual human not designed by a fellow computer science student and that responds to previous literature and previous findings on virtual human design.
\end{itemize}

\begin{figure}[!]
 \centering 
 \includegraphics[scale=1]{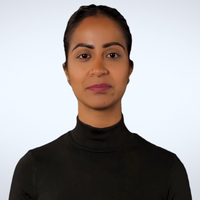}
 \caption{\centering Control Group - ``No Legacy": Virtual Human Appearance}
 \label{fig:nina}
\end{figure}

The virtual human was female because participants generally preferred a female agent: these preferences match what previous literature on virtual human genders has reported \cite{baylor, pedro, genderreview}. Based on Feijóo-García et al. \cite{TAP}, the virtual human featured the female agent ranked first in their research and a British-accented voice: the top preferred female voice from their research. Participants in the control group did not select any British-accented voice during part one (Section \ref{W3:part1}), nor selected the female appearance they interacted with in this second part (Figure \ref{fig:nina}). Therefore, the control group featured a virtual human whose appearance and voice have been reported to be engaging and whose gender responds to good practices on virtual human design. This is to compare the effectiveness of the designer's legacy in scenarios where participants interact with virtual humans designed by someone who shares their own demographics.

\subsubsection{Virtual humans generation} \label{W3 P2: VH Creation}
\begin{figure}[!]
 \centering 
 \includegraphics[scale=0.50]{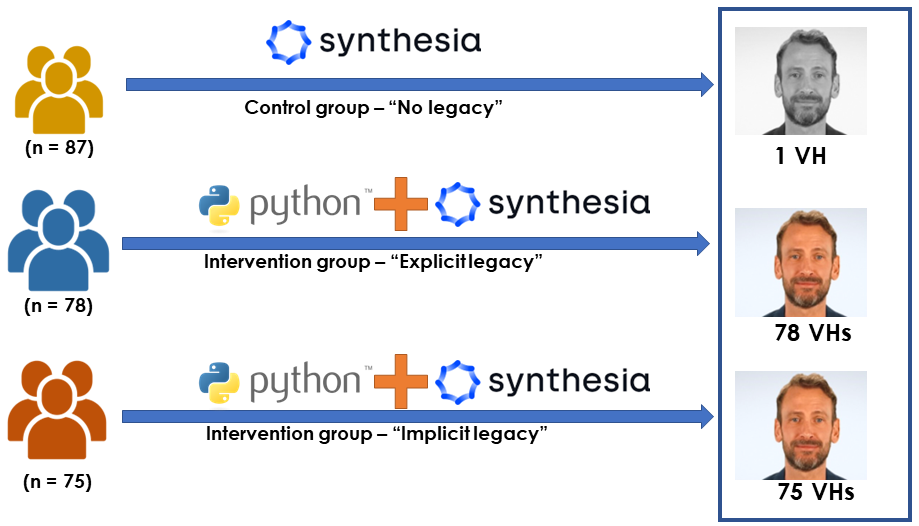}
 \caption{\centering Virtual Humans Generation}
 \label{fig:W3_Part2_Videos}
\end{figure}

A total of 154 virtual humans were created among the experimental groups: one for the control group and 153 for both intervention groups (Section \ref{W3 P2: Experimental Groups}). To do so, the artificial intelligence (AI) video generation platform Synthesia \cite{synthesia} was used in addition to its application programming interface (API) \cite{synthesiaAPI} for Python. For both intervention groups, appearances and voices were gathered from participants' input during the first part of the user study (Section \ref{W3:part1}). Figure \ref{fig:W3_Part2_Videos} illustrates the virtual human generation process.

The message among groups was alike. For the \textbf{Control Group -``No legacy"} and the \textbf{Intervention group - ``Implicit legacy"}, the message fully referred to gratitude journaling \cite{gratitude}. For the \textbf{Intervention group - ``Explicit legacy"}, the message started by indicating that the virtual human was designed by someone similar to the participant and then proceeded to disclose the designer's age, gender, ethnic groups, and place of origin: e.g., \textit{``Hello and welcome. I am a mental wellness virtual counselor at the University of Florida. I was designed by a fellow college student who shares similar characteristics to yours. He reported being Multiracial; with Latino and White roots, and is also from Vera. He is 20 years old. I want to share with you the power of gratitude in promoting mental health and life satisfaction. Gratitude as a..."}. The demographics of the designer were provided at the start of the interaction to ensure that the majority of participants in that group had access to this information before delving into the topic. Appendix \ref{study2:explicitScript} presents a sample of a script for an agent from the explicit group. Appendix \ref{study2:implicitScript} presents the script agents used in the control and implicit groups.

The gender of the designer was disclosed through pronouns, using \textit{He} for participants self-identified as \textit{male}, \textit{She} for those self-identified as \textit{female}, and \textit{They} for those who self-identified in other gender groups. The age was first rounded to the closest multiple of five and disclosed directly in the message: e.g., for an 33 years of age, the agent disclosed an age of 35.

\subsubsection{Intervention} \label{W3 P2: Intervention}
Participants were asked to respond to an online questionnaire, in which they were assessed on their intentions toward gratitude journaling \cite{gratitude} before and after interacting with a virtual human co-designed in the first part of the user study (Section \ref{W3:part1}). Participants were also asked to describe and explain their description of the designer behind the virtual human they interacted with, responding to questions concerning the designer's place of origin, ethnicity, age, and gender. Questions were organized in five blocks (Figure \ref{fig:W3Part2} illustrates the procedure):

\textbf{Block 1-\textit{Informed Consent}} provided an informed consent document with details of the study and two multiple-choice questions. Participants did not have to respond to any questions here, as they already gave their consent in the first part of the user study (Section \ref{W3:part1}).

\textbf{Block 2 - \textbf{Pre-TPB}} presented seven 7-point semantic differential scales based on the theory of planned behavior (TPB) \cite{mohan, tpb} as a pre-interaction assessment. All scales were continuous (included decimal numbers). These scales align with those previously outlined by Feijóo-García et al. \cite{TAP}, originally introduced by Zalake et al. \cite{mohan}.

\textbf{Block 3 - \textbf{Virtual Human Interaction}} introduced a virtual human to talk about gratitude journaling \cite{gratitude}. The interaction lasted no more than four minutes.

\textbf{Block 4 - \textbf{Post-TPB}} presented the same seven 7-point semantic differential scales from block 2 as a post-interaction assessment.

\textbf{Block 5 - \textbf{Participant's Perceptions of The Designer's Demographics}} asked participants to describe the \textbf{designer} who created the virtual human they interacted with. The block included five questions concerning the designer's demographic characteristics: age (open-ended numeric input), ethnicity (multiple-choice close-ended question based on \cite{pedro,iva2022}), place of origin (open-ended input as country-city), and gender (close-ended multiple-choice question based on \cite{gender}). The final open-ended question gathered participants' insights about their selections. 

 \textbf{Block 6 - \textbf{Participant Demographics}} focused on participants' demographics, presenting six questions. First, it asked about participants' age (open-ended numeric input), ethnicity (multiple-choice close-ended question based on \cite{pedro}), and gender (close-ended multiple-choice question based on \cite{gender}). Then, it followed by asking about participants' international student status (domestic--United States, or international) and their place of origin (open-ended input as country-city). Finally, it asked one open-ended question to have participants list all languages used to communicate.

\subsection{Data analysis} \label{W3_P2:Analysis}

Participants' responses were first analyzed concerning how they described the designers behind the virtual humans they interacted with. Responses were studied using descriptive statistics crossing participants' responses with their demographic data and the actual demographics of the designers behind the virtual humans. The demographics considered were gender, ethnicity, age, and place of origin.

Participants' rationale was qualitatively analyzed by coding participants' responses from the open-ended question on participants' insights about their selections to determine if they selected the designers' demographics based on the demographics of the virtual humans. Two classes were used for this process: \textit{Similarity} and \textit{Non-Similarity}. Cohen's kappa was used to calculate inter-rater reliability \cite{cohen1,cohen2}.

Then, the analysis centered on participants' intentions on doing gratitude journaling \cite{gratitude}, following the intervention previously described (Section \ref{W3 P2: Intervention}). The analysis considered participants' scores on the 7-point semantic differential scale \textbf{\textit{Intention:}} ``I intend to write down what I feel grateful about for the next week to have a good mental health" (1-false, 7-true).

\begin{equation}
\label{equationMohan}
  Score_N = \begin{cases}(post - pre)/(7-pre) & post  > pre \\(post - pre)/pre & post \leq pre\end{cases}
\end{equation}

Participants' responses were normalized based on their pre-interaction and post-interaction scores (Equation \ref{equationMohan}) \cite{mohan, TAP}. Due to the non-Gaussian distribution of responses, the non-parametric Kruskal-Wallis Test \cite{kruskal,kruskal1} was used to compare the three experimental groups and evaluate the following hypothesis:

\begin{itemize}
\item\textbf{H-``User-Designer Similarity Effect"}: \textit{Virtual humans co-designed by computer science students are more effective in promoting these students' intentions toward gratitude journaling when there is a higher degree of demographic similarity between the user and the designer and the virtual human explicitly communicates the designer's demographics.}
\end{itemize}

\subsection{Findings: Participants' perceptions and rationale on designers' demographics} \label{W3_P2:DesignerDemographics}
Participants from the \textbf{Intervention Group - ``Implicit Legacy"} (n=75) were assessed on their capacity to accurately tell the demographic characteristics of the designer behind the virtual human they interacted with. These virtual humans were initially designed by themselves in the first part of the user study (Section \ref{W3:part1}) and were intended to support students similar to them in mental health conversations.

Out of the four demographic characteristics of the designer, age was the characteristic that participants could better identify: 56.0\% of them correctly identified the designer's age. Two demographic characteristics followed, with 37.3\% of them correctly identifying gender, and 32.0\% ethnicity. The designer's country of origin was only identified by 21.3\% of participants.

Participants from the \textbf{Intervention Group - ``Explicit Legacy"} (n=78) were also assessed concerning the designer's demographic characteristics. For this group, the virtual human explicitly mentioned the designer's characteristics at the beginning of the interaction--no longer than four minutes. Surprisingly, participants from this group did not fully recall the four characteristics initially mentioned at the beginning of the interaction (Section \ref{W3 P2: VH Creation}). The designer's country of origin was the characteristics participants from this group mostly recalled (50.0\% of participants). It was followed by ethnicity (44.9\% of participants), age (43.9\% of participants) and gender (42.3\% of participants). 

To better understand the interplay between the agent, the designer, and the participant, especially for the \textbf{Intervention Group - ``Explicit Legacy"} (n=78), participants' perceptions were analyzed concerning three of the virtual humans' demographics they were asked to identify in the first part of the user study: age, gender, and ethnicity (Sections \ref{perceptionsMaleRace} for male appearances and \ref{perceptionsFemaleRace} for female ones). Concerning gender, 75.6\% of the participants from this group considered the agent's gender to match the designer's, while 64.1\% of them did so concerning the agent's ethnicity. Of the three demographic characteristics, age was the only one that participants did not generally relate to the designer: only 46.2\% of them considered the agent's age to match the designer's. Looking toward the participants from the \textbf{Intervention Group - ``Implicit Legacy"} (n=75), 80.0\% of them considered the agent's gender to match the designer's, and 78.7\% of them did so with the agent's ethnicity. Also, although lower, over half of that group related the agent's perceived age with the designer's (56.0\% of participants). 

Participants from the \textbf{Control Group -``No legacy"} (n=87) also guessed who was the designer behind the virtual human. Considering their perceptions of the agent's demographics, it was observed that 64.4\% of them related the agent's gender to the designer's, while 59.8\% did so for ethnicity. Also, 62.1\% of participants related the agent's and designer's age. 

Participants (n=240) were also asked to share their insights behind their responses to the four demographic characteristics they were asked to identify: gender, age, ethnicity, and place of origin. 

Participants' responses were coded into two classes to label if they used or not agent-designer similarity to determine the designer's demographic characteristics. Then, Inter-rater reliability was used to categorize participants' open-ended responses. For this, an independent researcher coded 20\% of the responses \cite{cohen1}. Following, another independent rater coded that sub-set of responses. Cohen's kappa was calculated to measure agreement among raters, getting a score of 0.81-- a high level of agreement between raters \cite{cohen2}.

\begin{figure}[!]
 \centering 
 \includegraphics[scale=1]{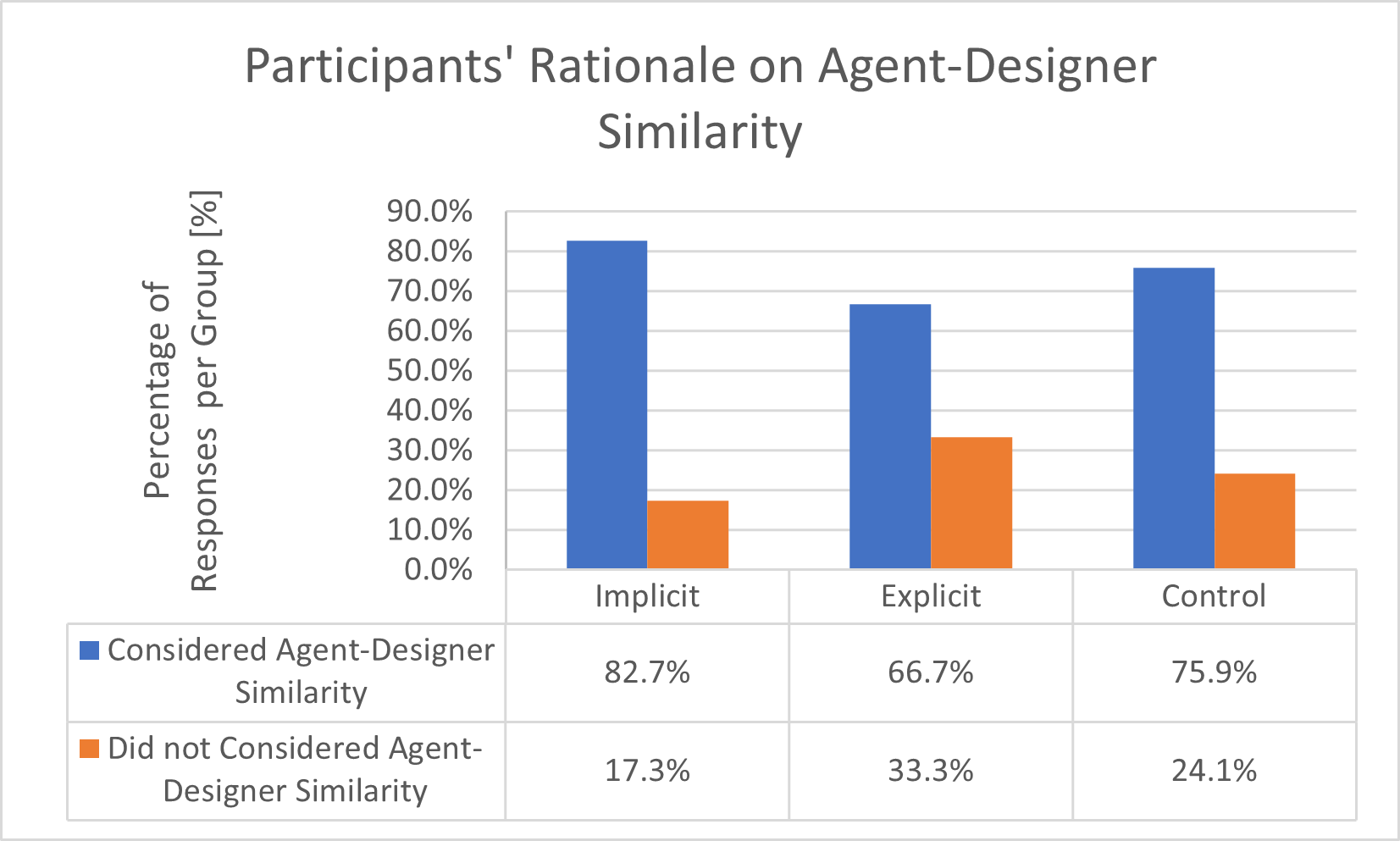}
 \caption{\centering Participants' Rationale on Agent-Designer Similarity}
 \label{fig:rationalSimilarity}
\end{figure}

In general, participants from the three experimental groups (Section \ref{W3 P2: Experimental Groups}) considered agent-designer similarity to identify who was behind the virtual human. As observed in Figure \ref{fig:rationalSimilarity}, both the \textbf{Intervention Group - ``Implicit Legacy"} and the \textbf{Control Group - ``No Legacy"} had over 75\% of their participants referring to agent-designer similarity. Close to them, 66.7\% of the participants from the \textbf{Intervention Group - ``Explicit Legacy"} also considered agent-designer similarity when determining the designer's demographic characteristics: e.g., \textit{``From her appearance, I could clearly figure out that she is in her 40s and belong to south-asian origin"} [DR028], \textit{``Ethnic group and gender were based on how they look. Place of origin is because it was in front of a UF background. Age was a guest, I'm not sure who is conducting this research, but I suppose maybe on the younger side"} [DR213], `\textit{`She could have been born anywhere I guess but based on her looks her ethnicity seemed to be South Asian so I said India. I didn't know what part so I just thought of the first city that came to mind. I was more confident about her gender and age, and I guessed 27 because she seemed young"} [DR073].

\subsection{Findings: Designer cues and participants' intentions towards gratitude journaling} \label{W3_P2:GratitudeJournaling}

Following the strategy presented in Section \ref{W3_P2:Analysis}, the three experimental groups (Section \ref{W3 P2: Experimental Groups}) were compared in regard to participants' intentions towards gratitude journaling \cite{gratitude}.

\begin{table}[!]
	\centering
	\caption{\centering \label{descriptivesNormalized} Descriptive Statistics - Normalized TPB Score of Intention}
	\label{tab:descriptives-NormalizedTPBScore:Intention}
	{
		\begin{tabular}{lrrrrr}
			\toprule
			Experimental Group & N & Mean & SD & SE & Coefficient of variation  \\
			\cmidrule[0.4pt]{1-6}
			Explicit Legacy & $78$ & $0.236$ & $0.37$ & $0.04$ & $1.65$  \\
			Implicit Legacy & $75$ & $0.20$ & $0.34$ & $0.04$ & $1.70$  \\
			No Legacy (Control) & $87$ & $0.24$ & $0.36$ & $0.04$ & $1.50$  \\
			\bottomrule
		\end{tabular}
	}
\end{table}

\begin{figure}[h]
 \centering 
 \includegraphics[scale=0.45]{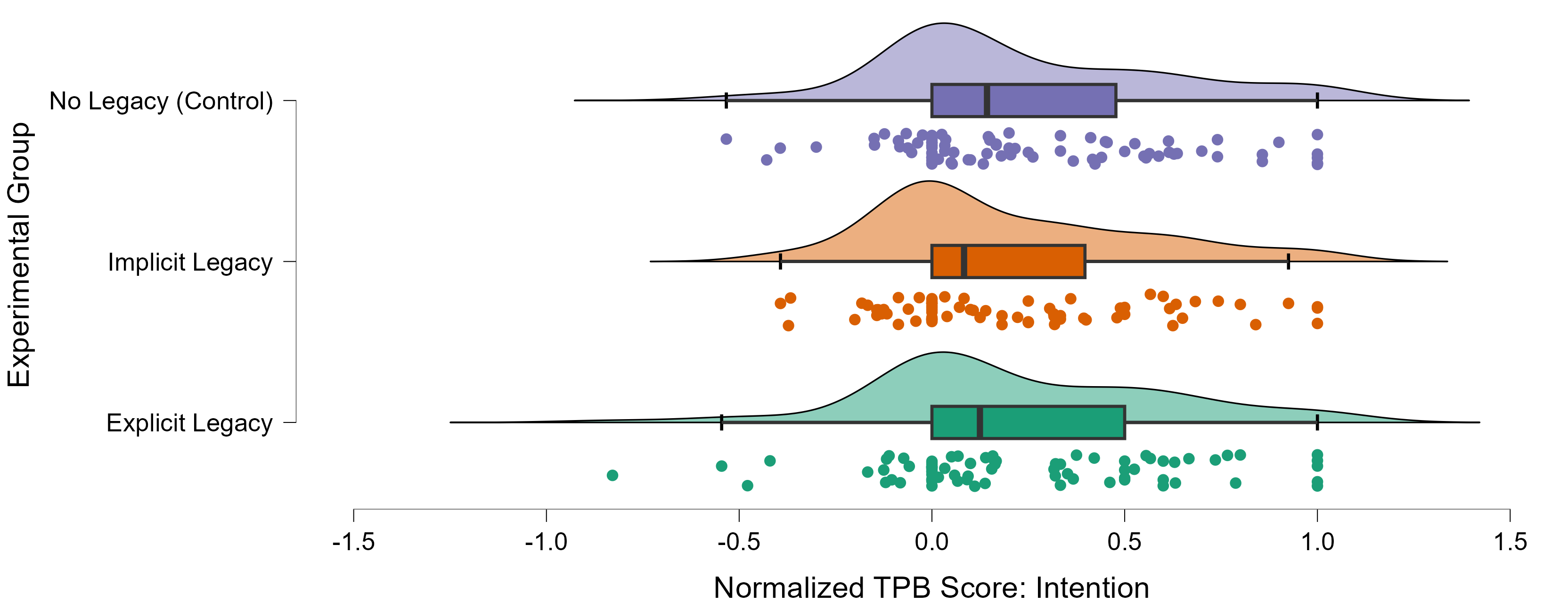}
 \caption{\centering Normalized Intentions per Experimental Group}
 \label{fig:normalized}
\end{figure}

Experimental groups were compared based on participants' normalized scores of intentions (Equation \ref{equationMohan}). The non-parametric Kruskal-Wallis \cite{kruskal,kruskal1} was used to validate the following hypothesis \textbf{H-``User-Designer Similarity Effect"}: \textit{Virtual humans co-designed by computer science students are more effective in promoting these students' intentions toward gratitude journaling when there is a higher degree of demographic similarity between the user and the designer and the virtual human explicitly communicates the designer's demographics.} The hypothesis was rejected, as no significant difference was observed among groups concerning participants' normalized scores of intention (Kruskal-Wallis Test, H(2) = 1.06, p=0.59). Figure \ref{fig:normalized} shows that the three groups presented a similar distribution with no visible differences among them (see Table \ref{descriptivesNormalized} for detailed descriptive statistics), with normalized scores generally over 0.0.

Experimental groups were also compared concerning participants' post-TPB scores on intention (Section \ref{W3 P2: Intervention}. No significant difference was observed among groups concerning post-TPB scores (Kruskal-Wallis Test, H(2) = 0.29, p=0.86).

\begin{figure}[h]
 \centering 
 \includegraphics[scale=0.45]{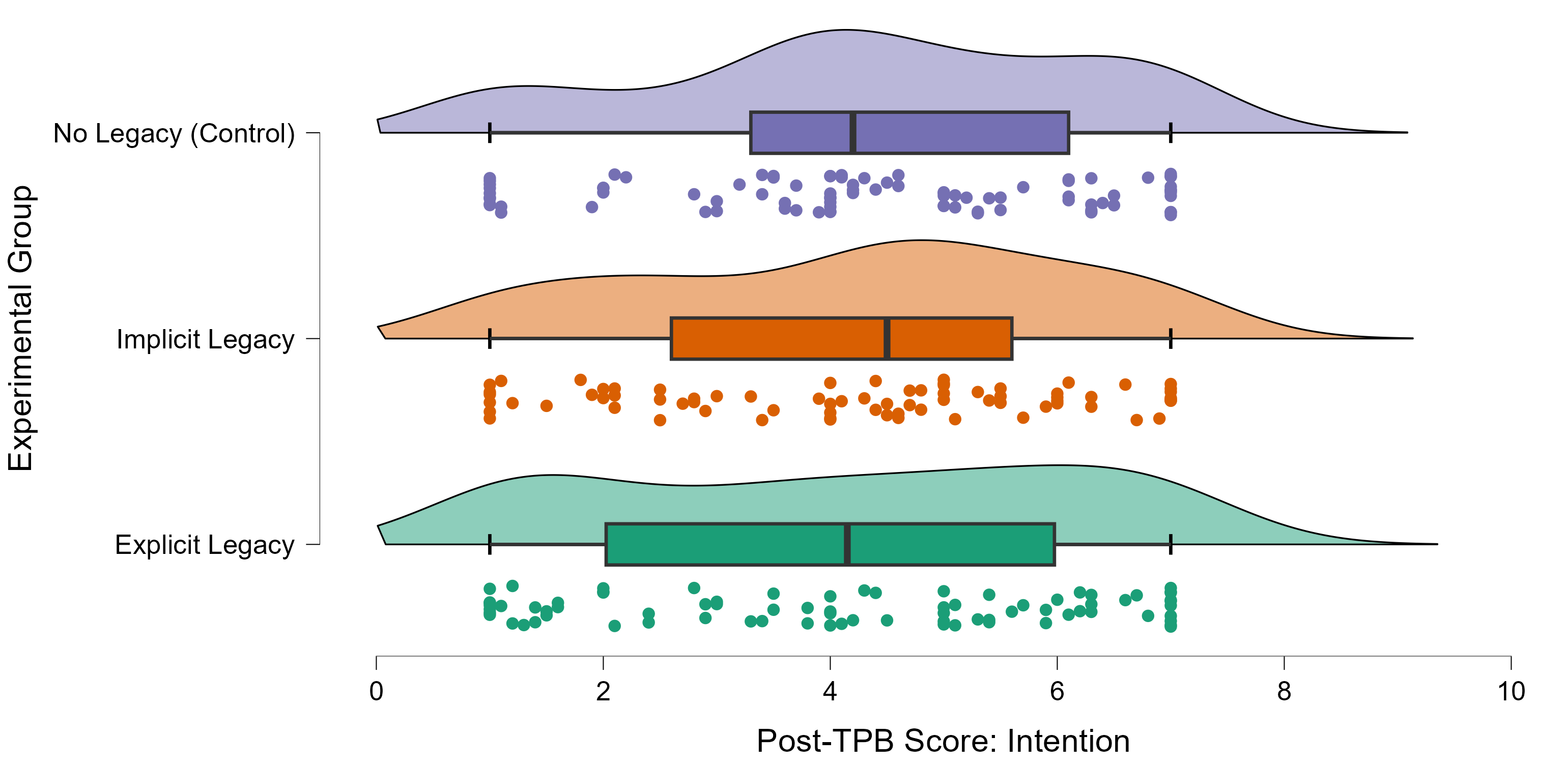}
 \caption{\centering Post-TPB Intentions per Experimental Group}
 \label{fig:postTPB}
\end{figure}

Just as with participants' normalized scores, the post-TPB scores were similarly distributed among groups (Figure \ref{fig:postTPB}) with means around 4.0 on a 7-point scale (Table \ref{postTPBTable}).

\begin{table}[h]
	\centering
	\caption{\centering \label{postTPBTable}Descriptives - Post-TPB Score: Intention}
	\label{tab:descriptives-Post-TPBScore:Intention}
	{
		\begin{tabular}{lrrrrr}
			\toprule
			Experimental Group & N & Mean & SD & SE & Coefficient of variation  \\
			\cmidrule[0.4pt]{1-6}
			Explicit Legacy & $78$ & $4.108$ & $2.080$ & $0.236$ & $0.506$  \\
			Implicit Legacy & $75$ & $4.197$ & $1.874$ & $0.216$ & $0.446$  \\
			No Legacy (Control) & $87$ & $4.318$ & $1.887$ & $0.202$ & $0.437$  \\
			\bottomrule
		\end{tabular}
	}
\end{table}

\section{Discussion and Future Work} \label{W3: Discussion}

 This section is structured to discuss our findings in regard to the three research questions initially posed in Section \ref{introduction}.

\subsection{Insights on agent-designer similarity in virtual humans co-design} \label{discussion:RQ1}
Responding to the research question \textbf{RQ-``Similarity Elucidation"}: \textit{How do similarity expectations influence the design decisions made by computer science students when co-designing virtual humans for mental health support?}, this work's findings provide valuable insights into the design of virtual humans to support college students in mental health conversations, especially those majoring in computer science. From the different factors identified in Section \ref{W3:agentDesignerDemographicMatching}, age was found to be an essential design factor, regardless of participants' genders or ethnic groups, for the design of virtual humans to support mental health. Although user-agent age matching was not as predominant as other demographic characteristics (e.g., place of origin), it was observed that participants' rationale generally considered age for both male and female appearance selection (Sections \ref{W3:AppearancesMaleRationale} and \ref{W3:femaleAppearancesRationale}), suggesting age can help promote effective virtual human interventions in assisting computer science students (and other potential populations) in mental health conversations.

Notably, the preference for ethnic matching differed between male and female appearances (Sections \ref{W3:AppearancesMaleRationale} and \ref{W3:femaleAppearancesRationale}). While ethnic matching was more prevalent in the selection of male appearances, it was less pronounced for female appearances. This suggests that computer science students may have different expectations and criteria regarding the ethnicity of male and female virtual humans. As female virtual humans were generally preferred among ethnic groups, male virtual humans may have triggered participants toward a need for user-agent ethnic similarity. These findings suggest that gender and ethnicity interact in a complex way and highlight the importance of considering user-agent similarity and their intersectionality in designing virtual humans for mental health support. More research is needed to explore these gender-specific preferences and the interplay between ethnicity and gender in virtual human design.

Regarding participants' preferences for virtual human voices, it was observed that appearance-voice matching played a crucial role, particularly in the design of male virtual humans (Section \ref{W3:AppearancesMaleRationale}). This observation was particularly seen among domestic participants, for whom ethnicity had more prevalence compared to international participants. They prioritized appearance-voice matching, voice clarity, and voice comfort (Table \ref{table:voiceCategories}). These findings support previous work by Feijóo-García et al. \cite{TAP}, stating the importance of appearance-voice matching in virtual humans' effectiveness in promoting mental health conversations.

Ethnicity also emerged as a notable criterion when selecting voices (Sections \ref{W3:maleVoices} and \ref{W3:femaleVoices}), with some participants even expressing a preference for voices that matched their own ethnic background. It was found that user-agent ethnic similarity has a major role in encouraging computer science students to engage in mental health discussions with male virtual humans, as participants preferred male virtual humans who shared age and ethnicity. This indicates that computer science students may feel more comfortable and open when interacting with male virtual humans who are closer in age and ethnic background. Hence, correctly modeling these factors may help in designing more effective virtual humans.

These findings contribute to understanding factors influencing computer science students' preferences and expectations regarding virtual humans in mental health conversations. Age, gender, ethnicity, appearance-voice matching, and user-agent similarity were found to be essential factors. Virtual human designers should aim to understand the preferences and expectations of their target population (e.g., computer science students) when creating virtual humans. The insights gained from this work can inform the design and development of virtual humans that effectively support and engage students in meaningful mental health conversations. Future research should aim to replicate and expand upon these findings with more extensive and diverse sample populations to understand better the complex interplay between age, gender, ethnicity, appearance, and voice in virtual human design for mental health contexts.

\subsection{Insights on the designer's legacy and users' perceptions of it in virtual human interactions} \label{discussion:RQ2}
Responding to the research question \textbf{RQ-``Designer Cues"}: \textit{Which characteristics of a virtual human's designer can be identified by computer science students during their interactions with the agent?}, this work's findings provide valuable insights into the connection between the demographic characteristics of virtual humans and those of their designers (Section \ref{W3_P2:DesignerDemographics}). 

First, participants tend to relate the demographics of virtual humans to those of their designers--designer-agent demographic similarity, particularly in regard to ethnicity and gender. Although relevant in the selection of virtual human appearances (Sections \ref{W3:AppearancesMaleRationale} and \ref{W3:femaleAppearancesRationale}), the designer's age was not identifiable nor recognizable through the virtual human. Same for the place of origin: the characteristic that most likely presented a similarity-attraction effect (Section \ref{W3:agentDesignerDemographicMatching}). Considering how participants were co-designers in the first part of the user study (Section \ref{W3:part1}), these findings suggest that computer science students acting as designers could be influenced by their own demographics when creating virtual humans.

Second, we observed that the mere awareness of the demographic characteristics of the virtual human's designer was insufficient for participants to accurately identify the virtual human's designer (Section \ref{W3_P2:DesignerDemographics}). As reported in Section \ref{W3_P2:DesignerDemographics}, participants' perceptions led them to perceive the designer as demographically similar to their agent--designer-agent demographic similarity. 

Finally, we observed a more intriguing phenomenon where the visual cues of the virtual human appeared to overshadow the verbally shared information regarding the designer's demographics. Specifically, participants who interacted with a virtual human that verbally shared the demographic information of its designer (see Section \ref{W3 P2: Experimental Groups}: \textbf{Intervention Group - ``Explicit Legacy"}) had a tough time recalling the designer's information. This observation suggests that the visual aspects of virtual human demographics hold more influence over users' perceptions compared to verbal cues, and it may be more pronounced when participants base their rationale on agent-designer similarity. This highlights the limitations of demographic information in determining the creators of virtual human designs. User-designer and agent-designer similarity interact in a complex manner, even when computer science students serve as designers and users of virtual humans. Future research should explore different ways to present designer cues, adding factors apart from demographic characteristics, appearances, and voices.

\subsection{Insights on the designer's legacy effectiveness in virtual human interactions} \label{discussion:RQ3}
Responding to the research question \textbf{RQ-``Designer's Legacy Effectiveness"}: \textit{How effective are virtual humans in promoting computer science students' intentions toward mental wellness conversations when co-designed by computer science students with similar demographic characteristics to their end users?}, it was found that virtual humans promoting computer science students' intentions toward mental wellness conversations are not more effective when co-designed by fellow computer science students with similar demographic characteristics to their end users (Section \ref{W3_P2:GratitudeJournaling}).

The lack of significance can be interpreted from several angles, considering the findings previously discussed in Section \ref{W3_P2:DesignerDemographics}. It is possible that the demographic characteristics of the designers and their impact on the design of the virtual humans were not as influential on participants' intentions toward gratitude journaling. The findings of this work indicate that the degree of demographic similarity between the user and the designer, and the virtual human's explicit communication of the designer's demographics, did not significantly differentiate the groups based on the participants' normalized intentions scores.

Second, this work's findings suggest that the demographic information of the designers, which the participants were aware of, did not provide sufficient cues to accurately identify the designer of the virtual human. Thus, this awareness may not have been strong enough to influence participants' intentions toward promoting gratitude journaling. Moreover, the fact that the \textbf{Intervention Group - ``Explicit Legacy"} (Section\ref{W3 P2: Experimental Groups}) struggled to consistently remember the designer's demographics initially provided could serve as an indication that the visual aspects of the virtual human's demographics might have dominated participants' perceptions, potentially affecting their intentions toward gratitude journaling. The visual aspects of virtual humans and users' biases could influence users' intentions regardless of 1) the demographic similarity between the user and the designer and 2) the virtual human explicitly communicating the designer's demographics. Future research is needed to understand how to tailor designer cues and assess their impact on promoting intentions and behaviors toward mental health conversations.

\section{Limitations} \label{W3:limitations}

Several limitations should be acknowledged when interpreting the findings from this study. Initially, participants used the artificial intelligence (AI) video generation platform Synthesia \cite{synthesia} and its application programming interface (API) \cite{synthesiaAPI} to create virtual humans. Although the platform provided a diverse catalog of female and male agents, the choices were restricted to the designs available in the catalog. Notably, this catalog only featured male and female agents, which did not accommodate non-binary gender expressions. Furthermore, participants, particularly those from Latin American backgrounds, found the options for male appearances limited. Future research should aim to expand the design capabilities of virtual humans, allowing for a more nuanced interplay of gender, race, ethnicity, and age to better reflect the voice and appearance of the virtual agents.

In addition to the limitations posed by the platform's catalog, the first part of our user study also faced challenges with the diversity and balance of its sample population. Although the number of participants was increased from previous efforts \cite{TAP} and had a larger population of female participants (n=182), the group remained predominantly male (n=281) and also lacked sufficient representation from some ethnic backgrounds. We acknowledge that we could have recruited a more balanced demographic sample through platforms that allow demographic recruitment specifications (e.g., Prolific \cite{prolific}). However, our user study depended on the student population we could recruit from our institution at the time and was limited to its socio-geographic context: we were limited by our resources and relied on a sample population (n=481) originating from the University of Florida. Also, we acknowledge that the decision to include only computer science students could limit the broader applicability of the findings, as it does not represent a cross-section of the general population. Please keep in mind that we limited our population to this context as previous literature suggests computer science students experience the highest rates of anxiety and depression among STEM college students \cite{brazil3}, with female computer science students being particularly more at risk \cite{porter, brazil1}.

Another possible limitation concerns the reuse of the same sample of participants across different stages of the study, which could have introduced biases, especially in the second part of the user study: Their previous experiences designing the virtual agents might have influenced their subsequent perceptions of the agents and their designers. Moreover, although we gave at least four days between the parts of the study, this period might not have been sufficient to prevent carryover effects, potentially affecting the outcomes.

The study also did not sufficiently explore the influence of factors beyond gender and ethnicity, such as facial expressions and physical attractiveness, on participants' perceptions of virtual humans. For example, the top-selected male virtual humans predominantly displayed happy expressions, which might have influenced participant choices and highlighted a potential oversight regarding the impact of facial cues.

Moreover, in the second part of the study, it was observed that participants from the \textbf{Intervention group - ``Explicit legacy"} (Section \ref{W3 P2: Experimental Groups}) struggled to fully recall the designer's demographic characteristics, possibly affecting participants’ intentions toward mental health conversations. This suggests that future research should explore alternative methods to mitigate this issue and consider that information on designer cues should not be limited to verbal messages.

\section{Conclusions}\label{conclusions}

Our work contributes insights concerning agent-user, agent-designer (which we refer as the designer's legacy), and user-designer demographic similarity in the co-design of virtual humans, reporting on \textit{how computer science students, when acting as designers, reflect their own characteristics in their virtual humans}. Moreover, our work contributes insights into \textit{how computer science students, when acting as users, expect virtual humans to resemble their designers}.

We found \textit{participants preferred voices that matched their own ethnic background}, suggesting that user-agent ethnic similarity could encourage more active engagement in mental health discussions. Moreover, we observed that age, gender, ethnicity, appearance-voice matching, and user-agent similarity emerged as critical factors in the design of virtual humans for mental health support.

Additionally, \textit{participants generally associated the demographic characteristics of virtual humans with those of their designers}, particularly in terms of ethnicity and gender. Our findings also suggest that the visual cues of the virtual human could overshadow the verbally shared information about the designer's demographics, indicating that visual aspects have more influence over users' perceptions.

Finally, virtual humans were not found to be more effective in promoting mental health conversations when co-designed by computer science students with similar demographic characteristics to their end users. However, as mentioned above, \textit{the visual cues of the virtual humans overshadowed the information verbally communicated about the designer's demographics}, possibly affecting participants' intentions toward mental health conversations. Thus, the designers' demographic information was insufficient to influence participants' intentions toward mental health conversations. This suggests that tailoring designer cues in virtual human design is a complex task that should be explored further, especially in sensitive scenarios such as mental health support.

Future directions should further explore the role of user-designer demographic similarity to understand how designer cues should be modeled for virtual human interactions. Also, future next steps should further investigate how user-agent and user-designer characteristics should be matched and featured in user-agent interactions and their impact on users' intentions and behaviors toward mental health conversations.

\begin{acks}
The authors wish to thank the students who actively participated in the user study. Additionally, we thank Dr. Sharon Chu, Dr. Laura Cruz Castro, Mr. Amanpreet Kapoor, Dr. Yingbo Ma, and Ms. Lisha Zhou for letting us recruit their students. This work was sponsored by the Fulbright Scholar Program and the Colombian Ministry of Science, Technology, and Innovation. This work was also supported by the National Science Foundation award numbers 1900961 and 1800947, the American foundation for Suicide Prevention (LSRG-1-050-18), and the National Institute of Mental Health (NIMH1R34MH119294-01).
\end{acks}

\bibliographystyle{ACM-Reference-Format}
\bibliography{sample-base}


\begin{thebibliography}{96}


\ifx \showCODEN    \undefined \def \showCODEN     #1{\unskip}     \fi
\ifx \showDOI      \undefined \def \showDOI       #1{#1}\fi
\ifx \showISBNx    \undefined \def \showISBNx     #1{\unskip}     \fi
\ifx \showISBNxiii \undefined \def \showISBNxiii  #1{\unskip}     \fi
\ifx \showISSN     \undefined \def \showISSN      #1{\unskip}     \fi
\ifx \showLCCN     \undefined \def \showLCCN      #1{\unskip}     \fi
\ifx \shownote     \undefined \def \shownote      #1{#1}          \fi
\ifx \showarticletitle \undefined \def \showarticletitle #1{#1}   \fi
\ifx \showURL      \undefined \def \showURL       {\relax}        \fi
\providecommand\bibfield[2]{#2}
\providecommand\bibinfo[2]{#2}
\providecommand\natexlab[1]{#1}
\providecommand\showeprint[2][]{arXiv:#2}

\bibitem[syn(2023)]%
        {synthesiaAPI}
 \bibinfo{year}{2023}\natexlab{}.
\newblock \bibinfo{title}{Synthesia API}.
\newblock \bibinfo{howpublished}{\url{https://docs.synthesia.io/docs}}.
\newblock


\bibitem[Ajzen(1985)]%
        {tpb}
\bibfield{author}{\bibinfo{person}{Icek Ajzen}.} \bibinfo{year}{1985}\natexlab{}.
\newblock \bibinfo{booktitle}{\emph{From Intentions to Actions: A Theory of Planned Behavior}}.
\newblock \bibinfo{publisher}{Springer Berlin Heidelberg}, \bibinfo{address}{Berlin, Heidelberg}, \bibinfo{pages}{11--39}.
\newblock
\showISBNx{978-3-642-69746-3}
\urldef\tempurl%
\url{https://doi.org/10.1007/978-3-642-69746-3_2}
\showDOI{\tempurl}


\bibitem[Ali et~al\mbox{.}(2020)]%
        {33}
\bibfield{author}{\bibinfo{person}{Mohammad~Rafayet Ali}, \bibinfo{person}{Seyedeh~Zahra Razavi}, \bibinfo{person}{Raina Langevin}, \bibinfo{person}{Abdullah Al~Mamun}, \bibinfo{person}{Benjamin Kane}, \bibinfo{person}{Reza Rawassizadeh}, \bibinfo{person}{Lenhart~K Schubert}, {and} \bibinfo{person}{Ehsan Hoque}.} \bibinfo{year}{2020}\natexlab{}.
\newblock \showarticletitle{A virtual conversational agent for teens with autism spectrum disorder: Experimental results and design lessons}. In \bibinfo{booktitle}{\emph{Proceedings of the 20th ACM International Conference on Intelligent Virtual Agents}}. \bibinfo{pages}{1--8}.
\newblock


\bibitem[Baylor et~al\mbox{.}(2003)]%
        {89}
\bibfield{author}{\bibinfo{person}{Amy Baylor}, \bibinfo{person}{E Shen}, {and} \bibinfo{person}{Xiaoxia Huang}.} \bibinfo{year}{2003}\natexlab{}.
\newblock \showarticletitle{Which pedagogical agent do learners choose? The effects of gender and ethnicity}. In \bibinfo{booktitle}{\emph{E-Learn: World Conference on E-Learning in Corporate, Government, Healthcare, and Higher Education}}. Association for the Advancement of Computing in Education (AACE), \bibinfo{pages}{1507--1510}.
\newblock


\bibitem[Baylor(2009)]%
        {baylor}
\bibfield{author}{\bibinfo{person}{Amy~L Baylor}.} \bibinfo{year}{2009}\natexlab{}.
\newblock \showarticletitle{Promoting motivation with virtual agents and avatars: role of visual presence and appearance}.
\newblock \bibinfo{journal}{\emph{Philosophical Transactions of the Royal Society B: Biological Sciences}} \bibinfo{volume}{364}, \bibinfo{number}{1535} (\bibinfo{year}{2009}), \bibinfo{pages}{3559--3565}.
\newblock


\bibitem[Baylor and Kim(2004)]%
        {86}
\bibfield{author}{\bibinfo{person}{Amy~L Baylor} {and} \bibinfo{person}{Yanghee Kim}.} \bibinfo{year}{2004}\natexlab{}.
\newblock \showarticletitle{Pedagogical agent design: The impact of agent realism, gender, ethnicity, and instructional role}. In \bibinfo{booktitle}{\emph{International conference on intelligent tutoring systems}}. Springer, \bibinfo{pages}{592--603}.
\newblock


\bibitem[Bearman et~al\mbox{.}(2001)]%
        {bearman2001}
\bibfield{author}{\bibinfo{person}{Margaret Bearman}, \bibinfo{person}{Branko Cesnik}, {and} \bibinfo{person}{Merilyn Liddell}.} \bibinfo{year}{2001}\natexlab{}.
\newblock \showarticletitle{Random comparison of ‘virtual patient’models in the context of teaching clinical communication skills}.
\newblock \bibinfo{journal}{\emph{Medical education}} \bibinfo{volume}{35}, \bibinfo{number}{9} (\bibinfo{year}{2001}), \bibinfo{pages}{824--832}.
\newblock


\bibitem[Bernier and Scassellati(2010)]%
        {bernier2010}
\bibfield{author}{\bibinfo{person}{Emily~P Bernier} {and} \bibinfo{person}{Brian Scassellati}.} \bibinfo{year}{2010}\natexlab{}.
\newblock \showarticletitle{The similarity-attraction effect in human-robot interaction}. In \bibinfo{booktitle}{\emph{2010 IEEE 9th International Conference on Development and Learning}}. IEEE, \bibinfo{pages}{286--290}.
\newblock


\bibitem[Beverland(2005)]%
        {wine}
\bibfield{author}{\bibinfo{person}{Michael~B Beverland}.} \bibinfo{year}{2005}\natexlab{}.
\newblock \showarticletitle{Managing the design innovation--brand marketing interface: Resolving the tension between artistic creation and commercial imperatives}.
\newblock \bibinfo{journal}{\emph{Journal of Product Innovation Management}} \bibinfo{volume}{22}, \bibinfo{number}{2} (\bibinfo{year}{2005}), \bibinfo{pages}{193--207}.
\newblock


\bibitem[Bhati(2014)]%
        {72}
\bibfield{author}{\bibinfo{person}{Kuldhir~S Bhati}.} \bibinfo{year}{2014}\natexlab{}.
\newblock \showarticletitle{Effect of client-therapist gender match on the therapeutic relationship: An exploratory analysis}.
\newblock \bibinfo{journal}{\emph{Psychological reports}} \bibinfo{volume}{115}, \bibinfo{number}{2} (\bibinfo{year}{2014}), \bibinfo{pages}{565--583}.
\newblock


\bibitem[Blow et~al\mbox{.}(2008)]%
        {blow2008}
\bibfield{author}{\bibinfo{person}{Adrian~J Blow}, \bibinfo{person}{Tina~M Timm}, {and} \bibinfo{person}{Ronald Cox}.} \bibinfo{year}{2008}\natexlab{}.
\newblock \showarticletitle{The role of the therapist in therapeutic change: Does therapist gender matter?}
\newblock \bibinfo{journal}{\emph{Journal of Feminist Family Therapy}} \bibinfo{volume}{20}, \bibinfo{number}{1} (\bibinfo{year}{2008}), \bibinfo{pages}{66--86}.
\newblock


\bibitem[Braun et~al\mbox{.}(2019)]%
        {6}
\bibfield{author}{\bibinfo{person}{Michael Braun}, \bibinfo{person}{Anja Mainz}, \bibinfo{person}{Ronee Chadowitz}, \bibinfo{person}{Bastian Pfleging}, {and} \bibinfo{person}{Florian Alt}.} \bibinfo{year}{2019}\natexlab{}.
\newblock \showarticletitle{At your service: Designing voice assistant personalities to improve automotive user interfaces}. In \bibinfo{booktitle}{\emph{Proceedings of the 2019 CHI Conference on Human Factors in Computing Systems}}. \bibinfo{pages}{1--11}.
\newblock


\bibitem[Byrne et~al\mbox{.}(1963)]%
        {52}
\bibfield{author}{\bibinfo{person}{Donn Byrne}, \bibinfo{person}{James Barry}, {and} \bibinfo{person}{Don Nelson}.} \bibinfo{year}{1963}\natexlab{}.
\newblock \showarticletitle{Relation of the revised Repression-Sensitization Scale to measures of self-description}.
\newblock \bibinfo{journal}{\emph{Psychological Reports}} \bibinfo{volume}{13}, \bibinfo{number}{2} (\bibinfo{year}{1963}), \bibinfo{pages}{323--334}.
\newblock


\bibitem[Byrne et~al\mbox{.}(1967)]%
        {67}
\bibfield{author}{\bibinfo{person}{Donn Byrne}, \bibinfo{person}{William Griffitt}, {and} \bibinfo{person}{Daniel Stefaniak}.} \bibinfo{year}{1967}\natexlab{}.
\newblock \showarticletitle{Attraction and similarity of personality characteristics.}
\newblock \bibinfo{journal}{\emph{Journal of Personality and social Psychology}} \bibinfo{volume}{5}, \bibinfo{number}{1} (\bibinfo{year}{1967}), \bibinfo{pages}{82}.
\newblock


\bibitem[Cabral and Smith(2011)]%
        {73}
\bibfield{author}{\bibinfo{person}{Raquel~R Cabral} {and} \bibinfo{person}{Timothy~B Smith}.} \bibinfo{year}{2011}\natexlab{}.
\newblock \showarticletitle{Racial/ethnic matching of clients and therapists in mental health services: a meta-analytic review of preferences, perceptions, and outcomes.}
\newblock \bibinfo{journal}{\emph{Journal of counseling psychology}} \bibinfo{volume}{58}, \bibinfo{number}{4} (\bibinfo{year}{2011}), \bibinfo{pages}{537}.
\newblock


\bibitem[Cambre et~al\mbox{.}(2020)]%
        {14}
\bibfield{author}{\bibinfo{person}{Julia Cambre}, \bibinfo{person}{Jessica Colnago}, \bibinfo{person}{Jim Maddock}, \bibinfo{person}{Janice Tsai}, {and} \bibinfo{person}{Jofish Kaye}.} \bibinfo{year}{2020}\natexlab{}.
\newblock \showarticletitle{Choice of voices: A large-scale evaluation of text-to-speech voice quality for long-form content}. In \bibinfo{booktitle}{\emph{Proceedings of the 2020 CHI Conference on Human Factors in Computing Systems}}. \bibinfo{pages}{1--13}.
\newblock


\bibitem[Chen and Kenrick(2002)]%
        {71}
\bibfield{author}{\bibinfo{person}{Fang~Fang Chen} {and} \bibinfo{person}{Douglas~T Kenrick}.} \bibinfo{year}{2002}\natexlab{}.
\newblock \showarticletitle{Repulsion or attraction? Group membership and assumed attitude similarity.}
\newblock \bibinfo{journal}{\emph{Journal of personality and social psychology}} \bibinfo{volume}{83}, \bibinfo{number}{1} (\bibinfo{year}{2002}), \bibinfo{pages}{111}.
\newblock


\bibitem[Chen(2009)]%
        {art1}
\bibfield{author}{\bibinfo{person}{Yu Chen}.} \bibinfo{year}{2009}\natexlab{}.
\newblock \showarticletitle{Possession and access: Consumer desires and value perceptions regarding contemporary art collection and exhibit visits}.
\newblock \bibinfo{journal}{\emph{Journal of Consumer Research}} \bibinfo{volume}{35}, \bibinfo{number}{6} (\bibinfo{year}{2009}), \bibinfo{pages}{925--940}.
\newblock


\bibitem[Clance and Imes(1978)]%
        {clance1978imposter}
\bibfield{author}{\bibinfo{person}{Pauline~Rose Clance} {and} \bibinfo{person}{Suzanne~Ament Imes}.} \bibinfo{year}{1978}\natexlab{}.
\newblock \showarticletitle{The imposter phenomenon in high achieving women: Dynamics and therapeutic intervention.}
\newblock \bibinfo{journal}{\emph{Psychotherapy: Theory, research \& practice}} \bibinfo{volume}{15}, \bibinfo{number}{3} (\bibinfo{year}{1978}), \bibinfo{pages}{241}.
\newblock


\bibitem[Clore and Baldridge(1968)]%
        {68}
\bibfield{author}{\bibinfo{person}{Gerald~L Clore} {and} \bibinfo{person}{Barbara Baldridge}.} \bibinfo{year}{1968}\natexlab{}.
\newblock \showarticletitle{Interpersonal attraction: The role of agreement and topic interest.}
\newblock \bibinfo{journal}{\emph{Journal of Personality and Social Psychology}} \bibinfo{volume}{9}, \bibinfo{number}{4} (\bibinfo{year}{1968}), \bibinfo{pages}{340}.
\newblock


\bibitem[Cohen et~al\mbox{.}(2009)]%
        {pearson}
\bibfield{author}{\bibinfo{person}{Israel Cohen}, \bibinfo{person}{Yiteng Huang}, \bibinfo{person}{Jingdong Chen}, \bibinfo{person}{Jacob Benesty}, \bibinfo{person}{Jacob Benesty}, \bibinfo{person}{Jingdong Chen}, \bibinfo{person}{Yiteng Huang}, {and} \bibinfo{person}{Israel Cohen}.} \bibinfo{year}{2009}\natexlab{}.
\newblock \showarticletitle{Pearson correlation coefficient}.
\newblock \bibinfo{journal}{\emph{Noise reduction in speech processing}} (\bibinfo{year}{2009}), \bibinfo{pages}{1--4}.
\newblock


\bibitem[Condon and Crano(1988)]%
        {69}
\bibfield{author}{\bibinfo{person}{John~W Condon} {and} \bibinfo{person}{William~D Crano}.} \bibinfo{year}{1988}\natexlab{}.
\newblock \showarticletitle{Inferred evaluation and the relation between attitude similarity and interpersonal attraction.}
\newblock \bibinfo{journal}{\emph{Journal of personality and social psychology}} \bibinfo{volume}{54}, \bibinfo{number}{5} (\bibinfo{year}{1988}), \bibinfo{pages}{789}.
\newblock


\bibitem[CSUSM(2022)]%
        {gender}
\bibfield{author}{\bibinfo{person}{CSUSM}.} \bibinfo{year}{2022}\natexlab{}.
\newblock \bibinfo{title}{Inclusive Language Guidelines: Gender Identity}.
\newblock
\newblock
\urldef\tempurl%
\url{https://www.csusm.edu/ipa/surveys/inclusive-language-guidelines.html}
\showURL{%
Retrieved June 20, 2022 from \tempurl}


\bibitem[Da~Silva et~al\mbox{.}(2015)]%
        {DaSilva}
\bibfield{author}{\bibinfo{person}{Odette Da~Silva}, \bibinfo{person}{Nathan Crilly}, {and} \bibinfo{person}{Paul Hekkert}.} \bibinfo{year}{2015}\natexlab{}.
\newblock \showarticletitle{How people’s appreciation of products is affected by their knowledge of the designers’ intentions}.
\newblock \bibinfo{journal}{\emph{International Journal of Design}} \bibinfo{volume}{9}, \bibinfo{number}{2} (\bibinfo{year}{2015}).
\newblock


\bibitem[Dahlb{\"a}ck et~al\mbox{.}(2007)]%
        {76}
\bibfield{author}{\bibinfo{person}{Nils Dahlb{\"a}ck}, \bibinfo{person}{QianYing Wang}, \bibinfo{person}{Clifford Nass}, {and} \bibinfo{person}{Jenny Alwin}.} \bibinfo{year}{2007}\natexlab{}.
\newblock \showarticletitle{Similarity is more important than expertise: Accent effects in speech interfaces}. In \bibinfo{booktitle}{\emph{Proceedings of the SIGCHI conference on Human factors in computing systems}}. \bibinfo{pages}{1553--1556}.
\newblock


\bibitem[Danowitz and Beddoes(2018)]%
        {brazil3}
\bibfield{author}{\bibinfo{person}{Andrew Danowitz} {and} \bibinfo{person}{Kacey Beddoes}.} \bibinfo{year}{2018}\natexlab{}.
\newblock \showarticletitle{Characterizing mental health and wellness in students across engineering disciplines}. In \bibinfo{booktitle}{\emph{2018 The Collaborative Network for Engineering and Computing Diversity Conference Proceedings}}.
\newblock


\bibitem[De~Visser et~al\mbox{.}(2016)]%
        {1}
\bibfield{author}{\bibinfo{person}{Ewart~J De~Visser}, \bibinfo{person}{Samuel~S Monfort}, \bibinfo{person}{Ryan McKendrick}, \bibinfo{person}{Melissa~AB Smith}, \bibinfo{person}{Patrick~E McKnight}, \bibinfo{person}{Frank Krueger}, {and} \bibinfo{person}{Raja Parasuraman}.} \bibinfo{year}{2016}\natexlab{}.
\newblock \showarticletitle{Almost human: Anthropomorphism increases trust resilience in cognitive agents.}
\newblock \bibinfo{journal}{\emph{Journal of Experimental Psychology: Applied}} \bibinfo{volume}{22}, \bibinfo{number}{3} (\bibinfo{year}{2016}), \bibinfo{pages}{331}.
\newblock


\bibitem[Evans et~al\mbox{.}(2018)]%
        {graduateHealth}
\bibfield{author}{\bibinfo{person}{Teresa~M Evans}, \bibinfo{person}{Lindsay Bira}, \bibinfo{person}{Jazmin~Beltran Gastelum}, \bibinfo{person}{L~Todd Weiss}, {and} \bibinfo{person}{Nathan~L Vanderford}.} \bibinfo{year}{2018}\natexlab{}.
\newblock \showarticletitle{Evidence for a mental health crisis in graduate education}.
\newblock \bibinfo{journal}{\emph{Nature biotechnology}} \bibinfo{volume}{36}, \bibinfo{number}{3} (\bibinfo{year}{2018}), \bibinfo{pages}{282--284}.
\newblock


\bibitem[Feij{\'o}o-Garc{\'\i}a et~al\mbox{.}(2023)]%
        {TAP}
\bibfield{author}{\bibinfo{person}{Pedro~Guillermo Feij{\'o}o-Garc{\'\i}a}, \bibinfo{person}{Chase Wrenn}, \bibinfo{person}{Jacob Stuart}, \bibinfo{person}{Alexandre~Gomes De~Siqueira}, {and} \bibinfo{person}{Benjamin Lok}.} \bibinfo{year}{2023}\natexlab{}.
\newblock \showarticletitle{Participatory Design of Virtual Humans for Mental Health Support Among North American Computer Science Students: Voice, Appearance, and the Similarity-attraction Effect}.
\newblock \bibinfo{journal}{\emph{ACM Transactions on Applied Perception}} \bibinfo{volume}{20}, \bibinfo{number}{3} (\bibinfo{year}{2023}), \bibinfo{pages}{1--27}.
\newblock


\bibitem[Feij\'{o}o-Garc\'{\i}a et~al\mbox{.}(2021)]%
        {pedro}
\bibfield{author}{\bibinfo{person}{Pedro~Guillermo Feij\'{o}o-Garc\'{\i}a}, \bibinfo{person}{Mohan Zalake}, \bibinfo{person}{Alexandre~Gomes de Siqueira}, \bibinfo{person}{Benjamin Lok}, {and} \bibinfo{person}{Felix Hamza-Lup}.} \bibinfo{year}{2021}\natexlab{}.
\newblock \bibinfo{booktitle}{\emph{Effects of Virtual Humans' Gender and Spoken Accent on Users' Perceptions of Expertise in Mental Wellness Conversations}}.
\newblock \bibinfo{publisher}{Association for Computing Machinery}, \bibinfo{address}{New York, NY, USA}, \bibinfo{pages}{68–75}.
\newblock
\showISBNx{9781450386197}
\urldef\tempurl%
\url{https://doi.org/10.1145/3472306.3478367}
\showURL{%
\tempurl}


\bibitem[Feij{\'o}o-Garc{\'\i}a et~al\mbox{.}(2022)]%
        {iva2022}
\bibfield{author}{\bibinfo{person}{Pedro~Guillermo Feij{\'o}o-Garc{\'\i}a}, \bibinfo{person}{Mohan Zalake}, \bibinfo{person}{Heng Yao}, \bibinfo{person}{Alexandre~Gomes de Siqueira}, {and} \bibinfo{person}{Benjamin Lok}.} \bibinfo{year}{2022}\natexlab{}.
\newblock \showarticletitle{Can we talk about bruno? exploring virtual human counselors' spoken accents and their impact on users' conversations}. In \bibinfo{booktitle}{\emph{Proceedings of the 22nd ACM International Conference on Intelligent Virtual Agents}}. \bibinfo{pages}{1--7}.
\newblock


\bibitem[Flinchbaugh et~al\mbox{.}(2012)]%
        {gratitude}
\bibfield{author}{\bibinfo{person}{Carol~L Flinchbaugh}, \bibinfo{person}{E~Whitney~G Moore}, \bibinfo{person}{Young~K Chang}, {and} \bibinfo{person}{Douglas~R May}.} \bibinfo{year}{2012}\natexlab{}.
\newblock \showarticletitle{Student well-being interventions: The effects of stress management techniques and gratitude journaling in the management education classroom}.
\newblock \bibinfo{journal}{\emph{Journal of Management Education}} \bibinfo{volume}{36}, \bibinfo{number}{2} (\bibinfo{year}{2012}), \bibinfo{pages}{191--219}.
\newblock


\bibitem[Franke et~al\mbox{.}(2012)]%
        {chisquare}
\bibfield{author}{\bibinfo{person}{Todd~Michael Franke}, \bibinfo{person}{Timothy Ho}, {and} \bibinfo{person}{Christina~A Christie}.} \bibinfo{year}{2012}\natexlab{}.
\newblock \showarticletitle{The chi-square test: Often used and more often misinterpreted}.
\newblock \bibinfo{journal}{\emph{American journal of evaluation}} \bibinfo{volume}{33}, \bibinfo{number}{3} (\bibinfo{year}{2012}), \bibinfo{pages}{448--458}.
\newblock


\bibitem[Frauenberger et~al\mbox{.}(2011)]%
        {good}
\bibfield{author}{\bibinfo{person}{Christopher Frauenberger}, \bibinfo{person}{Judith Good}, {and} \bibinfo{person}{Wendy Keay-Bright}.} \bibinfo{year}{2011}\natexlab{}.
\newblock \showarticletitle{Designing technology for children with special needs: bridging perspectives through participatory design}.
\newblock \bibinfo{journal}{\emph{CoDesign}} \bibinfo{volume}{7}, \bibinfo{number}{1} (\bibinfo{year}{2011}), \bibinfo{pages}{1--28}.
\newblock


\bibitem[Fuchs et~al\mbox{.}(2015)]%
        {Fuchs}
\bibfield{author}{\bibinfo{person}{Christoph Fuchs}, \bibinfo{person}{Martin Schreier}, {and} \bibinfo{person}{Stijn~MJ Van~Osselaer}.} \bibinfo{year}{2015}\natexlab{}.
\newblock \showarticletitle{The handmade effect: What's love got to do with it?}
\newblock \bibinfo{journal}{\emph{Journal of marketing}} \bibinfo{volume}{79}, \bibinfo{number}{2} (\bibinfo{year}{2015}), \bibinfo{pages}{98--110}.
\newblock


\bibitem[Ghosh et~al\mbox{.}(2023)]%
        {rashi}
\bibfield{author}{\bibinfo{person}{Rashi Ghosh}, \bibinfo{person}{Pedro~Guillermo Feijóo-García}, \bibinfo{person}{Jacob Stuart}, \bibinfo{person}{Chase Wrenn}, {and} \bibinfo{person}{Benjamin Lok}.} \bibinfo{year}{2023}\natexlab{}.
\newblock \showarticletitle{Evaluating face gender cues in virtual humans within and beyond the gender binary}.
\newblock \bibinfo{journal}{\emph{Frontiers in Virtual Reality}}  \bibinfo{volume}{4} (\bibinfo{year}{2023}).
\newblock
\showISSN{2673-4192}
\urldef\tempurl%
\url{https://doi.org/10.3389/frvir.2023.1251420}
\showDOI{\tempurl}


\bibitem[Gorenstein et~al\mbox{.}(1995)]%
        {brazil2}
\bibfield{author}{\bibinfo{person}{Clarice Gorenstein}, \bibinfo{person}{Sabine Pomp{\'e}ia}, {and} \bibinfo{person}{Laura Andrade}.} \bibinfo{year}{1995}\natexlab{}.
\newblock \showarticletitle{Scores of Brazilian university students on the Beck depression and the state-trait anxiety inventories}.
\newblock \bibinfo{journal}{\emph{Psychological Reports}} \bibinfo{volume}{77}, \bibinfo{number}{2} (\bibinfo{year}{1995}), \bibinfo{pages}{635--641}.
\newblock


\bibitem[Guadagno et~al\mbox{.}(2007)]%
        {guadagno2007virtual}
\bibfield{author}{\bibinfo{person}{Rosanna~E Guadagno}, \bibinfo{person}{Jim Blascovich}, \bibinfo{person}{Jeremy~N Bailenson}, {and} \bibinfo{person}{Cade McCall}.} \bibinfo{year}{2007}\natexlab{}.
\newblock \showarticletitle{Virtual humans and persuasion: The effects of agency and behavioral realism}.
\newblock \bibinfo{journal}{\emph{Media Psychology}} \bibinfo{volume}{10}, \bibinfo{number}{1} (\bibinfo{year}{2007}), \bibinfo{pages}{1--22}.
\newblock


\bibitem[Guadagno et~al\mbox{.}(2011)]%
        {90}
\bibfield{author}{\bibinfo{person}{Rosanna~E Guadagno}, \bibinfo{person}{Kimberly~R Swinth}, {and} \bibinfo{person}{Jim Blascovich}.} \bibinfo{year}{2011}\natexlab{}.
\newblock \showarticletitle{Social evaluations of embodied agents and avatars}.
\newblock \bibinfo{journal}{\emph{Computers in Human Behavior}} \bibinfo{volume}{27}, \bibinfo{number}{6} (\bibinfo{year}{2011}), \bibinfo{pages}{2380--2385}.
\newblock


\bibitem[Gwet(2008)]%
        {cohen1}
\bibfield{author}{\bibinfo{person}{Kilem~Li Gwet}.} \bibinfo{year}{2008}\natexlab{}.
\newblock \showarticletitle{Computing inter-rater reliability and its variance in the presence of high agreement}.
\newblock \bibinfo{journal}{\emph{Brit. J. Math. Statist. Psych.}} \bibinfo{volume}{61}, \bibinfo{number}{1} (\bibinfo{year}{2008}), \bibinfo{pages}{29--48}.
\newblock


\bibitem[Halan et~al\mbox{.}(2014)]%
        {halan2014}
\bibfield{author}{\bibinfo{person}{Shivashankar Halan}, \bibinfo{person}{Benjamin Lok}, \bibinfo{person}{Isaac Sia}, {and} \bibinfo{person}{Michael Crary}.} \bibinfo{year}{2014}\natexlab{}.
\newblock \showarticletitle{Virtual agent constructionism: experiences from health professions students creating virtual conversational agent representations of patients}. In \bibinfo{booktitle}{\emph{2014 IEEE 14th International Conference on Advanced Learning Technologies}}. IEEE, \bibinfo{pages}{249--253}.
\newblock


\bibitem[Halan et~al\mbox{.}(2015)]%
        {halan2015}
\bibfield{author}{\bibinfo{person}{Shivashankar Halan}, \bibinfo{person}{Isaac Sia}, \bibinfo{person}{Michael Crary}, {and} \bibinfo{person}{Benjamin Lok}.} \bibinfo{year}{2015}\natexlab{}.
\newblock \showarticletitle{Exploring the effects of healthcare students creating virtual patients for empathy training}. In \bibinfo{booktitle}{\emph{International Conference on Intelligent Virtual Agents}}. Springer, \bibinfo{pages}{239--249}.
\newblock


\bibitem[Hatcher et~al\mbox{.}(2005)]%
        {hatcher2005}
\bibfield{author}{\bibinfo{person}{Sherry~L Hatcher}, \bibinfo{person}{Todd~K Favorite}, \bibinfo{person}{Elizabeth~A Hardy}, \bibinfo{person}{Robert~L Goode}, \bibinfo{person}{Linda~A DeShetler}, {and} \bibinfo{person}{Rosa~M Thomas}.} \bibinfo{year}{2005}\natexlab{}.
\newblock \showarticletitle{An Analogue Study of Therapist Empathic Process: Working With Difference.}
\newblock \bibinfo{journal}{\emph{Psychotherapy: Theory, Research, Practice, Training}} \bibinfo{volume}{42}, \bibinfo{number}{2} (\bibinfo{year}{2005}), \bibinfo{pages}{198}.
\newblock


\bibitem[Hennig-Thurau et~al\mbox{.}(2007)]%
        {cinema}
\bibfield{author}{\bibinfo{person}{Thorsten Hennig-Thurau}, \bibinfo{person}{Mark~B Houston}, {and} \bibinfo{person}{Gianfranco Walsh}.} \bibinfo{year}{2007}\natexlab{}.
\newblock \showarticletitle{Determinants of motion picture box office and profitability: an interrelationship approach}.
\newblock \bibinfo{journal}{\emph{Review of Managerial Science}} \bibinfo{volume}{1}, \bibinfo{number}{1} (\bibinfo{year}{2007}), \bibinfo{pages}{65--92}.
\newblock


\bibitem[Hofstede(2010)]%
        {80}
\bibfield{author}{\bibinfo{person}{G Hofstede}.} \bibinfo{year}{2010}\natexlab{}.
\newblock \showarticletitle{Culture and Organizations: Software of the Mind: Intercultural Cooperation and its Importance for Survival--3rd ed.}
\newblock  (\bibinfo{year}{2010}).
\newblock


\bibitem[Iacobelli and Cassell(2007)]%
        {17}
\bibfield{author}{\bibinfo{person}{Francisco Iacobelli} {and} \bibinfo{person}{Justine Cassell}.} \bibinfo{year}{2007}\natexlab{}.
\newblock \showarticletitle{Ethnic identity and engagement in embodied conversational agents}. In \bibinfo{booktitle}{\emph{International Workshop on Intelligent Virtual Agents}}. Springer, \bibinfo{pages}{57--63}.
\newblock


\bibitem[Idemen et~al\mbox{.}(2021)]%
        {idemen}
\bibfield{author}{\bibinfo{person}{Elif Idemen}, \bibinfo{person}{Ayse~Banu Elmadag}, {and} \bibinfo{person}{Mehmet Okan}.} \bibinfo{year}{2021}\natexlab{}.
\newblock \showarticletitle{A qualitative approach to designer as a product cue: proposed conceptual model of consumers perceptions and attitudes}.
\newblock \bibinfo{journal}{\emph{Review of Managerial Science}} \bibinfo{volume}{15}, \bibinfo{number}{5} (\bibinfo{year}{2021}), \bibinfo{pages}{1281--1309}.
\newblock


\bibitem[Jan et~al\mbox{.}(2007)]%
        {20}
\bibfield{author}{\bibinfo{person}{Du{\v{s}}an Jan}, \bibinfo{person}{David Herrera}, \bibinfo{person}{Bilyana Martinovski}, \bibinfo{person}{David Novick}, {and} \bibinfo{person}{David Traum}.} \bibinfo{year}{2007}\natexlab{}.
\newblock \showarticletitle{A computational model of culture-specific conversational behavior}. In \bibinfo{booktitle}{\emph{International Workshop on Intelligent Virtual Agents}}. Springer, \bibinfo{pages}{45--56}.
\newblock


\bibitem[Jin and Youn(2021)]%
        {5}
\bibfield{author}{\bibinfo{person}{S~Venus Jin} {and} \bibinfo{person}{Seounmi Youn}.} \bibinfo{year}{2021}\natexlab{}.
\newblock \showarticletitle{Why do consumers with social phobia prefer anthropomorphic customer service chatbots? Evolutionary explanations of the moderating roles of social phobia}.
\newblock \bibinfo{journal}{\emph{Telematics and Informatics}}  \bibinfo{volume}{62} (\bibinfo{year}{2021}), \bibinfo{pages}{101644}.
\newblock


\bibitem[Joosse et~al\mbox{.}(2013)]%
        {9}
\bibfield{author}{\bibinfo{person}{Michiel Joosse}, \bibinfo{person}{Manja Lohse}, \bibinfo{person}{Jorge~Gallego P{\'e}rez}, {and} \bibinfo{person}{Vanessa Evers}.} \bibinfo{year}{2013}\natexlab{}.
\newblock \showarticletitle{What you do is who you are: The role of task context in perceived social robot personality}. In \bibinfo{booktitle}{\emph{2013 IEEE International Conference on Robotics and Automation}}. IEEE, \bibinfo{pages}{2134--2139}.
\newblock


\bibitem[Joy et~al\mbox{.}(2014)]%
        {fashion}
\bibfield{author}{\bibinfo{person}{Annamma Joy}, \bibinfo{person}{Jeff~Jianfeng Wang}, \bibinfo{person}{Tsang-Sing Chan}, \bibinfo{person}{John~F Sherry~Jr}, {and} \bibinfo{person}{Geng Cui}.} \bibinfo{year}{2014}\natexlab{}.
\newblock \showarticletitle{M (Art) worlds: consumer perceptions of how luxury brand stores become art institutions}.
\newblock \bibinfo{journal}{\emph{Journal of Retailing}} \bibinfo{volume}{90}, \bibinfo{number}{3} (\bibinfo{year}{2014}), \bibinfo{pages}{347--364}.
\newblock


\bibitem[Kim et~al\mbox{.}(2007)]%
        {91}
\bibfield{author}{\bibinfo{person}{Yanghee Kim}, \bibinfo{person}{Amy~L Baylor}, {and} \bibinfo{person}{Entong Shen}.} \bibinfo{year}{2007}\natexlab{}.
\newblock \showarticletitle{Pedagogical agents as learning companions: the impact of agent emotion and gender}.
\newblock \bibinfo{journal}{\emph{Journal of Computer Assisted Learning}} \bibinfo{volume}{23}, \bibinfo{number}{3} (\bibinfo{year}{2007}), \bibinfo{pages}{220--234}.
\newblock


\bibitem[Koda and Takeda(2018)]%
        {88}
\bibfield{author}{\bibinfo{person}{Tomoko Koda} {and} \bibinfo{person}{Yuuki Takeda}.} \bibinfo{year}{2018}\natexlab{}.
\newblock \showarticletitle{Perception of Culture-specific Gaze Behaviors of Agents and Gender Effects}. In \bibinfo{booktitle}{\emph{Proceedings of the 6th International Conference on Human-Agent Interaction}}. \bibinfo{pages}{138--143}.
\newblock


\bibitem[Kruskal and Wallis(1952)]%
        {kruskal}
\bibfield{author}{\bibinfo{person}{William~H Kruskal} {and} \bibinfo{person}{W~Allen Wallis}.} \bibinfo{year}{1952}\natexlab{}.
\newblock \showarticletitle{Use of ranks in one-criterion variance analysis}.
\newblock \bibinfo{journal}{\emph{Journal of the American statistical Association}} \bibinfo{volume}{47}, \bibinfo{number}{260} (\bibinfo{year}{1952}), \bibinfo{pages}{583--621}.
\newblock


\bibitem[Kulms et~al\mbox{.}(2011)]%
        {19}
\bibfield{author}{\bibinfo{person}{Philipp Kulms}, \bibinfo{person}{Nicole~C Kr{\"a}mer}, \bibinfo{person}{Jonathan Gratch}, {and} \bibinfo{person}{Sin-Hwa Kang}.} \bibinfo{year}{2011}\natexlab{}.
\newblock \showarticletitle{It’s in their eyes: A study on female and male virtual humans’ gaze}. In \bibinfo{booktitle}{\emph{International workshop on intelligent virtual agents}}. Springer, \bibinfo{pages}{80--92}.
\newblock


\bibitem[Lee et~al\mbox{.}(2000)]%
        {84}
\bibfield{author}{\bibinfo{person}{Eun~Ju Lee}, \bibinfo{person}{Clifford Nass}, {and} \bibinfo{person}{Scott Brave}.} \bibinfo{year}{2000}\natexlab{}.
\newblock \showarticletitle{Can computer-generated speech have gender? An experimental test of gender stereotype}. In \bibinfo{booktitle}{\emph{CHI'00 extended abstracts on Human factors in computing systems}}. \bibinfo{pages}{289--290}.
\newblock


\bibitem[Luaces et~al\mbox{.}(2012)]%
        {jaccard}
\bibfield{author}{\bibinfo{person}{Oscar Luaces}, \bibinfo{person}{Jorge D{\'\i}ez}, \bibinfo{person}{Jos{\'e} Barranquero}, \bibinfo{person}{Juan~Jos{\'e} del Coz}, {and} \bibinfo{person}{Antonio Bahamonde}.} \bibinfo{year}{2012}\natexlab{}.
\newblock \showarticletitle{Binary relevance efficacy for multilabel classification}.
\newblock \bibinfo{journal}{\emph{Progress in Artificial Intelligence}}  \bibinfo{volume}{1} (\bibinfo{year}{2012}), \bibinfo{pages}{303--313}.
\newblock


\bibitem[Lucas et~al\mbox{.}(2018)]%
        {38}
\bibfield{author}{\bibinfo{person}{Gale~M Lucas}, \bibinfo{person}{Jill Boberg}, \bibinfo{person}{David Traum}, \bibinfo{person}{Ron Artstein}, \bibinfo{person}{Jonathan Gratch}, \bibinfo{person}{Alesia Gainer}, \bibinfo{person}{Emmanuel Johnson}, \bibinfo{person}{Anton Leuski}, {and} \bibinfo{person}{Mikio Nakano}.} \bibinfo{year}{2018}\natexlab{}.
\newblock \showarticletitle{Culture, errors, and rapport-building dialogue in social agents}. In \bibinfo{booktitle}{\emph{Proceedings of the 18th International Conference on intelligent virtual agents}}. \bibinfo{pages}{51--58}.
\newblock


\bibitem[Lucas et~al\mbox{.}(2014)]%
        {lucasDisclosure}
\bibfield{author}{\bibinfo{person}{Gale~M. Lucas}, \bibinfo{person}{Jonathan Gratch}, \bibinfo{person}{Aisha King}, {and} \bibinfo{person}{Louis-Philippe Morency}.} \bibinfo{year}{2014}\natexlab{}.
\newblock \showarticletitle{It’s only a computer: Virtual humans increase willingness to disclose}.
\newblock \bibinfo{journal}{\emph{Computers in Human Behavior}}  \bibinfo{volume}{37} (\bibinfo{year}{2014}), \bibinfo{pages}{94--100}.
\newblock
\showISSN{0747-5632}
\urldef\tempurl%
\url{https://doi.org/10.1016/j.chb.2014.04.043}
\showDOI{\tempurl}


\bibitem[Lugrin et~al\mbox{.}(2018)]%
        {11}
\bibfield{author}{\bibinfo{person}{Birgit Lugrin}, \bibinfo{person}{Andrea Bartl}, \bibinfo{person}{Hendrik Striepe}, \bibinfo{person}{Jennifer Lax}, {and} \bibinfo{person}{Takashi Toriizuka}.} \bibinfo{year}{2018}\natexlab{}.
\newblock \showarticletitle{Do I act familiar? Investigating the Similarity-Attraction Principle on Culture-specific Communicative behaviour for Social Robots}. In \bibinfo{booktitle}{\emph{2018 IEEE/RSJ International Conference on Intelligent Robots and Systems (IROS)}}. IEEE, \bibinfo{pages}{2033--2039}.
\newblock


\bibitem[McKight and Najab(2010)]%
        {kruskal1}
\bibfield{author}{\bibinfo{person}{Patrick~E McKight} {and} \bibinfo{person}{Julius Najab}.} \bibinfo{year}{2010}\natexlab{}.
\newblock \showarticletitle{Kruskal-wallis test}.
\newblock \bibinfo{journal}{\emph{The corsini encyclopedia of psychology}} (\bibinfo{year}{2010}), \bibinfo{pages}{1--1}.
\newblock


\bibitem[Moulard et~al\mbox{.}(2014)]%
        {art2}
\bibfield{author}{\bibinfo{person}{Julie~Guidry Moulard}, \bibinfo{person}{Dan~Hamilton Rice}, \bibinfo{person}{Carolyn~Popp Garrity}, {and} \bibinfo{person}{Stephanie~M Mangus}.} \bibinfo{year}{2014}\natexlab{}.
\newblock \showarticletitle{Artist authenticity: How artists’ passion and commitment shape consumers’ perceptions and behavioral intentions across genders}.
\newblock \bibinfo{journal}{\emph{Psychology \& Marketing}} \bibinfo{volume}{31}, \bibinfo{number}{8} (\bibinfo{year}{2014}), \bibinfo{pages}{576--590}.
\newblock


\bibitem[Muller and Kuhn(1993)]%
        {muller}
\bibfield{author}{\bibinfo{person}{Michael~J Muller} {and} \bibinfo{person}{Sarah Kuhn}.} \bibinfo{year}{1993}\natexlab{}.
\newblock \showarticletitle{Participatory design}.
\newblock \bibinfo{journal}{\emph{Commun. ACM}} \bibinfo{volume}{36}, \bibinfo{number}{6} (\bibinfo{year}{1993}), \bibinfo{pages}{24--28}.
\newblock


\bibitem[Nag and Yal{\c{c}}{\i}n(2020)]%
        {40}
\bibfield{author}{\bibinfo{person}{Procheta Nag} {and} \bibinfo{person}{{\"O}zge~Nilay Yal{\c{c}}{\i}n}.} \bibinfo{year}{2020}\natexlab{}.
\newblock \showarticletitle{Gender stereotypes in virtual agents}. In \bibinfo{booktitle}{\emph{Proceedings of the 20th ACM International Conference on Intelligent Virtual Agents}}. \bibinfo{pages}{1--8}.
\newblock


\bibitem[Norouzi et~al\mbox{.}(2018)]%
        {36}
\bibfield{author}{\bibinfo{person}{Nahal Norouzi}, \bibinfo{person}{Kangsoo Kim}, \bibinfo{person}{Jason Hochreiter}, \bibinfo{person}{Myungho Lee}, \bibinfo{person}{Salam Daher}, \bibinfo{person}{Gerd Bruder}, {and} \bibinfo{person}{Greg Welch}.} \bibinfo{year}{2018}\natexlab{}.
\newblock \showarticletitle{A systematic survey of 15 years of user studies published in the intelligent virtual agents conference}. In \bibinfo{booktitle}{\emph{Proceedings of the 18th international conference on intelligent virtual agents}}. \bibinfo{pages}{17--22}.
\newblock


\bibitem[Oliveira et~al\mbox{.}(2020)]%
        {2}
\bibfield{author}{\bibinfo{person}{Raquel Oliveira}, \bibinfo{person}{Patr{\'\i}cia Arriaga}, \bibinfo{person}{Minja Axelsson}, {and} \bibinfo{person}{Ana Paiva}.} \bibinfo{year}{2020}\natexlab{}.
\newblock \showarticletitle{Humor-Robot Interaction: A Scoping Review of the Literature and Future Directions}.
\newblock \bibinfo{journal}{\emph{International Journal of Social Robotics}} (\bibinfo{year}{2020}), \bibinfo{pages}{1--15}.
\newblock


\bibitem[Palan and Schitter(2018)]%
        {prolific}
\bibfield{author}{\bibinfo{person}{Stefan Palan} {and} \bibinfo{person}{Christian Schitter}.} \bibinfo{year}{2018}\natexlab{}.
\newblock \showarticletitle{Prolific. ac—A subject pool for online experiments}.
\newblock \bibinfo{journal}{\emph{Journal of Behavioral and Experimental Finance}}  \bibinfo{volume}{17} (\bibinfo{year}{2018}), \bibinfo{pages}{22--27}.
\newblock


\bibitem[Parmar et~al\mbox{.}(2018)]%
        {30}
\bibfield{author}{\bibinfo{person}{Dhaval Parmar}, \bibinfo{person}{Stefan Olafsson}, \bibinfo{person}{Dina Utami}, {and} \bibinfo{person}{Timothy Bickmore}.} \bibinfo{year}{2018}\natexlab{}.
\newblock \showarticletitle{Looking the part: The effect of attire and setting on perceptions of a virtual health counselor}. In \bibinfo{booktitle}{\emph{Proceedings of the 18th international conference on intelligent virtual agents}}. \bibinfo{pages}{301--306}.
\newblock


\bibitem[Parviainen and S{\o}ndergaard(2020)]%
        {3}
\bibfield{author}{\bibinfo{person}{Emmi Parviainen} {and} \bibinfo{person}{Marie Louise~Juul S{\o}ndergaard}.} \bibinfo{year}{2020}\natexlab{}.
\newblock \showarticletitle{Experiential qualities of whispering with voice assistants}. In \bibinfo{booktitle}{\emph{Proceedings of the 2020 CHI Conference on Human Factors in Computing Systems}}. \bibinfo{pages}{1--13}.
\newblock


\bibitem[Pratt et~al\mbox{.}(2007)]%
        {85}
\bibfield{author}{\bibinfo{person}{Jean~A Pratt}, \bibinfo{person}{Karina Hauser}, \bibinfo{person}{Zsolt Ugray}, {and} \bibinfo{person}{Olga Patterson}.} \bibinfo{year}{2007}\natexlab{}.
\newblock \showarticletitle{Looking at human--computer interface design: Effects of ethnicity in computer agents}.
\newblock \bibinfo{journal}{\emph{Interacting with Computers}} \bibinfo{volume}{19}, \bibinfo{number}{4} (\bibinfo{year}{2007}), \bibinfo{pages}{512--523}.
\newblock


\bibitem[Rosenstein et~al\mbox{.}(2020)]%
        {porter}
\bibfield{author}{\bibinfo{person}{Adam Rosenstein}, \bibinfo{person}{Aishma Raghu}, {and} \bibinfo{person}{Leo Porter}.} \bibinfo{year}{2020}\natexlab{}.
\newblock \showarticletitle{Identifying the Prevalence of the Impostor Phenomenon Among Computer Science Students}. In \bibinfo{booktitle}{\emph{Proceedings of the 51st ACM Technical Symposium on Computer Science Education}} (Portland, OR, USA) \emph{(\bibinfo{series}{SIGCSE '20})}. \bibinfo{publisher}{Association for Computing Machinery}, \bibinfo{address}{New York, NY, USA}, \bibinfo{pages}{30–36}.
\newblock
\showISBNx{9781450367936}
\urldef\tempurl%
\url{https://doi.org/10.1145/3328778.3366815}
\showDOI{\tempurl}


\bibitem[Rossen et~al\mbox{.}(2008)]%
        {18}
\bibfield{author}{\bibinfo{person}{Brent Rossen}, \bibinfo{person}{Kyle Johnsen}, \bibinfo{person}{Adeline Deladisma}, \bibinfo{person}{Scott Lind}, {and} \bibinfo{person}{Benjamin Lok}.} \bibinfo{year}{2008}\natexlab{}.
\newblock \showarticletitle{Virtual humans elicit skin-tone bias consistent with real-world skin-tone biases}. In \bibinfo{booktitle}{\emph{International Workshop on Intelligent Virtual Agents}}. Springer, \bibinfo{pages}{237--244}.
\newblock


\bibitem[Seaborn et~al\mbox{.}(2021)]%
        {13}
\bibfield{author}{\bibinfo{person}{Katie Seaborn}, \bibinfo{person}{Norihisa~P Miyake}, \bibinfo{person}{Peter Pennefather}, {and} \bibinfo{person}{Mihoko Otake-Matsuura}.} \bibinfo{year}{2021}\natexlab{}.
\newblock \showarticletitle{Voice in Human--Agent Interaction: A Survey}.
\newblock \bibinfo{journal}{\emph{ACM Computing Surveys (CSUR)}} \bibinfo{volume}{54}, \bibinfo{number}{4} (\bibinfo{year}{2021}), \bibinfo{pages}{1--43}.
\newblock


\bibitem[Simonsen and Robertson(2013)]%
        {simonsen}
\bibfield{author}{\bibinfo{person}{Jesper Simonsen} {and} \bibinfo{person}{Toni Robertson}.} \bibinfo{year}{2013}\natexlab{}.
\newblock \bibinfo{booktitle}{\emph{Routledge international handbook of participatory design}}. Vol.~\bibinfo{volume}{711}.
\newblock \bibinfo{publisher}{Routledge New York}.
\newblock


\bibitem[Singh(1974)]%
        {70}
\bibfield{author}{\bibinfo{person}{Ramadhar Singh}.} \bibinfo{year}{1974}\natexlab{}.
\newblock \showarticletitle{Reinforcement and attraction specifying the effects of affective states}.
\newblock \bibinfo{journal}{\emph{Journal of Research in Personality}} \bibinfo{volume}{8}, \bibinfo{number}{3} (\bibinfo{year}{1974}), \bibinfo{pages}{294--305}.
\newblock


\bibitem[Smith et~al\mbox{.}(2016)]%
        {music}
\bibfield{author}{\bibinfo{person}{Rosanna~K Smith}, \bibinfo{person}{George~E Newman}, {and} \bibinfo{person}{Ravi Dhar}.} \bibinfo{year}{2016}\natexlab{}.
\newblock \showarticletitle{Closer to the creator: Temporal contagion explains the preference for earlier serial numbers}.
\newblock \bibinfo{journal}{\emph{Journal of Consumer Research}} \bibinfo{volume}{42}, \bibinfo{number}{5} (\bibinfo{year}{2016}), \bibinfo{pages}{653--668}.
\newblock


\bibitem[Soares~Passos et~al\mbox{.}(2020)]%
        {brazil1}
\bibfield{author}{\bibinfo{person}{L\'{\i}gia~Maria Soares~Passos}, \bibinfo{person}{Christian Murphy}, \bibinfo{person}{Rita Zhen~Chen}, \bibinfo{person}{Marcos Gon\c{c}alves~de Santana}, {and} \bibinfo{person}{Giselle Soares~Passos}.} \bibinfo{year}{2020}\natexlab{}.
\newblock \showarticletitle{The Prevalence of Anxiety and Depression Symptoms among Brazilian Computer Science Students}. In \bibinfo{booktitle}{\emph{Proceedings of the 51st ACM Technical Symposium on Computer Science Education}} (Portland, OR, USA) \emph{(\bibinfo{series}{SIGCSE '20})}. \bibinfo{publisher}{Association for Computing Machinery}, \bibinfo{address}{New York, NY, USA}, \bibinfo{pages}{316–322}.
\newblock
\showISBNx{9781450367936}
\urldef\tempurl%
\url{https://doi.org/10.1145/3328778.3366836}
\showDOI{\tempurl}


\bibitem[Sutton et~al\mbox{.}(2019)]%
        {15}
\bibfield{author}{\bibinfo{person}{Selina~Jeanne Sutton}, \bibinfo{person}{Paul Foulkes}, \bibinfo{person}{David Kirk}, {and} \bibinfo{person}{Shaun Lawson}.} \bibinfo{year}{2019}\natexlab{}.
\newblock \showarticletitle{Voice as a design material: Sociophonetic inspired design strategies in human-computer interaction}. In \bibinfo{booktitle}{\emph{Proceedings of the 2019 CHI Conference on Human Factors in Computing Systems}}. \bibinfo{pages}{1--14}.
\newblock


\bibitem[Synthesia(2022)]%
        {synthesia}
\bibfield{author}{\bibinfo{person}{Synthesia}.} \bibinfo{year}{2022}\natexlab{}.
\newblock \bibinfo{title}{Synthesia: AI video generation platform}.
\newblock
\newblock
\urldef\tempurl%
\url{https://www.synthesia.io/}
\showURL{%
Retrieved June 20, 2022 from \tempurl}


\bibitem[{ter Stal} et~al\mbox{.}(2020)]%
        {genderreview}
\bibfield{author}{\bibinfo{person}{Silke {ter Stal}}, \bibinfo{person}{Lean~Leonie Kramer}, \bibinfo{person}{Monique Tabak}, \bibinfo{person}{Harm {op den Akker}}, {and} \bibinfo{person}{Hermie Hermens}.} \bibinfo{year}{2020}\natexlab{}.
\newblock \showarticletitle{Design Features of Embodied Conversational Agents in eHealth: a Literature Review}.
\newblock \bibinfo{journal}{\emph{International Journal of Human-Computer Studies}}  \bibinfo{volume}{138} (\bibinfo{year}{2020}), \bibinfo{pages}{102409}.
\newblock
\showISSN{1071-5819}
\urldef\tempurl%
\url{https://doi.org/10.1016/j.ijhcs.2020.102409}
\showDOI{\tempurl}


\bibitem[ter Stal et~al\mbox{.}(2020)]%
        {RaganGender}
\bibfield{author}{\bibinfo{person}{Silke ter Stal}, \bibinfo{person}{Monique Tabak}, \bibinfo{person}{Harm op~den Akker}, \bibinfo{person}{Tessa Beinema}, {and} \bibinfo{person}{Hermie Hermens}.} \bibinfo{year}{2020}\natexlab{}.
\newblock \showarticletitle{Who do you prefer? The effect of age, gender and role on users’ first impressions of embodied conversational agents in eHealth}.
\newblock \bibinfo{journal}{\emph{International Journal of Human--Computer Interaction}} \bibinfo{volume}{36}, \bibinfo{number}{9} (\bibinfo{year}{2020}), \bibinfo{pages}{881--892}.
\newblock


\bibitem[USA(2017)]%
        {demographics}
\bibfield{author}{\bibinfo{person}{DATA USA}.} \bibinfo{year}{2017}\natexlab{}.
\newblock \bibinfo{title}{US Counselors Demographics}.
\newblock
\newblock
\urldef\tempurl%
\url{https://datausa.io/profile/soc/counselors}
\showURL{%
Retrieved April 17, 2022 from \tempurl}


\bibitem[Van~Mechelen et~al\mbox{.}(2021)]%
        {participatoryDesignLiteratureReview}
\bibfield{author}{\bibinfo{person}{Maarten Van~Mechelen}, \bibinfo{person}{Line Have~Musaeus}, \bibinfo{person}{Ole~Sejer Iversen}, \bibinfo{person}{Christian Dindler}, {and} \bibinfo{person}{Arthur Hjorth}.} \bibinfo{year}{2021}\natexlab{}.
\newblock \showarticletitle{A Systematic Review of Empowerment in Child-Computer Interaction Research}. In \bibinfo{booktitle}{\emph{Interaction Design and Children}} (Athens, Greece) \emph{(\bibinfo{series}{IDC '21})}. \bibinfo{publisher}{Association for Computing Machinery}, \bibinfo{address}{New York, NY, USA}, \bibinfo{pages}{119–130}.
\newblock
\showISBNx{9781450384520}
\urldef\tempurl%
\url{https://doi.org/10.1145/3459990.3460701}
\showDOI{\tempurl}


\bibitem[Vasalou et~al\mbox{.}(2021)]%
        {vasalou}
\bibfield{author}{\bibinfo{person}{Asimina Vasalou}, \bibinfo{person}{Seray Ibrahim}, \bibinfo{person}{Michael Clarke}, {and} \bibinfo{person}{Yvonne Griffiths}.} \bibinfo{year}{2021}\natexlab{}.
\newblock \showarticletitle{On power and participation: Reflections from design with developmentally diverse children}.
\newblock \bibinfo{journal}{\emph{International Journal of Child-Computer Interaction}}  \bibinfo{volume}{27} (\bibinfo{year}{2021}), \bibinfo{pages}{100241}.
\newblock


\bibitem[Viera et~al\mbox{.}(2005)]%
        {cohen2}
\bibfield{author}{\bibinfo{person}{Anthony~J Viera}, \bibinfo{person}{Joanne~M Garrett}, {et~al\mbox{.}}} \bibinfo{year}{2005}\natexlab{}.
\newblock \showarticletitle{Understanding interobserver agreement: the kappa statistic}.
\newblock \bibinfo{journal}{\emph{Fam med}} \bibinfo{volume}{37}, \bibinfo{number}{5} (\bibinfo{year}{2005}), \bibinfo{pages}{360--363}.
\newblock


\bibitem[Wang et~al\mbox{.}(2020)]%
        {61}
\bibfield{author}{\bibinfo{person}{Isaac Wang}, \bibinfo{person}{Lea Buchweitz}, \bibinfo{person}{Jesse Smith}, \bibinfo{person}{Lara-Sophie Bornholdt}, \bibinfo{person}{Jonas Grund}, \bibinfo{person}{Jaime Ruiz}, {and} \bibinfo{person}{Oliver Korn}.} \bibinfo{year}{2020}\natexlab{}.
\newblock \showarticletitle{Wow, You Are Terrible at This! An Intercultural Study on Virtual Agents Giving Mixed Feedback}. In \bibinfo{booktitle}{\emph{Proceedings of the 20th ACM International Conference on Intelligent Virtual Agents}}. \bibinfo{pages}{1--8}.
\newblock


\bibitem[Whittaker et~al\mbox{.}(2021)]%
        {whittaker2021}
\bibfield{author}{\bibinfo{person}{Steve Whittaker}, \bibinfo{person}{Yvonne Rogers}, \bibinfo{person}{Elena Petrovskaya}, {and} \bibinfo{person}{Hongbin Zhuang}.} \bibinfo{year}{2021}\natexlab{}.
\newblock \showarticletitle{Designing personas for expressive robots: Personality in the new breed of moving, speaking, and colorful social home robots}.
\newblock \bibinfo{journal}{\emph{ACM Transactions on Human-Robot Interaction (THRI)}} \bibinfo{volume}{10}, \bibinfo{number}{1} (\bibinfo{year}{2021}), \bibinfo{pages}{1--25}.
\newblock


\bibitem[Wintersteen et~al\mbox{.}(2005)]%
        {wintersteen2005}
\bibfield{author}{\bibinfo{person}{Matthew~B Wintersteen}, \bibinfo{person}{Janell~L Mensinger}, {and} \bibinfo{person}{Guy~S Diamond}.} \bibinfo{year}{2005}\natexlab{}.
\newblock \showarticletitle{Do gender and racial differences between patient and therapist affect therapeutic alliance and treatment retention in adolescents?}
\newblock \bibinfo{journal}{\emph{Professional Psychology: Research and Practice}} \bibinfo{volume}{36}, \bibinfo{number}{4} (\bibinfo{year}{2005}), \bibinfo{pages}{400}.
\newblock


\bibitem[Wobbrock et~al\mbox{.}(2009)]%
        {wobbrock}
\bibfield{author}{\bibinfo{person}{Jacob~O. Wobbrock}, \bibinfo{person}{Meredith~Ringel Morris}, {and} \bibinfo{person}{Andrew~D. Wilson}.} \bibinfo{year}{2009}\natexlab{}.
\newblock \showarticletitle{User-Defined Gestures for Surface Computing}. In \bibinfo{booktitle}{\emph{Proceedings of the SIGCHI Conference on Human Factors in Computing Systems}} (Boston, MA, USA) \emph{(\bibinfo{series}{CHI '09})}. \bibinfo{publisher}{Association for Computing Machinery}, \bibinfo{address}{New York, NY, USA}, \bibinfo{pages}{1083–1092}.
\newblock
\showISBNx{9781605582467}
\urldef\tempurl%
\url{https://doi.org/10.1145/1518701.1518866}
\showDOI{\tempurl}


\bibitem[Wu et~al\mbox{.}(2017)]%
        {wu}
\bibfield{author}{\bibinfo{person}{Freeman Wu}, \bibinfo{person}{Adriana Samper}, \bibinfo{person}{Andrea~C Morales}, {and} \bibinfo{person}{Gavan~J Fitzsimons}.} \bibinfo{year}{2017}\natexlab{}.
\newblock \showarticletitle{It’s too pretty to use! When and how enhanced product aesthetics discourage usage and lower consumption enjoyment}.
\newblock \bibinfo{journal}{\emph{Journal of Consumer Research}} \bibinfo{volume}{44}, \bibinfo{number}{3} (\bibinfo{year}{2017}), \bibinfo{pages}{651--672}.
\newblock


\bibitem[Yao et~al\mbox{.}(2020)]%
        {43}
\bibfield{author}{\bibinfo{person}{Heng Yao}, \bibinfo{person}{Alexandre~Gomes de Siqueira}, \bibinfo{person}{Adriana Foster}, \bibinfo{person}{Igor Galynker}, {and} \bibinfo{person}{Benjamin Lok}.} \bibinfo{year}{2020}\natexlab{}.
\newblock \showarticletitle{Toward Automated Evaluation of Empathetic Responses in Virtual Human Interaction Systems for Mental Health Scenarios}. In \bibinfo{booktitle}{\emph{Proceedings of the 20th ACM International Conference on Intelligent Virtual Agents}}. \bibinfo{pages}{1--8}.
\newblock


\bibitem[Yin et~al\mbox{.}(2010)]%
        {bickmoreLinguistics}
\bibfield{author}{\bibinfo{person}{Langxuan Yin}, \bibinfo{person}{Timothy Bickmore}, {and} \bibinfo{person}{Dharma~E Cort{\'e}s}.} \bibinfo{year}{2010}\natexlab{}.
\newblock \showarticletitle{The impact of linguistic and cultural congruity on persuasion by conversational agents}. In \bibinfo{booktitle}{\emph{Intelligent Virtual Agents: 10th International Conference, IVA 2010, Philadelphia, PA, USA, September 20-22, 2010. Proceedings 10}}. Springer, \bibinfo{pages}{343--349}.
\newblock


\bibitem[Zalake et~al\mbox{.}(2021a)]%
        {mohan}
\bibfield{author}{\bibinfo{person}{Mohan Zalake}, \bibinfo{person}{Alexandre~Gomes de Siqueira}, \bibinfo{person}{Krishna Vaddiparti}, \bibinfo{person}{Pavlo Antonenko}, {and} \bibinfo{person}{Benjamin Lok}.} \bibinfo{year}{2021}\natexlab{a}.
\newblock \showarticletitle{Towards Understanding How Virtual Human's Verbal Persuasion Strategies Influence User Intentions To Perform Health Behavior}. In \bibinfo{booktitle}{\emph{Proceedings of the 21st ACM International Conference on Intelligent Virtual Agents}}. \bibinfo{pages}{216--223}.
\newblock


\bibitem[Zalake et~al\mbox{.}(2021b)]%
        {31}
\bibfield{author}{\bibinfo{person}{Mohan Zalake}, \bibinfo{person}{Fatemeh Tavassoli}, \bibinfo{person}{Kyle Duke}, \bibinfo{person}{Thomas George}, \bibinfo{person}{Francois Modave}, \bibinfo{person}{Jordan Neil}, \bibinfo{person}{Janice Krieger}, {and} \bibinfo{person}{Benjamin Lok}.} \bibinfo{year}{2021}\natexlab{b}.
\newblock \showarticletitle{Internet-based tailored virtual human health intervention to promote colorectal cancer screening: design guidelines from two user studies}.
\newblock \bibinfo{journal}{\emph{Journal on Multimodal User Interfaces}} \bibinfo{volume}{15}, \bibinfo{number}{2} (\bibinfo{year}{2021}), \bibinfo{pages}{147--162}.
\newblock


\bibitem[Zane and Ku(2014)]%
        {74}
\bibfield{author}{\bibinfo{person}{Nolan Zane} {and} \bibinfo{person}{Helen Ku}.} \bibinfo{year}{2014}\natexlab{}.
\newblock \showarticletitle{Effects of ethnic match, gender match, acculturation, cultural identity, and face concern on self-disclosure in counseling for Asian Americans.}
\newblock \bibinfo{journal}{\emph{Asian American Journal of Psychology}} \bibinfo{volume}{5}, \bibinfo{number}{1} (\bibinfo{year}{2014}), \bibinfo{pages}{66}.
\newblock


\bibitem[Zlotnick et~al\mbox{.}(1998)]%
        {zlotnick1998}
\bibfield{author}{\bibinfo{person}{Caron Zlotnick}, \bibinfo{person}{Irene Elkin}, {and} \bibinfo{person}{M~Tracie Shea}.} \bibinfo{year}{1998}\natexlab{}.
\newblock \showarticletitle{Does the gender of a patient or the gender of a therapist affect the treatment of patients with major depression?}
\newblock \bibinfo{journal}{\emph{Journal of Consulting and Clinical Psychology}} \bibinfo{volume}{66}, \bibinfo{number}{4} (\bibinfo{year}{1998}), \bibinfo{pages}{655}.
\newblock


\end{thebibliography}

\newpage
\appendix
\section{Part 1: Additional Tables}\label{study1:additional}

\begin{table}[H]\centering
\caption{\centering\label{W3:Places}Part 1: Participants' Countries of Origin}
\begin{tabular}{lclc}
\hline
\multicolumn{4}{|c|}{\textit{\textbf{Participants' Countries of Origin}}} \\ \hline
\multicolumn{1}{|l|}{\textit{\textbf{Country}}} & \multicolumn{1}{l|}{\textit{\textbf{Participants}}} & \multicolumn{1}{l|}{\textit{\textbf{Country}}} & \multicolumn{1}{l|}{\textit{\textbf{Participants}}} \\ \hline
\multicolumn{1}{|l|}{\textit{ALBANIA}} & \multicolumn{1}{c|}{\textit{1}} & \multicolumn{1}{l|}{\textit{KOREA}} & \multicolumn{1}{c|}{\textit{2}} \\ \hline
\multicolumn{1}{|l|}{\textit{ARGENTINA}} & \multicolumn{1}{c|}{\textit{1}} & \multicolumn{1}{l|}{\textit{MEXICO}} & \multicolumn{1}{c|}{\textit{1}} \\ \hline
\multicolumn{1}{|l|}{\textit{BANGLADESH}} & \multicolumn{1}{c|}{\textit{1}} & \multicolumn{1}{l|}{\textit{NETHERLANDS}} & \multicolumn{1}{c|}{\textit{1}} \\ \hline
\multicolumn{1}{|l|}{\textit{BARBADOS}} & \multicolumn{1}{c|}{\textit{1}} & \multicolumn{1}{l|}{\textit{PERU}} & \multicolumn{1}{c|}{\textit{1}} \\ \hline
\multicolumn{1}{|l|}{\textit{BRAZIL}} & \multicolumn{1}{c|}{\textit{7}} & \multicolumn{1}{l|}{\textit{PHILIPPINES}} & \multicolumn{1}{c|}{\textit{3}} \\ \hline
\multicolumn{1}{|l|}{\textit{CANADA}} & \multicolumn{1}{c|}{\textit{1}} & \multicolumn{1}{l|}{\textit{POLAND}} & \multicolumn{1}{c|}{\textit{1}} \\ \hline
\multicolumn{1}{|l|}{\textit{CHINA}} & \multicolumn{1}{c|}{\textit{16}} & \multicolumn{1}{l|}{\textit{PUERTO   RICO}} & \multicolumn{1}{c|}{\textit{1}} \\ \hline
\multicolumn{1}{|l|}{\textit{COLOMBIA}} & \multicolumn{1}{c|}{\textit{2}} & \multicolumn{1}{l|}{\textit{SOUTH   KOREA}} & \multicolumn{1}{c|}{\textit{1}} \\ \hline
\multicolumn{1}{|l|}{\textit{COSTA   RICA}} & \multicolumn{1}{c|}{\textit{1}} & \multicolumn{1}{l|}{\textit{SPAIN}} & \multicolumn{1}{c|}{\textit{2}} \\ \hline
\multicolumn{1}{|l|}{\textit{CUBA}} & \multicolumn{1}{c|}{\textit{6}} & \multicolumn{1}{l|}{\textit{TAIWAN}} & \multicolumn{1}{c|}{\textit{1}} \\ \hline
\multicolumn{1}{|l|}{\textit{DOMINICAN   REPUBLIC}} & \multicolumn{1}{c|}{\textit{1}} & \multicolumn{1}{l|}{\textit{TANZANIA}} & \multicolumn{1}{c|}{\textit{1}} \\ \hline
\multicolumn{1}{|l|}{\textit{EGYPT}} & \multicolumn{1}{c|}{\textit{1}} & \multicolumn{1}{l|}{\textit{THAILAND}} & \multicolumn{1}{c|}{\textit{1}} \\ \hline
\multicolumn{1}{|l|}{\textit{GERMANY}} & \multicolumn{1}{c|}{\textit{1}} & \multicolumn{1}{l|}{\textit{TURKEY}} & \multicolumn{1}{c|}{\textit{1}} \\ \hline
\multicolumn{1}{|l|}{\textit{INDIA}} & \multicolumn{1}{c|}{\textit{57}} & \multicolumn{1}{l|}{\textit{UNITED   KINGDOM}} & \multicolumn{1}{c|}{\textit{1}} \\ \hline
\multicolumn{1}{|l|}{\textit{INDONESIA}} & \multicolumn{1}{c|}{\textit{1}} & \multicolumn{1}{l|}{\textit{UNITED STATES}} & \multicolumn{1}{c|}{\textit{354}} \\ \hline
\multicolumn{1}{|l|}{\textit{JAMAICA}} & \multicolumn{1}{c|}{\textit{1}} & \multicolumn{1}{l|}{\textit{VENEZUELA}} & \multicolumn{1}{c|}{\textit{7}} \\ \hline
\multicolumn{1}{|l|}{\textit{JAPAN}} & \multicolumn{1}{c|}{\textit{1}} & \multicolumn{1}{l|}{\textit{VIETNAM}} & \multicolumn{1}{c|}{\textit{2}} \\ \hline
\textit{} & \textit{} &  & \multicolumn{1}{l}{}
\end{tabular}
\end{table}

\begin{table}[H]\centering
\caption{\centering\label{W3:ethnicitiesAndGenders}Part 1: Participants' Preferences on Gender Based on Ethnic Groups}
\begin{tabular}{|l|l|ccc|}
\hline
\multicolumn{1}{|c|}{\multirow{2}{*}{\textit{\textbf{Ethnic Group}}}} & \multicolumn{1}{c|}{\multirow{2}{*}{\textit{\textbf{Total}}}} & \multicolumn{3}{c|}{\textit{\textbf{Preferred Gender}}} \\ \cline{3-5} 
\multicolumn{1}{|c|}{} & \multicolumn{1}{c|}{} & \multicolumn{1}{c|}{\textit{\textbf{Female}}} & \multicolumn{1}{c|}{\textit{\textbf{Male}}} & \textit{\textbf{Other}} \\ \hline
\textit{Asian} & \textit{n=217} & \multicolumn{1}{c|}{\textit{57.6\%}} & \multicolumn{1}{c|}{\textit{35.5\%}} & \textit{6.9\%} \\ \hline
\textit{Afro/Black American} & \textit{n=28} & \multicolumn{1}{c|}{\textit{53.6\%}} & \multicolumn{1}{c|}{\textit{32.1\%}} & \textit{14.3\%} \\ \hline
\textit{Latin American} & \textit{n=100} & \multicolumn{1}{c|}{\textit{56.0\%}} & \multicolumn{1}{c|}{\textit{32.0\%}} & \textit{12.0\%} \\ \hline
\textit{Middle Eastern/North African} & \textit{n=14} & \multicolumn{1}{c|}{\textit{64.3\%}} & \multicolumn{1}{c|}{\textit{21.4\%}} & \textit{14.3\%} \\ \hline
\textit{Native American/Alaska Native} & \textit{n=2} & \multicolumn{1}{c|}{\textit{50.0\%}} & \multicolumn{1}{c|}{\textit{50.0\%}} & \textit{0.0\%} \\ \hline
\textit{Pacific Islander} & \textit{n=8} & \multicolumn{1}{c|}{\textit{62.5\%}} & \multicolumn{1}{c|}{\textit{25.0\%}} & \textit{12.5\%} \\ \hline
\textit{White (non-Latin American)} & \textit{n=202} & \multicolumn{1}{c|}{\textit{45.0\%}} & \multicolumn{1}{c|}{\textit{41.1\%}} & \textit{13.9\%} \\ \hline
\end{tabular}
\end{table}

\begin{table}[H]\centering
\caption{\centering\label{appearance:Cohen}Part 1: Inter-rater Reliability Scores for Female and Male Appearance Selection}
\begin{tabular}{|c|l|c|c|}
\hline
\multirow{2}{*}{\textit{\textbf{ID}}} & \multicolumn{1}{c|}{\multirow{2}{*}{\textit{\textbf{Theme}}}} & \textit{\textbf{\begin{tabular}[c]{@{}c@{}}Male Appearances\\      (Jaccard Index= 0.78)\end{tabular}}} & \textit{\textbf{\begin{tabular}[c]{@{}c@{}}Female Appearances\\      (Jaccard Index= 0.81)\end{tabular}}} \\ \cline{3-4} 
 & \multicolumn{1}{c|}{} & \textit{\textbf{Cohen's kappa}} & \textit{\textbf{Cohen's kappa}} \\ \hline
\textit{A1} & \textit{Age} & \textit{0.99} & \textit{0.98} \\ \hline
\textit{A2} & \textit{Approachability} & \textit{0.91} & \textit{0.88} \\ \hline
\textit{A3} & \textit{Attire} & \textit{0.92} & \textit{0.98} \\ \hline
\textit{A4} & \textit{Ethnicity} & \textit{0.98} & \textit{1.00} \\ \hline
\textit{A5} & \textit{Expertise} & \textit{0.94} & \textit{0.89} \\ \hline
\textit{A6} & \textit{Gender} & \textit{0.98} & \textit{1.00} \\ \hline
\textit{A7} & \textit{Non-verbal Language} & \textit{0.92} & \textit{0.95} \\ \hline
\textit{A8} & \textit{Random} & \textit{0.96} & \textit{0.98} \\ \hline
\textit{A9} & \textit{Self-similarity} & \textit{0.91} & \textit{0.91} \\ \hline
\textit{A10} & \textit{Similarity to Trusted Agent} & \textit{0.97} & \textit{0.96} \\ \hline
\end{tabular}
\end{table}

\begin{table}[H]\centering
\caption{\centering\label{table:appearanceThemesPerEthnicity}Part 1: Thematic Analysis per Ethnic Group for Male Appearances' Selection}
\begin{tabular}{|c|c|c|c|c|c|c|c|c|}
\hline
\textit{\textbf{ID}} & \textit{\textbf{Count}} & \textit{\textbf{\begin{tabular}[c]{@{}c@{}}Asian\\      (n=217)\end{tabular}}} & \textit{\textbf{\begin{tabular}[c]{@{}c@{}}Afro/Black\\      American\\      (n=28)\end{tabular}}} & \textit{\textbf{\begin{tabular}[c]{@{}c@{}}Latin \\ American\\      (n=100)\end{tabular}}} & \textit{\textbf{\begin{tabular}[c]{@{}c@{}}Middle \\ Eastern/\\      North \\ African\\      (n=14)\end{tabular}}} & \textit{\textbf{\begin{tabular}[c]{@{}c@{}}Native \\ American/\\      Alaska \\ Native\\      (n=2)\end{tabular}}} & \textit{\textbf{\begin{tabular}[c]{@{}c@{}}Pacific\\      Islander\\      (n=8)\end{tabular}}} & \textit{\textbf{\begin{tabular}[c]{@{}c@{}}White\\      (n=202)\end{tabular}}} \\ \hline
\textit{A1} & \textit{262} & \textit{54.8\%} & \textit{46.4\%} & \textit{48.0\%} & \textit{64.3\%} & \textit{0.0\%} & \textit{62.5\%} & \textit{55.9\%} \\ \hline
\textit{A2} & \textit{264} & \textit{48.4\%} & \textit{39.3\%} & \textit{59.0\%} & \textit{71.4\%} & \textit{50.0\%} & \textit{37.5\%} & \textit{58.9\%} \\ \hline
\textit{A3} & \textit{75} & \textit{15.2\%} & \textit{17.9\%} & \textit{15.0\%} & \textit{21.4\%} & \textit{0.0\%} & \textit{12.5\%} & \textit{18.3\%} \\ \hline
\textit{A4} & \textit{46} & \textit{16.1\%} & \textit{10.7\%} & \textit{7.0\%} & \textit{0.0\%} & \textit{0.0\%} & \textit{12.5\%} & \textit{2.5\%} \\ \hline
\textit{A5} & \textit{110} & \textit{20.7\%} & \textit{17.9\%} & \textit{24.0\%} & \textit{21.4\%} & \textit{100.0\%} & \textit{12.5\%} & \textit{25.7\%} \\ \hline
\textit{A6} & \textit{13} & \textit{1.8\%} & \textit{3.6\%} & \textit{2.0\%} & \textit{7.1\%} & \textit{0.0\%} & \textit{12.5\%} & \textit{3.5\%} \\ \hline
\textit{A7} & \textit{74} & \textit{15.2\%} & \textit{10.7\%} & \textit{11.0\%} & \textit{42.9\%} & \textit{0.0\%} & \textit{12.5\%} & \textit{15.8\%} \\ \hline
\textit{A8} & \textit{8} & \textit{0.9\%} & \textit{0.0\%} & \textit{4.0\%} & \textit{0.0\%} & \textit{0.0\%} & \textit{0.0\%} & \textit{1.5\%} \\ \hline
\textit{A9} & \textit{119} & \textit{27.2\%} & \textit{32.1\%} & \textit{29.0\%} & \textit{21.4\%} & \textit{0.0\%} & \textit{62.5\%} & \textit{20.8\%} \\ \hline
\textit{A10} & \textit{70} & \textit{14.3\%} & \textit{25.0\%} & \textit{14.0\%} & \textit{7.1\%} & \textit{50.0\%} & \textit{25.0\%} & \textit{13.4\%} \\ \hline
\end{tabular}
\end{table}

\begin{table}[H]\centering
\caption{\centering\label{table:femaleAppearanceThemesPerEthnicity}Part 1: Thematic Analysis per Ethnic Group for Female Appearances' Selection}
\begin{tabular}{|c|l|c|c|c|c|c|c|c|}
\hline
\textit{\textbf{ID}} & \multicolumn{1}{c|}{\textit{\textbf{Count}}} & \textit{\textbf{\begin{tabular}[c]{@{}c@{}}Asian\\      (n=217)\end{tabular}}} & \textit{\textbf{\begin{tabular}[c]{@{}c@{}}Afro/Black\\      American\\      (n=28)\end{tabular}}} & \textit{\textbf{\begin{tabular}[c]{@{}c@{}}Latin \\ American\\      (n=100)\end{tabular}}} & \textit{\textbf{\begin{tabular}[c]{@{}c@{}}Middle \\ Eastern/\\      North \\ African\\      (n=14)\end{tabular}}} & \textit{\textbf{\begin{tabular}[c]{@{}c@{}}Native \\ American/\\      Alaska \\ Native\\      (n=2)\end{tabular}}} & \textit{\textbf{\begin{tabular}[c]{@{}c@{}}Pacific\\      Islander\\      (n=8)\end{tabular}}} & \textit{\textbf{\begin{tabular}[c]{@{}c@{}}White\\      (n=202)\end{tabular}}} \\ \hline
\textit{A1} & \textit{254} & \textit{51.2\%} & \textit{46.4\%} & \textit{53.0\%} & \textit{57.1\%} & \textit{0.0\%} & \textit{87.5\%} & \textit{54.0\%} \\ \hline
\textit{A2} & \textit{355} & \textit{61.8\%} & \textit{50.0\%} & \textit{67.0\%} & \textit{71.4\%} & \textit{100.0\%} & \textit{50.0\%} & \textit{61.4\%} \\ \hline
\textit{A3} & \textit{54} & \textit{8.3\%} & \textit{10.7\%} & \textit{6.0\%} & \textit{14.3\%} & \textit{0.0\%} & \textit{0.0\%} & \textit{12.4\%} \\ \hline
\textit{A4} & \textit{30} & \textit{8.3\%} & \textit{3.6\%} & \textit{10.0\%} & \textit{0.0\%} & \textit{0.0\%} & \textit{0.0\%} & \textit{0.5\%} \\ \hline
\textit{A5} & \textit{141} & \textit{27.6\%} & \textit{32.1\%} & \textit{22.0\%} & \textit{7.1\%} & \textit{50.0\%} & \textit{25.0\%} & \textit{22.8\%} \\ \hline
\textit{A6} & \textit{11} & \textit{2.8\%} & \textit{0.0\%} & \textit{0.0\%} & \textit{7.1\%} & \textit{0.0\%} & \textit{0.0\%} & \textit{2.0\%} \\ \hline
\textit{A7} & \textit{117} & \textit{21.2\%} & \textit{17.9\%} & \textit{19.0\%} & \textit{35.7\%} & \textit{0.0\%} & \textit{25.0\%} & \textit{19.8\%} \\ \hline
\textit{A8} & \textit{16} & \textit{1.4\%} & \textit{7.1\%} & \textit{4.0\%} & \textit{0.0\%} & \textit{0.0\%} & \textit{12.5\%} & \textit{3.0\%} \\ \hline
\textit{A9} & \textit{122} & \textit{20.3\%} & \textit{21.4\%} & \textit{25.0\%} & \textit{28.6\%} & \textit{0.0\%} & \textit{25.0\%} & \textit{20.3\%} \\ \hline
\textit{A10} & \textit{91} & \textit{14.7\%} & \textit{10.7\%} & \textit{20.0\%} & \textit{7.1\%} & \textit{50.0\%} & \textit{37.5\%} & \textit{15.3\%} \\ \hline
\end{tabular}
\end{table}

\begin{table}[H]
\centering
\caption{\centering\label{voice:Cohen}Part 1: Inter-rater Reliability Scores for Voice Selection}
\begin{tabular}{|c|l|c|c|}
\hline
\multirow{2}{*}{\textit{\textbf{ID}}} & \multicolumn{1}{c|}{\multirow{2}{*}{\textit{\textbf{Theme}}}} & \textbf{\begin{tabular}[c]{@{}c@{}}Male Voices\\ (Jaccard Index=0.73)\end{tabular}} & \textit{\textbf{\begin{tabular}[c]{@{}c@{}}Female Voices\\ (Jaccard Index=0.75)\end{tabular}}} \\ \cline{3-4} 
 & \multicolumn{1}{c|}{} & \textbf{Cohen's kappa} & \textbf{Cohen's kappa} \\ \hline
\textit{V1} & \textit{Clarity} & 0.86 & \textit{0.92} \\ \hline
\textit{V2} & \textit{Comfort} & 0.86 & \textit{0.86} \\ \hline
\textit{V3} & \textit{Ethnicity} & 0.84 & \textit{0.84} \\ \hline
\textit{V4} & \textit{Familiarity} & 0.97 & \textit{0.95} \\ \hline
\textit{V5} & \textit{Matching} & 0.93 & \textit{0.90} \\ \hline
\textit{V6} & \textit{Random} & 0.98 & \textit{0.97} \\ \hline
\textit{V7} & \textit{Self-similarity} & 0.94 & \textit{0.96} \\ \hline
\end{tabular}
\end{table}

\begin{table}[H] \centering
\caption{\centering \label{W3:maleVoicesTable}Part 1: Male Voices Preferences}
\begin{tabular}{|lllcc|}
\hline
\multicolumn{5}{|c|}{\textit{\textbf{Male Voices}}} \\ \hline
\multicolumn{1}{|l|}{\textit{\textbf{ID}}} & \multicolumn{1}{l|}{\textit{\textbf{Name}}} & \multicolumn{1}{l|}{\textit{\textbf{Accent Origin}}} & \multicolumn{1}{l|}{\textit{\textbf{\begin{tabular}[c]{@{}l@{}}\% participants selecting \\ it in their top four\end{tabular}}}} & \multicolumn{1}{l|}{\textit{\textbf{\begin{tabular}[c]{@{}l@{}}\% participants selecting \\ it as their top one\end{tabular}}}} \\ \hline
\multicolumn{1}{|l|}{\textit{AU\_M}} & \multicolumn{1}{l|}{\textit{AU Natural}} & \multicolumn{1}{l|}{\textit{Australia}} & \multicolumn{1}{c|}{\textit{49.9\%}} & \textit{12.1\%} \\ \hline
\multicolumn{1}{|l|}{\textit{CA\_M}} & \multicolumn{1}{l|}{\textit{CA Natural}} & \multicolumn{1}{l|}{\textit{Canada}} & \multicolumn{1}{c|}{\textit{62.0\%}} & \textit{21.2\%} \\ \hline
\multicolumn{1}{|l|}{\textit{IN\_M}} & \multicolumn{1}{l|}{\textit{IN Natural 4}} & \multicolumn{1}{l|}{\textit{India}} & \multicolumn{1}{c|}{\textit{19.5\%}} & \textit{1.2\%} \\ \hline
\multicolumn{1}{|l|}{\textit{IE\_M}} & \multicolumn{1}{l|}{\textit{IE Natural}} & \multicolumn{1}{l|}{\textit{Ireland}} & \multicolumn{1}{c|}{\textit{32.6\%}} & \textit{5.8\%} \\ \hline
\multicolumn{1}{|l|}{\textit{KE\_M}} & \multicolumn{1}{l|}{\textit{KE Natural}} & \multicolumn{1}{l|}{\textit{Kenya}} & \multicolumn{1}{c|}{\textit{21.0\%}} & \textit{3.1\%} \\ \hline
\multicolumn{1}{|l|}{\textit{NZ\_M}} & \multicolumn{1}{l|}{\textit{NZ Natural}} & \multicolumn{1}{l|}{\textit{New Zealand}} & \multicolumn{1}{c|}{\textit{25.2\%}} & \textit{4.4\%} \\ \hline
\multicolumn{1}{|l|}{\textit{NG\_M}} & \multicolumn{1}{l|}{\textit{NG Natural}} & \multicolumn{1}{l|}{\textit{Nigeria}} & \multicolumn{1}{c|}{\textit{19.3\%}} & \textit{2.5\%} \\ \hline
\multicolumn{1}{|l|}{\textit{PH\_M}} & \multicolumn{1}{l|}{\textit{PH Natural}} & \multicolumn{1}{l|}{\textit{Philippines}} & \multicolumn{1}{c|}{\textit{33.9\%}} & \textit{14.6\%} \\ \hline
\multicolumn{1}{|l|}{\textit{ZA\_M}} & \multicolumn{1}{l|}{\textit{ZA Natural}} & \multicolumn{1}{l|}{\textit{South Africa}} & \multicolumn{1}{c|}{\textit{16.4\%}} & \textit{3.3\%} \\ \hline
\multicolumn{1}{|l|}{\textit{TZ\_M}} & \multicolumn{1}{l|}{\textit{TZ Natural}} & \multicolumn{1}{l|}{\textit{Tanzania}} & \multicolumn{1}{c|}{\textit{14.1\%}} & \textit{0.6\%} \\ \hline
\multicolumn{1}{|l|}{\textit{GB\_M}} & \multicolumn{1}{l|}{\textit{GB Natural}} & \multicolumn{1}{l|}{\textit{United Kingdom}} & \multicolumn{1}{c|}{\textit{55.1\%}} & \textit{9.1\%} \\ \hline
\multicolumn{1}{|l|}{\textit{US\_M}} & \multicolumn{1}{l|}{\textit{US Natural}} & \multicolumn{1}{l|}{\textit{United States}} & \multicolumn{1}{c|}{\textit{55.9\%}} & \textit{23.3\%} \\ \hline
\end{tabular}
\end{table}

\begin{table}[H] \centering
\caption{\centering \label{W3:femaleVoicesTable}Female Voices Preferences}
\begin{tabular}{|lllcc|}
\hline
\multicolumn{5}{|c|}{\textit{\textbf{Part 1: Female Voices}}} \\ \hline
\multicolumn{1}{|l|}{\textit{\textbf{ID}}} & \multicolumn{1}{l|}{\textit{\textbf{Name}}} & \multicolumn{1}{l|}{\textit{\textbf{Accent Origin}}} & \multicolumn{1}{l|}{\textit{\textbf{\begin{tabular}[c]{@{}l@{}}\% participants selecting \\ it in their top four\end{tabular}}}} & \multicolumn{1}{l|}{\textit{\textbf{\begin{tabular}[c]{@{}l@{}}\% participants selecting \\ it as their top one\end{tabular}}}} \\ \hline
\multicolumn{1}{|l|}{\textit{AU\_F}} & \multicolumn{1}{l|}{\textit{AU Natural}} & \multicolumn{1}{l|}{\textit{Australia}} & \multicolumn{1}{c|}{\textit{39.7\%}} & \textit{6.0\%} \\ \hline
\multicolumn{1}{|l|}{\textit{CA\_F}} & \multicolumn{1}{l|}{\textit{CA Natural}} & \multicolumn{1}{l|}{\textit{Canada}} & \multicolumn{1}{c|}{\textit{52.2\%}} & \textit{9.4\%} \\ \hline
\multicolumn{1}{|l|}{\textit{IN\_F}} & \multicolumn{1}{l|}{\textit{IN Natural}} & \multicolumn{1}{l|}{\textit{India}} & \multicolumn{1}{c|}{\textit{27.9\%}} & \textit{5.6\%} \\ \hline
\multicolumn{1}{|l|}{\textit{IE\_F}} & \multicolumn{1}{l|}{\textit{IE Natural}} & \multicolumn{1}{l|}{\textit{Ireland}} & \multicolumn{1}{c|}{\textit{49.9\%}} & \textit{18.7\%} \\ \hline
\multicolumn{1}{|l|}{\textit{KE\_F}} & \multicolumn{1}{l|}{\textit{KE Natural}} & \multicolumn{1}{l|}{\textit{Kenya}} & \multicolumn{1}{c|}{\textit{18.3\%}} & \textit{1.7\%} \\ \hline
\multicolumn{1}{|l|}{\textit{NZ\_F}} & \multicolumn{1}{l|}{\textit{NZ Natural}} & \multicolumn{1}{l|}{\textit{New Zealand}} & \multicolumn{1}{c|}{\textit{24.7\%}} & \textit{4.0\%} \\ \hline
\multicolumn{1}{|l|}{\textit{NG\_F}} & \multicolumn{1}{l|}{\textit{NG Natural}} & \multicolumn{1}{l|}{\textit{Nigeria}} & \multicolumn{1}{c|}{\textit{21.2\%}} & \textit{2.3\%} \\ \hline
\multicolumn{1}{|l|}{\textit{PH\_F}} & \multicolumn{1}{l|}{\textit{PH Natural}} & \multicolumn{1}{l|}{\textit{Philippines}} & \multicolumn{1}{c|}{\textit{38.5\%}} & \textit{15.6\%} \\ \hline
\multicolumn{1}{|l|}{\textit{ZA\_F}} & \multicolumn{1}{l|}{\textit{ZA Natural}} & \multicolumn{1}{l|}{\textit{South Africa}} & \multicolumn{1}{c|}{\textit{26.8\%}} & \textit{3.7\%} \\ \hline
\multicolumn{1}{|l|}{\textit{TZ\_F}} & \multicolumn{1}{l|}{\textit{TZ Natural}} & \multicolumn{1}{l|}{\textit{Tanzania}} & \multicolumn{1}{c|}{\textit{12.3\%}} & \textit{1.2\%} \\ \hline
\multicolumn{1}{|l|}{\textit{GB\_F}} & \multicolumn{1}{l|}{\textit{GB Natural}} & \multicolumn{1}{l|}{\textit{United Kingdom}} & \multicolumn{1}{c|}{\textit{41.4\%}} & \textit{7.5\%} \\ \hline
\multicolumn{1}{|l|}{\textit{US\_F}} & \multicolumn{1}{l|}{\textit{US Natural}} & \multicolumn{1}{l|}{\textit{United States}} & \multicolumn{1}{c|}{\textit{60.5\%}} & \textit{27.2\%} \\ \hline
\end{tabular}
\end{table}

\section{Part 2: Agent Script for the Explicit Group}\label{study2:implicitScript}
\textit{``Hello and welcome. I am a mental wellness virtual counselor at the University of Florida. I want to share with you the power of gratitude in promoting mental health and life satisfaction.}

\textit{Gratitude as a discipline helps protect us from destructive feelings such as envy, resentment, greed, and bitterness. One of the best ways to cultivate gratitude is to establish a daily practice, gratitude, journaling. When we are grateful, we affirm that sources of goodness exist in our lives. By writing each day, we magnify and expand on these sources of goodness.
Setting aside time on a daily basis to recall moments of gratitude associated with even mundane or ordinary events. Personal attributes one has or valued people, one encounters has the potential to weave together a sustainable life theme of gratefulness. Just as it nourishes a fundamentally affirming life stance. Generally, gratitude, journaling needs 15 minutes per day, at least three times per week for at least two weeks.}

\textit{There is no wrong way to keep a gratitude journal. But here are some general instructions that can help you get started, write down up to five things for which you feel grateful. The physical record is important, not just in your head, the things you list can be relatively small in importance. The goal of the exercise is to remember a good event, experience, person or thing in your life as you write down in the journal consider eight important tips.}

\textit{First, be specific as possible. Specificity is a key to fostering gratitude. Second, go for depth over breadth, elaborate in detail about a particular person or thing for which you are grateful. Third, get personal, focusing on people to whom you are grateful has more of an impact than focusing on things for which you are grateful. Fourth try, subtraction, not addition, consider what your life would be like without certain people or things rather than just tallying up all the good stuff. See good things as gifts, try to relish and savor the gifts you have received. Fifth savor surprises, try to record events that were unexpected or surprising. Sixth revise, if you repeat writing about some of the same people and things is ok. But think aspects that appear too finally write regularly, whether you write daily or every other day, commit to a regular time to journal. }

\textit{I hope you find gratitude. Journaling useful. As you may see, it is not a time consuming task and it brings with it many benefits. To your mental health."}

\section{Part 2: Agent Script for the Control and Implicit Groups}\label{study2:explicitScript}
\textit{``Hello and welcome. I am a mental wellness virtual counselor at the University of Florida. \textbf{I was designed by a fellow college student who shares similar characteristics to yours. She reported being multiracial with Latina and White roots and is also from Gainesville. She is 20 years old.} I want to share with you the power of gratitude in promoting mental health and life satisfaction.}

\textit{Gratitude as a discipline helps protect us from destructive feelings such as envy, resentment, greed, and bitterness. One of the best ways to cultivate gratitude is to establish a daily practice, gratitude, journaling. When we are grateful, we affirm that sources of goodness exist in our lives. By writing each day, we magnify and expand on these sources of goodness.
Setting aside time on a daily basis to recall moments of gratitude associated with even mundane or ordinary events. Personal attributes one has or valued people, one encounters has the potential to weave together a sustainable life theme of gratefulness. Just as it nourishes a fundamentally affirming life stance. Generally, gratitude, journaling needs 15 minutes per day, at least three times per week for at least two weeks.}

\textit{There is no wrong way to keep a gratitude journal. But here are some general instructions that can help you get started, write down up to five things for which you feel grateful. The physical record is important, not just in your head, the things you list can be relatively small in importance. The goal of the exercise is to remember a good event, experience, person or thing in your life as you write down in the journal consider eight important tips.}

\textit{First, be specific as possible. Specificity is a key to fostering gratitude. Second, go for depth over breadth, elaborate in detail about a particular person or thing for which you are grateful. Third, get personal, focusing on people to whom you are grateful has more of an impact than focusing on things for which you are grateful. Fourth try, subtraction, not addition, consider what your life would be like without certain people or things rather than just tallying up all the good stuff. See good things as gifts, try to relish and savor the gifts you have received. Fifth savor surprises, try to record events that were unexpected or surprising. Sixth revise, if you repeat writing about some of the same people and things is ok. But think aspects that appear too finally write regularly, whether you write daily or every other day, commit to a regular time to journal. }

\textit{I hope you find gratitude. Journaling useful. As you may see, it is not a time consuming task and it brings with it many benefits. To your mental health."}

\end{document}